\documentclass[11pt]{amsart}%
\usepackage{amsmath}
\usepackage{amsfonts}
\usepackage{amssymb}
\usepackage{graphicx}%
\providecommand{\U}[1]{\protect\rule{.1in}{.1in}}
\numberwithin{equation}{section}
\newtheorem{theorem}{Theorem}
\newtheorem{acknowledgement}[theorem]{Acknowledgement}

\email{chams@aub.edu.lb}
\email{alain@connes.org}
\begin{document}

\begin{titlepage}
\vspace{.3cm} \vspace{1cm}
\begin{center}
\baselineskip=16pt \centerline{\Large\bf Noncommutative Geometry as a }
\vspace{.5cm}
\centerline{\Large\bf Framework for Unification of all Fundamental }
\vspace{.5cm}
\centerline{\Large\bf Interactions including Gravity. Part I. }

\vspace{1truecm} \centerline{\large\bf Ali H.
Chamseddine$^{1,3}$\ , \ Alain Connes$^{2,3,4}$\ \ } \vspace{.5truecm}
\emph{\centerline{$^{1}$Physics Department, American University of Beirut, Lebanon}}
\emph{\centerline{$^{2}$College de France, 3 rue Ulm, F75005, Paris, France}}
\emph{\centerline{$^{3}$I.H.E.S. F-91440 Bures-sur-Yvette, France}}
\emph{\centerline{$^{4}$Department of Mathematics, Vanderbilt University, Nashville, TN 37240 USA}}
\end{center}
\vspace{1cm}
\begin{center}
{\bf Abstract}
\end{center}
We examine the hypothesis that space-time is a product of a continuous four-dimensional manifold times a   finite
space. A new tensorial notation is developed to present the various
constructs of noncommutative geometry. In particular,  this notation is used to determine the spectral data of the
standard model. The particle spectrum with all of its symmetries is derived, almost uniquely, under the assumption of irreducibility and of dimension 6 modulo 8 for the finite space.
The reduction from the natural symmetry group $SU(2)\times SU(2)\times SU(4)$ to $U(1)\times SU(2)\times SU(3)$ is a consequence of the hypothesis that the two layers of space-time are finite distance apart but is non-dynamical. The square of the Dirac
operator, and all geometrical invariants that appear in the
calculation of the heat kernel expansion are evaluated. We re-derive the leading order terms in the spectral action.
The geometrical action yields unification of all fundamental interactions
including gravity at very high energies. We make the following predictions: (i) The number of fermions per family is 16.
(ii) The symmetry group is $U(1)\times SU(2)\times SU(3)$. (iii) There are quarks and leptons in the correct representations. (iv) There is
a doublet Higgs that breaks the electroweak symmetry to $U(1)$. (v) Top quark mass of 170-175 Gev.
(v) There is a right-handed neutrino with a see-saw
mechanism.
Moreover, the zeroth order spectral action obtained with a cut-off function is consistent with experimental
data up to few
percent. We discuss a number of open issues.  We prepare the ground for computing higher order corrections since the predicted mass
of the Higgs field is quite sensitive to the higher order corrections. We speculate on the nature of the noncommutative space at
Planckian energies and the possible role of the fundamental group for the problem of generations.
\end{titlepage}\bigskip

\tableofcontents

\section{Introduction}

One of the basic problems facing theoretical physics is to determine the
nature of space-time. This is intimately related to the problem of unifying
all the fundamental interactions including gravity, and thus is not
independent of solving the problem of quantum gravity. In a series of
papers we have
made important understanding uncovering a first approximation of the hidden
structure of space-time. Our assumption is that at
energies below the planck scale, space-time can be approximated as a product of
a continuous four-dimensional manifold by a finite space.  We were able to
show in \cite{cc5} that finite spaces  satisfying the axioms of
noncommutative geometry are severely restricted, and the corresponding
irreducible representations on Hilbert spaces can only have dimensions which
are the square of  integers, or the double of such a square. The second possibility
is the only one allowed when the finite space has dimension $6$ modulo $8$ (in the sense of
 $K$-theory or more pragmatically of the periodicity of Clifford algebras)  as imposed by the need to have the total dimension $2=4+6$ modulo $8$ in order to be able to write down the Fermionic part of the action. Together with the
restriction of imposing a unitary--symplectic structure and grading on the finite
noncommutative space, this singles out $4^{2}=16$ as the number of physical
fermions per generation. Then, in the same way as was shown in \cite{mc2}, this predicts the existence of right-handed neutrinos,   and the see-saw mechanism. Our present framework
 using the classification of finite spaces is stronger and the symmetries of
the standard model emerge, rather than assumed and put in by hand. This
construction, using the spectral action principle, predicts certain relations
between the coupling constants, that can only hold at very high energies of the order of the unification scale.
The spectral action principle is the simple statement that the physical action
is determined by the spectrum of the Dirac operator $D$. This has now been tested in many interesting models including Superstring theory \cite{chams}, noncommutative tori \cite{Ioch}, Moyal planes \cite{Gayral}, 4D-Moyal space
\cite{Wulk1}, manifolds with boundary \cite{ccboundary}, in the presence of dilatons \cite{cc3}, for supersymmetric models \cite{Broek}
and torsion cases \cite{Stef}. The additivity of the action forces it to be of the form
$
\mathrm{Trace}\,f\left(  D/\Lambda\right).  %
$
  In
the approximation where the spectral function $f$ is a cut-off function, the
relations given by the spectral action are used as boundary conditions and the couplings are then allowed to run from unification scale to low energy using the renormalization group equations.
 The equations show, when fitted to the low energy boundary conditions, that
the three gauge coupling constants and the Newton constant nearly meet (within
few percent) at very high energies, two or three orders from the Planck scale.
This might be a coincidence but it can also be an indication that a more fundamental theory exists at unification scale and manifests itself at low scale through integration of the
intermediate modes, as in the Wilson understanding of renormalization.

In Part II we shall investigate  higher order
terms in the perturbative expansion of the spectral function. We shall show
that these can give important contributions which effects the low-energy form
of the spectral action. A prediction of the Higgs mass is sensitive to these
higher order contributions.

In this paper we will present our analysis in a transparent setting, geared
towards physicists, spelling out the very few assumptions we make, and thus
allowing for an exhaustive treatment. This will help us to pave the road for
future investigations, and hopefully be of help for students to learn and apply this topic. We
shall follow a new and simple tensorial notation to allow physicists to follow
our analysis with ease. This will be true for most of the calculations,
although we will not re-derive some of the abstract proofs because in this
case the tensorial notation is not very practical. Armed with this
simplification in our analysis we will evaluate the spectral action rederiving
old results in a simple way. The calculation will be extended in Part II, to include
higher order terms in a perturbative expansion in function of the inverse of
the unification scale. At the end of this Part I, we shall discuss a number of important issues which are:
\begin{itemize}
  \item The variant of the
Einstein-Yang Mills system  obtained with the algebra $A_F=M_{2}\left(  \mathbb{H}%
\right)  \oplus M_{4}\left(  \mathbb{C}\right)$, its relation with supersymmetry, and with the unimodularity condition (\S \ref{conclu}).
  \item The geometric role of $M_4(\mathbb{C})$ (\S \ref{mfourc}).
  \item The possible geometric meaning of several generations (\S \ref{generations}).
  \item Unification of couplings (\S \ref{unif}).
  \item Mass of the Higgs (\S \ref{Higgs}).
  \item New particles (\S \ref{New}).
  \item Quantum Level (\S \ref{quant}).
\end{itemize}
In the appendices we develop the computational tools which will be used in Part II to handle the higher order terms. We also compute an explicit concrete example to check the  sign in front of the Yang-Mills interaction.

\bigskip

\section{Determining the finite noncommutative space}

We first give a very brief summary of the properties of noncommutative spaces.
The basic idea is based on physics. The modern way of measuring distances is
spectral. The unit of distance is taken as the wavelength of atomic spectra.
To adopt this geometrically we have to replace the notion of real variable
which one takes as a function $f$  on a set $X$, $f:X\rightarrow\mathbb{R}$
to be  given now by a self adjoint operator in a Hilbert space as in quantum
mechanics. The space $X$ is described by the algebra $\mathcal{A}$ of
coordinates which is represented as operators in a fixed Hilbert space
$\mathcal{H}$. There is no a priori requirement that this algebra $\mathcal{A}$
 is commutative since Hilbert space operators model perfectly the lack of commutativity.
 In fact if $\mathcal{A}$ is the algebra of functions on a space $X$ and one replaces it
 by the algebra $\mathcal{B}=M_n(\mathcal{A})$ of matrices of functions, one obtains that
 the natural gauge invariance group $\mathcal{G}$ of gravity coupled with an $SU(n)$ Yang-Mills theory on $X$, which is the semi-direct product of the group ${\rm Map}(X,SU(n))$ by the group of diffeomorphisms ${\rm Diff}(X)$,
 $$
1\rightarrow{\rm Map}(X,SU(n))\rightarrow \mathcal G\rightarrow
{\rm Diff}(X)\rightarrow 1 .
$$
is (locally) nothing else than the group of automorphisms of $\mathcal{B}$
$$
1\rightarrow {\rm Int}(\mathcal{B})\rightarrow {\rm Aut}(\mathcal{B})\rightarrow
{\rm Out}(\mathcal{B})\rightarrow 1
$$
It is rather satisfying that this completely general decomposition of automorphisms of a
noncommutative algebra into inner ones (forming the normal subgroup ${\rm Int}(\mathcal{B})$) and outer ones (forming the quotient group ${\rm Out}(\mathcal{B})$) corresponds in the above simplest example to the decomposition of the gauge symmetries in the internal ones ${\rm Map}(X,SU(n))$ and the group  ${\rm Diff}(X)$ of diffeomorphisms. We have shown in \cite{cc2}  that the study of pure gravity on the ``space" associated to the algebra $\mathcal{B}$   yields
Einstein gravity on $X$ minimally coupled with Yang-Mills theory for the gauge
group $SU(n)$. The Yang-Mills gauge potential appears as the inner part of the
metric, in the same way as the group of gauge transformations (for the gauge
group $SU(n)$) appears as the group of inner diffeomorphisms.

The meaning of ``pure gravity" in the general noncommutative framework comes from Dirac's solution of the extraction of the square root in Riemann's formula for the distance between two points
$$
d(a,b)=\,{\rm Inf}\,\int_\gamma\,\sqrt{g_{\mu\,\nu}\,dx^\mu\,dx^\nu}
$$
which can be reexpressed, in terms of Dirac's operator $D$ on the Hilbert space $\mathcal{H}$ of spinors, in the form
$$
d(a,b)=\,{\rm Sup}\, \{ \vert f(a) - f(b) \vert \,;\, f \in {\mathcal
A}\, , \ \Vert [D,f] \Vert \leq 1\,\}
$$
This shows that giving the Dirac operator acting in the same Hilbert space $\mathcal{H}$ as the algebra $\mathcal{A}$ of coordinates provides an elegant way of giving the geometry of the associated space $X$. Moreover this way immediately permits the passage to noncommutative algebras. Thus the
  geometry of a noncommutative space is determined in
terms of the spectral data $\left(  {\mathcal{A},\mathcal{H},D,J,\gamma
}\right)  $, where the last two: $J,\gamma$ should be considered as decorations on the main structure, encoded by the spectral triple $\left(  \mathcal{A},\mathcal{H},D\right)  $. In practice  a {\emph{real, even spectral triple} } is defined by

\begin{itemize}
\item {$\mathcal{A}$ an associative algebra with unit 1 and involution {$\ast
$}.}

\item {$\mathcal{H}$ is a complex Hilbert space carrying a faithful
representation {$\pi$} of the algebra.}

\item $D${ is a self-adjoint operator on {$\mathcal{H}$} with the resolvent
$\left(  D-\lambda\right)  ^{-1},\lambda\notin\mathbb{R}$ of $D$ compact.}

\item {$J$ \ is an anti--unitary operator on {$\mathcal{H}$}, a real structure
(charge conjugation.)}

\item {$\gamma$ is a unitary operator on {$\mathcal{H}$}, the chirality.}
\end{itemize}

We require the following conditions to hold:

\begin{itemize}
\item {$J^{\,2}=\epsilon$} , ($\varepsilon=1$ in zero dimensions and
$\varepsilon=-1$ in 4 dimensions).

\item {$[a,b^{o}]=0$} for all {$a,b\in\mathcal{A}$, $b^{o}=Jb^{\ast}J^{-1}.$
This is the zeroth order condition and is needed to define the right action of the algebra  on
elements of $\mathcal{H}$} : $\zeta b=b^{o}\zeta.$

\item {$DJ=\varepsilon^{\prime}JD,\quad J\,\gamma=\varepsilon^{\prime\prime
}\gamma J,\quad D\gamma=-\gamma D$} where {$\varepsilon,\varepsilon^{\prime
},\varepsilon^{\prime\prime}\in\left\{  -1,1\right\}  .$} These reality
conditions resemble the conditions of existence of Majorana (real) fermions.

\item {$[[D,a],b^{o}]=0$} for all {$a,b\in\mathcal{A}$}. This is the first
order condition.

\item {$\gamma^{2}=1$} and {$[\gamma,a]=0$} for all {$a\in\mathcal{A}$}. Thus
$\gamma$ is the chirality operator and this gives the decomposition
\newline{$\mathcal{H}=\mathcal{H}_{L}\oplus\mathcal{H}_{R}$}.
\end{itemize}

It then follows from the above properties that:

\begin{itemize}
\item {$\mathcal{H}$ is endowed with an {$\mathcal{A-}$} bimodule structure
$a\,\zeta b=ab^{o}\zeta.$}

\item $\mathcal{A}$ has a well defined unitary group
\begin{equation}
\mathcal{U}=\left\{  u\in\mathcal{A};\quad u\,u^{\ast}=u^{\ast}u=1\right\}
\end{equation}
{The natural adjoint action of of $\mathcal{U}$ on $\mathcal{H}$ is given by
$\zeta\rightarrow u\zeta u^{\ast}=u\,J\,u\,J^{\ast}\zeta\quad\forall\zeta
\in\mathcal{H}.$} Then {%
\begin{equation}
\left\langle \zeta,D\zeta\right\rangle
\end{equation}
} is not invariant under the above transformation but one has: {%
\begin{equation}
\left(  u\,J\,u\,J^{\ast}\right)  D\left(  u\,J\,u\,J^{\ast}\right)  ^{\ast
}=D+u\left[  D,u^{\ast}\right]  +{\varepsilon^{\prime}}J\left(  u\left[
D,u^{\ast}\right]  \right)  J^{\ast}%
\end{equation}
}

\item The action $\left\langle \zeta,D_{A}\zeta\right\rangle $ is invariant
where
\begin{equation}
{D_{A}=D+A+\varepsilon^{\prime}JAJ^{-1},\quad A=%
{\displaystyle\sum\limits_{i}}
a^{i}\left[  D,b^{i}\right]  }%
\end{equation}
and $A=A^{\ast}$ is self-adjoint. \ This is similar to the appearance of the
interaction term for the photon with the electrons
\begin{equation}
{i\overline{\psi}\gamma^{\mu}\partial_{\mu}\psi\rightarrow i\overline{\psi
}\gamma^{\mu}\left(  \partial_{\mu}+ieA_{\mu}\right)  \psi}%
\end{equation}
to maintain invariance under the variations {$\psi\rightarrow u\psi
=e^{i\alpha\left(  x\right)  }\psi.$}
\end{itemize}

One then extends the familiar geometric notions to this framework:

\begin{itemize}
\item The notion of dimension is governed by the growth of eigenvalues of $D$,
and may be {fractal and involve complex numbers}.

\item The antilinear isometry {$J:\mathcal{H}\rightarrow\mathcal{H}$} gives a
real structure of {$KO$-dimension $n\in\mathbb{Z}/8$} on a spectral triple
{$(\mathcal{A},\mathcal{H},D)$} {
\begin{equation}
J^{\,\,2}=\varepsilon,\ \ \ \ JD=\varepsilon^{\prime}%
DJ,\ \ \hbox{and}\ \ J\,\gamma=\varepsilon^{\prime\prime}\gamma
J\,\hbox{(even case)}.
\end{equation}
}
\end{itemize}

The numbers {$\varepsilon,\varepsilon^{\prime},\varepsilon^{\prime\prime}%
\in\{-1,1\}$} are a function of $n$ mod $8$ given by \vspace{0.25in}

\begin{center}
{%
\begin{tabular}
[c]{|c|rrrrrrrr|}\hline
\textbf{n } & 0 & 1 & 2 & 3 & 4 & 5 & 6 & 7\\\hline\hline
$\varepsilon$ & 1 & 1 & -1 & -1 & -1 & -1 & 1 & 1\\
$\varepsilon^{\prime}$ & 1 & -1 & 1 & 1 & 1 & -1 & 1 & 1\\
$\varepsilon^{\prime\prime}$ & 1 &  & -1 &  & 1 &  & -1 & \\\hline
\end{tabular}
}

\end{center}

\vspace{0.5in}

Our starting point is the model:  \emph{space-time is a product of a
continuous four-dimensional manifold $M$ times a finite space $F$}. And of course we do not assume that the finite space $F$ is commutative. It is described by a  spectral data $\left(  {\mathcal{A_F},\mathcal{H_F},D_F,J_F,\gamma_F
}\right)  $ where all ingredients are \emph{ finite dimensional}.

The algebra {$\mathcal{A}$} for the product space is a tensor product. The spectral geometry of {$\mathcal{A}$} is
given by the product rule
\begin{equation}
\mathcal{A}=C^{\infty}\left(  M\right)  \otimes\mathcal{A}_{F}%
\end{equation}
\begin{equation}
\ {\mathcal{H}=L^{2}\left(  M,S\right)  \otimes\mathcal{H}_{F},}%
\end{equation}%
\begin{equation}
{\quad D=D_{M}\otimes1+\gamma_{5}\otimes D_{F},}%
\end{equation}
where {$L^{2}\left(  M,S\right)  $} is the Hilbert space of {$L^{2}$} spinors,
and {$D_{M}$} is the Dirac operator of the Levi-Civita spin connection on the
four manifold $M$,
\begin{equation}
{{D_{M}=\gamma^{\mu}\left(  \partial_{\mu}+\omega_{\mu}\right)  .}}%
\end{equation}
The chirality operator is {$\gamma=\gamma_{5}\otimes\gamma_{F}.$} The real structure $J$ is $J_M\otimes J_F$ where $J_M$ is charge conjugation.

In order to avoid the fermion doubling problem so that $\zeta,\zeta^{c}%
,\zeta^{\ast},\zeta^{c\ast}$ where $\zeta\in\mathcal{H},$ are not all
independent, it was shown in \cite{mc2} that the finite dimensional space must be taken to
be of K-theoretic dimension $6$ modulo $8$, where in this case $\left(  \varepsilon
,\varepsilon^{\prime},\varepsilon^{\prime\prime}\right)  =(1,1,-1)$. This makes the
total K-theoretic dimension of the noncommutative space to be $10$ and would
allow to impose the reality (Majorana) condition and the Weyl condition
simultaneously in the Minkowskian continued form, a situation very familiar in
ten-dimensional supersymmetry. In the Euclidean version, the use of the $J$
\ in the fermionic action, would give for the chiral fermions in the path
integral, a Pfaffian instead of determinant, and will thus cut the
fermionic degrees of freedom by 2. In other words, to have the fermionic
sector free of the fermionic doubling problem we  must make the choice
\begin{equation}
J_{F}^{\,2}=1,\qquad J_{F}D_{F}=D_{F}J_{F},\qquad J_{F}\,\gamma_{F}%
=-\gamma_{F}J_{F}%
\end{equation}
 In what follows we will restrict our attention to determination of the
finite algebra, and will omit the subscript $F$.

\section{Classification of the Finite Space}

There are two main constraints on the algebra from the axioms of
noncommutative geometry. We first look for involutive algebras ${\mathcal{A}}$
of operators in {${\mathcal{H}}$} such that,
\begin{equation}
\lbrack a,b^{0}]=0\,,\quad\forall\,a,b\in{\mathcal{A}}\,
\end{equation}
  where for any operator  $a$ in ${\mathcal{H}}$, $a^{0}=Ja^{\ast}J^{\,\,-1}%
$.  This is called the order zero condition. We now look for representations
of ${\mathcal{A}}$ and $J$ in {${\mathcal{H}}$ which are irreducible. Assume
that }$e\neq1$ is a projection in the center $Z\left(  {\mathcal{A}}\right)  $
of ${\mathcal{A}}$, where $e^{2}=e=e^*,$ $ea=ae$ $\forall a\in{\mathcal{A}}.$ We
then have for the projection $\left(  eJeJ^{\,\,-1}\right)  ^{2}%
=eJeJ^{\,-1},$
\begin{align}
\left[  eJeJ^{\,-1},a\right]   &  =eJeJ^{\,-1}a-aeJeJ^{-1}\\
&  =eaJeJ^{-1}-aeJeJ^{-1}=0\nonumber
\end{align}
where we have used the  order zero condition $\left[  a,JeJ^{\,-1}\right]  =0$.
We also have
\begin{align}
\left[  eJeJ^{\,-1},J\right]   &  =eJeJ^{-1}J-JJeJ^{-1}e\\
&  =eJe-\epsilon \, eJ^{-1}e=0.\nonumber
\end{align}
Thus, the projection $eJeJ^{\,-1}$ commutes with $\mathcal{A}$ and $J$ and,    by irreducibility, is equal to $0$ or $1.$ \ But the later
choice contradicts $e\neq1$ since the range of $eJeJ^{\,-1}$ is contained in
the range of $e$. Thus we have
\begin{equation}
eJeJ^{\,-1}=0.
\end{equation}
Similarly if we have two projections $e_{1}$ and $e_{2}$ in the center $Z\left(  {\mathcal{A}}\right)  $
of ${\mathcal{A}}$, such that $e_{1}%
e_{2}=0$, then a simple calculation as  above  shows that the
projection
\begin{equation}
\left(  e_{1}Je_{2}J^{\,-1}+e_{2}Je_{1}J^{\,-1}\right)  ^{2}=e_{1}%
Je_{2}J^{\,-1}+e_{2}Je_{1}J^{\,-1}%
\end{equation}
satisfies
\begin{align}
\left[  e_{1}Je_{2}J^{\,-1}+e_{2}Je_{1}J^{\,-1},a\right]   &  =0\\
\left[  e_{1}Je_{2}J\,^{-1}+e_{2}Je_{1}J^{\,-1},J\right]   &  =0
\end{align}
and thus by irreducibility is equal to $0$ or $1$. Assume that the center $Z\left(  {\mathcal{A}}\right)  $ allows
for more than two projections then $%
{\displaystyle\sum\limits_{j}}
e_{j}=1$ where
\begin{equation}
e_{i}^{2}=e_{i}=e_{i}^*,\qquad\forall i,\qquad e_{i}e_{j}=0,\qquad i\neq j.
\end{equation}
Thus one gets
\begin{align}
1 &  =%
{\displaystyle\sum\limits_{i}}
e_{i}J\left(
{\displaystyle\sum\limits_{j}}
e_{j}\right)  J^{-1}\\
&  =%
{\displaystyle\sum\limits_{j\neq i}}
e_{i}Je_{j}J^{-1}\qquad\text{since }e_{i}Je_{i}J^{-1}=0\nonumber\\
&  =\left(  e_{1}Je_{2}J^{-1}+e_{2}Je_{1}J^{-1}\right)  +\left(  e_{1}%
Je_{3}J^{-1}+e_{3}Je_{1}J^{-1}\right)  +\cdots\nonumber
\end{align}
and therefore only one combination (say $e_{1}$ and $e_{2}$) can be equal to
$1,$ the others being zero
\begin{align}
e_{1}Je_{2}J^{-1}+e_{2}Je_{1}J^{-1} &  =1\\
e_{i}Je_{j}J^{-1}+e_{j}Je_{i}J^{-1} &  =0\qquad i\neq1,2,\qquad\forall j.
\end{align}
From this we have that for $i\notin\left\{  1,2\right\},  $ $e_{i}Je_{j}J^{-1}=0$ for all $j$ and thus
\begin{align}
e_{i} &  =e_{i}J\left(
{\displaystyle\sum\limits_{j}}
e_{j}\right)  J^{-1}\\
&  =0\nonumber
\end{align}
Thus $e_i=0$ for $i\notin\left\{  1,2\right\},  $ $e_{1}+e_{2}=1$ and one can easily show that
\begin{equation}
Je_{1}J^{-1}=e_{2},\qquad Je_{2}J^{-1}=e_{1}%
\end{equation}
In general we only assume that the algebra ${\mathcal{A}}$ is real and preserved by the involution $x\mapsto x^*$, but the above argument applies to the complexified extension ${\mathcal{A}}_{{\mathbb{C}}}$.
The surprising result is that the classification of irreducible
representations of ${\mathcal{A}}$ and $J$ \ in {${\mathcal{H}}$ splits into
two cases only. The center for the complexified extension of the algebra can
only be }$Z({\mathcal{A}}_{{\mathbb{C}}})={\mathbb{C}}$ for $e=1$ or
{{$Z\left(  \mathcal{A}_{\mathbb{C}}\right)  =\mathbb{C\oplus C}$ for }}%
$e_{1}+e_{2}=1$ with $Je_{1}J^{-1}=e_{2}.$

Let $\mathcal{H}$ be a Hilbert space of dimension $n$. Then an irreducible
solution with $Z\left(  \mathcal{A}_{\mathbb{C}}\right)  =$ $\mathbb{C}$
exists iff $n=k^{2}$ is a square. It is given by $\mathcal{A}_{\mathbb{C}%
}=M_{k}\left(  \mathbb{C}\right)  $ acting by left multiplication on itself
and antilinear involution $J\left(  x\right)  =x^{\ast},\quad\forall x\in
M_{k}\left(  \mathbb{C}\right)  .$ The irreducible representation $\beta$ is
given by
\begin{equation}
{\mathcal{A}}_{\mathbb{C}}\otimes{\mathcal{A}}_{\mathbb{C}}^{0}\rightarrow{\mathcal{L}%
}({\mathcal{H}})\,,\ \beta(x\otimes y)=xy^{0}\,,\quad\forall\,x,y\in
{\mathcal{A}}_{\mathbb{C}}\,,
\end{equation}
which is injective since ${\mathcal{A}}_{\mathbb{C}}\otimes{\mathcal{A}}_{\mathbb{C}}^{0}\sim
M_{k^{2}}({\mathbb{C}})$. Since ${\mathcal{A}}_{\mathbb{C}}\otimes{\mathcal{A}}_{\mathbb{C}}%
^{0}\sim M_{k^{2}}({\mathbb{C}})$ then $n=k^{2}$ is a square. This determines
$\mathcal{A}_{\mathbb{C}}$ and its representations in $\left(  \mathcal{H}%
,J\right)  $ and allows only for three possibilities for $\mathcal{A}$. These
are $\mathcal{A=}M_{k}\left(  \mathbb{C}\right)  ,$ $M_{k}\left(
\mathbb{R}\right)  $ and $M_{a}\left(  \mathbb{H}\right)  $ for even $k=2a,$
where $\mathbb{H}$ is the field of quaternions. These correspond respectively
to the unitary, orthogonal and symplectic case. It can be shown that the case
$Z({\mathcal{A}}_{{\mathbb{C}}})={\mathbb{C}}$ is incompatible with the
commutation relation $J\gamma=-\gamma J$ and hence with the K-theoretic
dimension $6$ necessary to impose the reality condition on the spinors to
avoid fermion doubling. This implies that the only realistic case to consider
is the second possibility.

We thus have to assume that {{$Z\left(  \mathcal{A}_{\mathbb{C}}\right)
=\mathbb{C\oplus C}$}}. \ Then there exists $k_{j}\in{\mathbb{N}}$ such that
${\mathcal{A}}_{{\mathbb{C}}}=M_{k_{1}}({\mathbb{C}})\oplus M_{k_{2}%
}({\mathbb{C}})$ as an involutive algebra over ${\mathbb{C}}$. We let $e_{j}$
be the minimal projections $e_{j}\in Z({\mathcal{A}}_{{\mathbb{C}}})$ with
$e_{j}$ corresponding to the component $M_{k_{j}}({\mathbb{C}})$. There is a
corresponding decomposition
\begin{equation}
{\mathcal{H}}=e_{1}{\mathcal{H}}\oplus e_{2}{\mathcal{H}}={\mathcal{H}}%
_{1}\oplus{\mathcal{H}}_{2}\,,\qquad\ (x_{1},x_{2})(\xi_{1},\xi_{2})=(x_{1}%
\xi_{1},x_{2}\xi_{2})
\end{equation}
One can show (under the natural hypothesis that there is a separating vector
in ${\mathcal{H)}}$ that $k_{1}=k_{2}=k,$ the dimension $n$ of the Hilbert
space ${\mathcal{H}}$ is $n=2k^{2}$ and that the action of $J$ is given by
\begin{equation}
J\left(  x,y\right)  =\left(  y^{\ast},x^{\ast}\right)
\end{equation}
We then have six possibilities for the algebra ${\mathcal{A}}$
\begin{equation}
\left\{  M_{k}\left(  \mathbb{C}\right)  \text{ or }M_{k}\left(
\mathbb{R}\right)  \text{ or }M_{a}\left(  \mathbb{H}\right)  \right\}
\oplus\left\{  M_{k}\left(  \mathbb{C}\right)  \text{ or }M_{k}\left(
\mathbb{R}\right)  \text{ or }M_{a}\left(  \mathbb{H}\right)  \right\}  .
\end{equation}
We shall show, at the end of section five, that four of these possibilities
can be ruled out immediately and that the choice of
\begin{equation}
{\mathcal{A=}}M_{k}\left(  \mathbb{C}\right)  \oplus M_{k}\left(
\mathbb{C}\right)
\end{equation}
when $k=4$ suffers from $U(1)$ anomalies. We thus proceed to make the
assumption of imposing an antilinear isometry $I$ \ such that $I^{2}=-1$ on
one of the algebras and no condition on the other forcing ${\mathcal{A}}$ to
be
\begin{equation}
{\mathcal{A=}}M_{a}\left(  \mathbb{H}\right)  \oplus M_{k}\left(
\mathbb{C}\right)  ,\qquad k=2a
\end{equation}
The dimension of the Hilbert space $n=2k^{2}$ gives $k^{2}$ independent
fermions, where $k$ is an even integer, because of the reality condition. To
have a non-trivial grading on $M_{a}\left(  \mathbb{H}\right)  $ requires $a$
to be at least $2.$ Thus the simplest possibility is
\begin{equation}
{\mathcal{A=}}M_{2}\left(  \mathbb{H}\right)  \oplus M_{4}\left(
\mathbb{C}\right)
\end{equation}
and the grading $\gamma$ reduces $M_{2}\left(  \mathbb{H}\right)  $ to
$\mathbb{H}\oplus\mathbb{H}.$ This corresponds to a Hilbert space of $16$ fermions.

We next examine the order one condition
\begin{equation}
\left[  \left[  D,a\right]  ,b^{o}\right]  =0,\text{\qquad}\forall
a,b\in{\mathcal{A}}%
\end{equation}
First if the Dirac operator commutes with {{$Z\left(  \mathcal{A}\right)  $ }}%
\begin{equation}
\left[  D,{{Z\left(  \mathcal{A}\right)  }}\right]  =0
\end{equation}
then one can show that the Dirac operator has no non-diagonal elements that
connects the two pieces of the algebra $\mathcal{A}$ and thus $e_{1}De_{2}=0.$
This will correspond to unbroken color group $SU(4)$ and with only Dirac
masses for the neutrinos. On the other hand if there is a non-trivial mixing
such that
\begin{equation}
\left[  D,{{Z\left(  \mathcal{A}\right)  }}\right]  \neq0
\end{equation}
then the non-diagonal operator $T=e_{1}De_{2}:$ ${\mathcal{H}}_{1}%
\rightarrow{\mathcal{H}}_{2}$ must be of rank $1$ and thus can only have a
singlet non-zero entry forcing elements of \ the algebra to take the form
\begin{equation}
\left(  \lambda,\overline{\lambda},q\right)  \oplus\left(  \lambda,m\right)
,\qquad\lambda\in\mathbb{C},\quad q\in\mathbb{H},\quad m\in M_{3}\left(
\mathbb{C}\right)
\end{equation}
thus reducing the algebra to
\begin{equation}
\mathbb{C}\oplus\mathbb{H}\oplus M_{3}\left(  \mathbb{C}\right)
\end{equation}
These last steps will be made more transparent in the next section.

\section{Tensorial notation}

To acquaint ourselves with the abstract quantities defined so far, it is
useful to use the tensorial notation familiar to physicists. The main
advantage of this method is that it can be implemented using computer programs
with algebraic manipulations such as Mathematica and Maple. We will restrict
to the case where {{$Z\left(  \mathcal{A}_{\mathbb{C}}\right)
=\mathbb{C\oplus C}.$}}

An element of the Hilbert space $\Psi\in{\mathcal{H}}$ is represented by
\begin{equation}
\Psi_{M}=\left(
\begin{array}
[c]{c}%
\psi_{A}\\
\psi_{A^{^{\prime}}}%
\end{array}
\right)  ,\quad\psi_{A^{\prime}}=\psi_{A}^{c}%
\end{equation}
where $\psi_{A}^{c}$ is the conjugate spinor to $\psi_{A}.$ It is acted on by
both the left algebra $M_{2}\left(  \mathbb{H}\right)  $ and the right algebra
$M_{4}\left(  \mathbb{C}\right)  $. Therefore the index $A$ can take $16$
values and is represented by
\begin{equation}
A=\alpha I
\end{equation}
where the index $\alpha$ is acted on by the quaternionic matrices and the
index $I$ \ by the $M_{4}\left(  \mathbb{C}\right)  $ matrices. Moreover, when
grading breaks $M_{2}\left(  \mathbb{H}\right)  $ into $\mathbb{H}%
\oplus\mathbb{H}$ the index $\alpha$ is decomposed to $\alpha=\overset{.}%
{a},a$ where $\overset{.}{a}=\overset{.}{1},\overset{.}{2}$ is acted on by the
first quaternionic algebra $\ \mathbb{H}_{R}$ and $a=1,2$ is acted on by the
second quaternionic algebra $\ \mathbb{H}_{L}$ . Also when $M_{4}\left(
\mathbb{C}\right)  $ breaks into $\mathbb{C}\oplus M_{3}\left(  \mathbb{C}%
\right)  $ the index $I$ is decomposed into $I=1,i$ where the $1$ is acted on
by the $\mathbb{C}$ and the $i$ by $M_{3}\left(  \mathbb{C}\right)  .$
Therefore the various components of the spinor $\psi_{A}$ are
\begin{align}
\psi_{\overset{.}{1}1} &  =\nu_{R}\\
\psi_{\overset{.}{2}1} &  =e_{R}\\
\psi_{a1} &  =l_{a}=\left(
\begin{array}
[c]{c}%
\nu_{L}\\
e_{L}%
\end{array}
\right)  \\
\psi_{\overset{.}{1}i} &  =u_{iR}\\
\psi_{\overset{.}{2}i} &  =d_{iR}\\
\psi_{ai} &  =q_{ia}=\left(
\begin{array}
[c]{c}%
u_{iL}\\
d_{iL}%
\end{array}
\right)
\end{align}
The Dirac action then take the form
\begin{equation}
\Psi_{M}^{\ast}D_{M}^{N}\Psi_{N}%
\end{equation}
which we can expand to give
\begin{equation}
\psi_{A}^{\ast}D_{A}^{B}\psi_{B}+\psi_{A^{\prime}}^{\ast}D_{A^{\prime}}%
^{B}\psi_{B}+\psi_{A}^{\ast}D_{A}^{B^{^{\prime}}}\psi_{B^{^{\prime}\prime}%
}+\psi_{A^{\prime}}^{\ast}D_{A^{\prime}}^{B^{\prime}}\psi_{B^{\prime}}%
\end{equation}
The Dirac operator can be written in matrix form%
\begin{equation}
D=\left(
\begin{array}
[c]{cc}%
D_{A}^{B} & D_{A}^{B^{^{\prime}}}\\
D_{A^{^{\prime}}}^{B} & D_{A^{^{\prime}}}^{B^{^{\prime}}}%
\end{array}
\right)  ,
\end{equation}
where $\quad$%
\begin{align}
A &  =\alpha I,\quad\alpha=1,\cdots,4,\quad I=1,\cdots,4\\
\quad A^{\prime} &  =\alpha^{\prime}I^{\prime},\quad\alpha^{\prime}=1^{\prime
},\cdots,4^{\prime},\quad I=1^{\prime},\cdots,4^{\prime}%
\end{align}
Thus $D_{A}^{B}=D_{\alpha I}^{\beta J}$ . We start with the algebra
\begin{equation}
\mathcal{A}=M_{4}\left(  \mathbb{C}\right)  \oplus M_{4}\left(  \mathbb{C}%
\right)
\end{equation}
and write
\begin{equation}
a=\left(
\begin{array}
[c]{cc}%
X_{\alpha}^{\beta}\delta_{I}^{J} & 0\\
0 & \delta_{\alpha^{\prime}}^{\beta^{\prime}}Y_{I^{\prime}}^{J^{\prime}}%
\end{array}
\right)
\end{equation}
For $J^{2}=1$ we have
\begin{equation}
J=\left(
\begin{array}
[c]{cc}%
0 & \delta_{\alpha}^{\beta^{\prime}}\delta_{I}^{J^{\prime}}\\
\delta_{\alpha^{\prime}}^{\beta}\delta_{I^{\prime}}^{J} & 0
\end{array}
\right)  \times\text{\textrm{complex conjugation}}%
\end{equation}
In this form
\begin{equation}
a^{o}=Ja^{\ast}J^{-1}=\left(
\begin{array}
[c]{cc}%
\delta_{\alpha}^{\beta}Y_{I}^{tJ} & 0\\
0 & X_{\alpha^{\prime}}^{t\beta^{\prime}}\delta_{^{I^{\prime}\prime}%
}^{J^{\prime}}%
\end{array}
\right)
\end{equation}
where the superscript $t$ denotes the transpose matrix. This clearly satisfies
the commutation relation
\begin{equation}
\left[  a,b^{o}\right]  =0.
\end{equation}
The order one condition is
\begin{equation}
\left[  \left[  D,a\right]  ,b^{o}\right]  =0
\end{equation}
Writing
\begin{equation}
b=\left(
\begin{array}
[c]{cc}%
Z_{\alpha}^{t\beta}\delta_{I}^{J} & 0\\
0 & \delta_{\alpha^{\prime}}^{\beta^{\prime}}W_{I^{\prime}}^{tJ^{\prime}}%
\end{array}
\right)
\end{equation}
then
\begin{equation}
b^{o}=\left(
\begin{array}
[c]{cc}%
\delta_{\alpha}^{\beta}W_{I}^{J} & 0\\
0 & Z_{\alpha^{\prime}}^{\beta^{\prime}}\delta_{^{I^{\prime}}}^{J^{\prime}}%
\end{array}
\right)
\end{equation}
and so $\left[  \left[  D,a\right]  ,b^{o}\right]  $ is equal to
\begin{equation}
\left(
\begin{array}
[c]{cc}%
\left[  \left[  D,X\right]  ,W\right]  _{A}^{B} & \left(  \left(
DY-XD\right)  Z-W\left(  DY-XD\right)  \right)  _{A}^{B^{\prime}}\\
\left(  \left(  DX-YD\right)  W-Z\left(  DX-YD\right)  \right)  _{A^{\prime}%
}^{B} & \left[  \left[  D,Y\right]  ,Z\right]  _{A^{\prime}}^{B^{\prime}}%
\end{array}
\right)
\end{equation}
The first two equations can be made explicit by writing:
\begin{align}
\left(  D_{\alpha I}^{\gamma K}X_{\gamma}^{\beta}-X_{\alpha}^{\gamma}D_{\gamma
I}^{\beta K}\right)  W_{K}^{J}-W_{I}^{K}\left(  D_{\alpha K}^{\gamma
J}X_{\gamma}^{\beta}-X_{\alpha}^{\gamma}D_{\gamma K}^{\beta J}\right)   &
=0\\
\left(  D_{\alpha I}^{\gamma^{\prime}K^{\prime}}Y_{K^{\prime}}^{J^{\prime}%
}-X_{\alpha}^{\gamma}D_{\gamma I}^{\gamma^{\prime}K}\right)  Z_{\gamma
^{\prime}}^{\beta^{\prime}}-W_{I}^{K}\left(  D_{\alpha K}^{\beta^{\prime
}K^{\prime}}Y_{K^{\prime}}^{J^{\prime}}-X_{\alpha}^{\gamma}D_{\gamma K}%
^{\beta^{\prime}J^{\prime}}\right)   &  =0
\end{align}
Here we have two classes of solutions. First, if \ all of the $D_{\alpha
I}^{\beta^{\prime}K^{\prime}}$ are zero, implying that there is no mixing
between the fermions and their conjugates. In this case one can easily show
that the color group is $SU(4)$ and not $SU(3)$ and that there will be no
breaking of the left-right symmetry in the leptonic sector. If some of the
$D_{\alpha I}^{\beta^{\prime}K^{\prime}}$are non-zero, we have shown that the
only solution of the second equation is for $D_{\alpha I}^{\beta^{\prime
}K^{\prime}}$to have only one non-zero entry,
\begin{equation}
D_{\alpha I}^{\beta^{\prime}K^{\prime}}=\delta_{\alpha}^{\overset{.}{1}}%
\delta_{\overset{.}{1^{\prime}}}^{\beta^{\prime}}\delta_{I}^{1}\delta
_{1^{\prime}}^{K^{\prime}}k^{\ast\nu_{R}}\sigma
\end{equation}
where the $k^{\ast\nu_{R}}$ are matrices in generation space which will be
assumed to be $3\times3.$ We shall discuss the role of families below in \S \ref{generations}. We thus can write
\begin{align}
D_{\alpha I}^{\beta J} &  =D_{\alpha\left(  l\right)  }^{\beta}\delta_{I}%
^{1}\delta_{1}^{J}+D_{\alpha\left(  q\right)  }^{\beta}\delta_{I}^{i}%
\delta_{j}^{J}\delta_{i}^{j}\\
Y_{I^{\prime}}^{J^{\prime}} &  =\delta_{I^{\prime}}^{1^{\prime}}%
\delta_{1^{\prime}}^{J^{\prime}}Y_{1^{\prime}}^{1^{\prime}}+\delta_{I^{\prime
}}^{i^{\prime}}\delta_{j^{\prime}}^{J^{\prime}}Y_{i^{\prime}}^{j^{\prime}%
}\label{con1}\\
X_{\overset{.}{1}}^{\overset{.}{1}} &  =Y_{1^{\prime}}^{1^{\prime}},\text{
}X_{\overset{.}{1}}^{\alpha}=0,\quad\alpha\neq\overset{.}{1}\label{con2}%
\end{align}
We will be using the notation
\begin{equation}
\alpha=\overset{.}{1},\overset{.}{2},a\text{ \ where }a=1,2
\end{equation}
We further impose the condition of symplectic isometry on the first
$M_{4}\left(  \mathbb{C}\right)  $
\begin{equation}
\left(  \sigma_{2}\otimes1\right)  \ \left(  \overline{a}\right)  \left(
\sigma_{2}\otimes1\right)  \ =a,\quad a\in M_{4}\left(  \mathbb{C}\right)
\end{equation}
reduces it to $M_{2}\left(  \mathbb{H}\right)  .$ From the property of
commutation of the grading operator
\begin{align}
g_{\alpha}^{\beta} &  =\left(
\begin{array}
[c]{cc}%
1_{2} & 0\\
0 & -1_{2}%
\end{array}
\right)  \\
\left[  g,a\right]   &  =0\quad a\in M_{2}\left(  \mathbb{H}\right)
\end{align}
the algebra $M_{2}\left(  \mathbb{H}\right)  $ reduces to $\mathbb{H\oplus
H}.$ This, together with the  conditions \ \ref{con1} and \ref{con2}  implies
that
\begin{align}
X_{\alpha}^{\beta} &  =\delta_{\alpha}^{\overset{.}{1}}\delta_{\overset{.}{1}%
}^{\beta}X_{\overset{.}{1}}^{\overset{.}{1}}+\delta_{\alpha}^{\overset{.}{2}%
}\delta_{\overset{.}{2}}^{\beta^{\prime}}\overline{X}_{\overset{.}{1}%
}^{\overset{.}{1}}+\delta_{\alpha}^{a}\delta_{b}^{\beta}X_{a}^{b}\\
Y_{I^{\prime}}^{J^{\prime}} &  =\delta_{1^{\prime}}^{I^{\prime}}%
\delta_{J^{\prime}}^{1^{\prime}}Y_{1^{\prime}}^{1^{\prime}}+\delta_{i^{\prime
}}^{I^{\prime}}\delta_{J^{\prime}}^{j^{\prime}}Y_{i^{\prime}}^{j^{\prime}%
},\qquad X_{\overset{.}{1}}^{\overset{.}{1}}=Y_{1^{\prime}}^{1^{\prime}}%
\end{align}
and the algebra $\mathbb{H\oplus H}\oplus M_{4}\left(  \mathbb{C}\right)  $
reduces to
\begin{equation}
\mathbb{C\oplus H}\oplus M_{3}\left(  \mathbb{C}\right)
\end{equation}
Thus an element of the algebra, to be \ compatible with the axioms of
noncommutative geometry, and the few assumptions we made, must be restricted
to the form
\begin{equation}
a=\left(
\begin{array}
[c]{ccccc}%
X &  &  &  & \\
& \overline{X} &  &  & \\
&  & q &  & \\
&  &  & X & \\
&  &  &  & m
\end{array}
\right)  ,\qquad X\in\mathbb{C},\quad q\in\mathbb{H},\quad m\in M_{3}\left(
\mathbb{C}\right)  .
\end{equation}
We also note that the property that $DJ=JD$ implies that
\begin{equation}
D_{A^{\prime}}^{B^{\prime}}=\overline{D}_{A}^{B}%
\end{equation}
and that $D_{\alpha I}^{\beta^{\prime}K^{\prime}}$ is symmetric matrix, thus
$k^{\ast\nu_{R}}$ is symmetric so that $k^{\ast\nu_{R}}=\overline{k}^{\nu_{R}%
}.$ Further restriction is obtained on the form of the Dirac operator $D$ from
the property
\begin{equation}
D\gamma=-\gamma D
\end{equation}
where $\gamma$ is the grading operator. Writing
\begin{equation}
\gamma=\left(
\begin{array}
[c]{cc}%
G_{A}^{B} & 0\\
0 & -\overline{G}_{A^{^{\prime}}}^{B^{^{\prime}}}%
\end{array}
\right)  \quad G^{2}=1
\end{equation}
we obtain
\begin{equation}
\left(  GDG\right)  _{A}^{B}=-D_{A}^{B}%
\end{equation}
The grading operator acts only on the first algebra, thus
\begin{equation}
G_{A}^{B}=g_{\alpha}^{\beta}\delta_{I}^{J}%
\end{equation}
which implies that
\begin{equation}
g_{\alpha}^{\gamma}D_{\gamma\left(  l\right)  }^{\delta}g_{\delta}^{\beta
}=-D_{\alpha\left(  l\right)  }^{\beta}%
\end{equation}
thus
\begin{align}
D_{\alpha1}^{\beta1} &  =\left(
\begin{array}
[c]{cc}%
0 & D_{a1}^{\overset{.}{b}1}\\
D_{\overset{.}{a}1}^{b1} & 0
\end{array}
\right)  ,\qquad D_{a1}^{\overset{.}{b}1}=\left(  D_{\overset{.}{a}1}%
^{b1}\right)  ^{\ast}\equiv D_{a\left(  l\right)  }^{\overset{.}{b}}\\
D_{\alpha i}^{\beta j} &  =\left(
\begin{array}
[c]{cc}%
0 & D_{a\left(  q\right)  }^{\overset{.}{b}}\delta_{i}^{j}\\
D_{\overset{.}{a}\left(  q\right)  }^{b}\delta_{i}^{j} & 0
\end{array}
\right)  ,\qquad D_{\overset{.}{a}\left(  q\right)  }^{b}=\left(  D_{a\left(
q\right)  }^{\overset{.}{b}}\right)  ^{\ast}%
\end{align}
To summarize, {the matrix form for }$D_{A}^{B}$ {is given by
\begin{align}
&  \qquad\qquad\quad\left(
\begin{array}
[c]{cccccc}%
\begin{array}
[c]{c}%
\overset{.}{1}1\\
v_{R}%
\end{array}
&
\begin{array}
[c]{c}%
\overset{.}{2}1\\
e_{R}%
\end{array}
&
\begin{array}
[c]{c}%
a1\\
l_{a}%
\end{array}
&
\begin{array}
[c]{c}%
\overset{.}{1}i\\
u_{iR}%
\end{array}
&
\begin{array}
[c]{c}%
\overset{.}{2}i\\
d_{iR}%
\end{array}
&
\begin{array}
[c]{c}%
ai\\
q_{iL}%
\end{array}
\end{array}
\right)  \nonumber\\
&  \left(
\begin{array}
[c]{c}%
\overset{.}{1}1\\
\overset{.}{2}1\\
b1\\
\overset{.}{1}j\\
\overset{.}{2}j\\
bj
\end{array}
\right)  \left(
\begin{array}
[c]{cccccc}%
\left(  D\right)  _{\overset{.}{1}1}^{\overset{.}{1}1} & 0 & \left(  D\right)
_{\overset{.}{1}1}^{a1} & 0 & 0 & 0\\
0 & \left(  D\right)  _{\overset{.}{2}1}^{\overset{.}{2}1} & \left(  D\right)
_{\overset{.}{2}1}^{a1} & 0 & 0 & 0\\
\left(  D\right)  _{b1}^{\overset{.}{1}1} & \left(  D\right)  _{b1}%
^{\overset{.}{2}1} & \left(  D\right)  _{a1}^{b1} & 0 & 0 & 0\\
0 & 0 & 0 & \left(  D\right)  _{\overset{.}{1}j}^{\overset{.}{1}i} & 0 &
\left(  D\right)  _{\overset{.}{1}j}^{ai}\\
0 & 0 & 0 & 0 & \left(  D\right)  _{\overset{.}{2}j}^{\overset{.}{2}i} &
\left(  D\right)  _{\overset{.}{2}j}^{ai}\\
0 & 0 & 0 & \left(  D\right)  _{bj}^{\overset{.}{1}i} & \left(  D\right)
_{bj}^{\overset{.}{2}i} & \left(  D\right)  _{bj}^{ai}%
\end{array}
\right)
\end{align}
where the entries above and along the rows and columns denote the
corresponding fermion. }Finally we require the Dirac operator of the finite
space to commute with the element $C\subset\mathbb{C\oplus H}\oplus
M_{3}\left(  \mathbb{C}\right)  $ where
\begin{equation}
C=\left(
\begin{array}
[c]{cccc}%
\lambda &  &  & \\
& \overline{\lambda} &  & \\
&  & \lambda & \\
&  &  & \overline{\lambda}%
\end{array}
\right)
\end{equation}
This condition will ensure that the photon and not another vector will remain
massless. This also reduces $D_{a1}^{\overset{.}{b}1}$ to the form
\begin{equation}
D_{a1}^{\overset{.}{b}1}=D_{a\left(  l\right)  }^{\overset{.}{b}}=\left(
\begin{array}
[c]{cc}%
k^{\ast\nu} & 0\\
0 & k^{\ast e}%
\end{array}
\right)  ,\qquad a=1,2,\quad\overset{.}{b}=\overset{.}{1},\overset{.}{2}%
\end{equation}
and%
\begin{equation}
D_{a\left(  q\right)  }^{\overset{.}{b}}=\left(
\begin{array}
[c]{cc}%
k^{\ast u} & 0\\
0 & k^{\ast d}%
\end{array}
\right)
\end{equation}
To summarize the finite space Dirac operator is given by
\begin{align}
\left(  D_{F}\right)  _{\alpha I}^{\beta J} &  =\left(  \delta_{\alpha}%
^{1}\delta_{\overset{.}{1}}^{\beta}k^{\ast\nu}+\delta_{\alpha}^{\overset{.}%
{1}}\delta_{1}^{\beta}k^{\nu}+\delta_{\alpha}^{2}\delta_{\overset{.}{2}%
}^{\beta}k^{\ast e}+\delta_{\alpha}^{\overset{.}{2}}\delta_{2}^{\beta}%
k^{e}\right)  \delta_{I}^{1}\delta_{1}^{J}\\
&  +\left(  \delta_{\alpha}^{1}\delta_{\overset{.}{1}}^{\beta}k^{\ast
u}+\delta_{\alpha}^{\overset{.}{1}}\delta_{1}^{\beta}k^{u}+\delta_{\alpha}%
^{2}\delta_{\overset{.}{2}}^{\beta}k^{\ast d}+\delta_{\alpha}^{\overset{.}{2}%
}\delta_{2}^{\beta}k^{d}\right)  \delta_{I}^{i}\delta_{j}^{J}\delta_{i}%
^{j}\nonumber\\
\left(  D_{F}\right)  _{\alpha I}^{\beta^{\prime}K^{\prime}} &  =\delta
_{\alpha}^{\overset{.}{1}}\delta_{\overset{.}{1}^{\prime}}^{\beta^{\prime}%
}\delta_{I}^{1}\delta_{1^{\prime}}^{K^{\prime}}k^{\ast\nu_{R}}\sigma
\end{align}
We now form the Dirac operator of the product space of this finite space
times \ a four-dimensional Riemannian manifold
\begin{equation}
D=D_{M}\otimes1+\gamma_{5}\otimes D_{F}%
\end{equation}
Since $D_{F}$ is a $32\times32$ matrix tensored with the $3\times3$ matrices
of generation space, and the $4\times4$ Clifford algebra, $D$ is
$384\times384$ matrix.

In order for the Dirac action to be invariant under fluctuations of the inner
automorphisms of the algebra $\mathcal{A}$, the operator $D$ must be replaced
with the operator
\begin{equation}
D_{A}=D+A+JAJ^{-1}%
\end{equation}
where
\begin{align}
A &  =%
{\displaystyle\sum}
a\left[  D,b\right]  \\
a &  =\left(
\begin{array}
[c]{cc}%
X_{\alpha}^{\beta}\delta_{I}^{J} & 0\\
0 & \delta_{\alpha^{\prime}}^{\beta^{\prime}}Y_{I^{\prime}}^{J^{\prime}}%
\end{array}
\right)  \\
b &  =\left(
\begin{array}
[c]{cc}%
Z_{\alpha}^{\beta}\delta_{I}^{J} & 0\\
0 & \delta_{\alpha^{\prime}}^{\beta^{\prime}}W_{I^{\prime}}^{J^{\prime}}%
\end{array}
\right)
\end{align}
To calculate $A$ we write%
\begin{equation}
A_{A}^{B}=%
{\displaystyle\sum}
a_{A}^{C}\left(  D_{C}^{D}b_{D}^{B}-b_{C}^{D}D_{D}^{B}\right)
\end{equation}
(there are no mixing terms like $D_{C}^{D^{\prime}}b_{D^{\prime}}^{B}$ because
the matrix $b$ is block diagonal). Or%
\begin{equation}
A_{\alpha I}^{\beta J}=%
{\displaystyle\sum}
a_{\alpha I}^{\gamma K}\left(  D_{\gamma K}^{\delta L}b_{\delta L}^{\beta
J}-b_{\gamma K}^{\delta L}D_{\delta L}^{\beta J}\right)
\end{equation}
Enumerating all possibilities for $\alpha I$ and $\beta J$, where $I=1,i$ and
$J=1,j,$
\begin{align}
A_{\alpha1}^{\beta1} &  =%
{\displaystyle\sum}
X_{\alpha}^{\gamma}\left(  D_{\gamma\left(  l\right)  }^{\delta}Z_{\delta
}^{\beta}-Z_{\gamma}^{\delta}D_{\delta\left(  l\right)  }^{\beta}\right)  \\
A_{\alpha i}^{\beta j} &  =\delta_{i}^{j}%
{\displaystyle\sum}
X_{\alpha}^{\gamma}\left(  D_{\gamma\left(  q\right)  }^{\delta}Z_{\delta
}^{\beta}-Z_{\gamma k}^{\delta l}D_{\delta\left(  q\right)  }^{\beta}\right)
\\
A_{\alpha i}^{\beta1} &  =A_{\alpha1}^{\beta j}=0
\end{align}
with the mixing terms vanishing. Next we evaluate these, component by
component, by taking $\alpha=\overset{.}{1},$ $\overset{.}{2},$ $a,$ and
$\beta=\overset{.}{1},$ $\overset{.}{2},$ $b:$%
\begin{align}
A_{\overset{.}{1}1}^{\overset{.}{1}1} &  =%
{\displaystyle\sum}
X_{\overset{.}{1}}^{\overset{.}{1}}\left(  D_{\overset{.}{1\left(  l\right)
}}^{\overset{.}{1}}Z_{\overset{.}{1}}^{\overset{.}{1}}-Z_{\overset{.}{1}%
}^{\overset{.}{1}}D_{\overset{.}{1}\left(  l\right)  }^{\overset{.}{1}%
}\right)  \\
&  =%
{\displaystyle\sum}
X_{\overset{.}{1}}^{\overset{.}{1}}\gamma^{\mu}\partial_{\mu}Z_{\overset{.}%
{1}}^{\overset{.}{1}}\equiv-\frac{i}{2}g_{1}\gamma^{\mu}B_{\mu}\nonumber
\end{align}%
\begin{align}
A_{\overset{.}{2}1}^{\overset{.}{2}1} &  =%
{\displaystyle\sum}
X_{\overset{.}{2}}^{\overset{.}{2}}\left(  D_{\overset{.}{2}\left(  l\right)
}^{\overset{.}{2}}Z_{\overset{.}{2}}^{\overset{.}{2}}-Z_{\overset{.}{2}%
}^{\overset{.}{2}}D_{\overset{.}{2}\left(  l\right)  }^{\overset{.}{2}%
}\right)  \\
&  =%
{\displaystyle\sum}
X_{\overset{.}{2}}^{\overset{.}{2}}\gamma^{\mu}\partial_{\mu}Z_{\overset{.}%
{2}}^{\overset{.}{2}}\nonumber\\
&  =%
{\displaystyle\sum}
\overline{X}_{\overset{.}{1}}^{\overset{.}{1}}\gamma^{\mu}\partial_{\mu
}\overline{Z}_{\overset{.}{1}}^{\overset{.}{1}}=\frac{i}{2}g_{1}\gamma^{\mu
}B_{\mu}\nonumber
\end{align}%
\begin{align}
A_{\overset{.}{1}1}^{11} &  =%
{\displaystyle\sum}
X_{\overset{.}{1}}^{\overset{.}{1}}\left(  D_{\overset{.}{1}\left(  l\right)
}^{\overset{.}{1}}Z_{1}^{1}-Z_{\overset{.}{1}}^{\overset{.}{1}}D_{\overset
{.}{1}\left(  l\right)  }^{1}\right)  \\
&  =\gamma_{5}k^{\ast\nu}%
{\displaystyle\sum}
X_{\overset{.}{1}}^{\overset{.}{1}}\left(  Z_{1}^{1}-Z_{\overset{.}{1}%
}^{\overset{.}{1}}\right)  \nonumber\\
&  \equiv\gamma_{5}k^{\ast\nu}H_{2}\nonumber
\end{align}%
\begin{align}
A_{\overset{.}{1}1}^{21} &  =%
{\displaystyle\sum}
a_{\overset{.}{1}1}^{\overset{.}{1}1}\left(  D_{\overset{.}{1}1}^{\overset
{.}{1}1}b_{11}^{21}\right)  \\
&  =\gamma_{5}k^{\ast\nu}%
{\displaystyle\sum}
X_{\overset{.}{1}}^{\overset{.}{1}}\left(  Z_{1}^{2}\right)  \nonumber\\
&  \equiv\gamma_{5}k^{\ast\nu}\left(  -H_{1}\right)  \nonumber
\end{align}%
\begin{align}
A_{\overset{.}{2}1}^{11} &  =%
{\displaystyle\sum}
a_{\overset{.}{2}1}^{\overset{.}{2}1}\left(  D_{\overset{.}{2}1}^{\overset
{.}{2}1}b_{21}^{11}\right)  \\
&  =\gamma_{5}k^{\ast e}%
{\displaystyle\sum}
X_{\overset{.}{2}}^{\overset{.}{2}}\left(  Z_{2}^{1}\right)  \nonumber\\
&  =\gamma_{5}k^{\ast e}\overline{H}^{1}\nonumber
\end{align}%
\begin{align}
A_{\overset{.}{2}1}^{21} &  =%
{\displaystyle\sum}
X_{\overset{.}{2}}^{\overset{.}{2}}\left(  D_{\overset{.}{2}\left(  l\right)
}^{2}Z_{2}^{2}-Z_{\overset{.}{2}}^{\overset{.}{2}}D_{\overset{.}{2}\left(
l\right)  }^{2}\right)  \\
&  =\gamma_{5}k^{\ast e}%
{\displaystyle\sum}
X_{\overset{.}{2}}^{\overset{.}{2}}\left(  Z_{2}^{2}-Z_{\overset{.}{2}%
}^{\overset{.}{2}}\right)  \nonumber\\
&  =\gamma_{5}k^{\ast e}\overline{H}^{2}\nonumber
\end{align}
where we have used the relations $X_{\overset{.}{2}}^{\overset{.}{2}%
}=\overline{X_{\overset{.}{1}}^{\overset{.}{1}}}$ and $Z_{2}^{1}%
=-\overline{Z_{1}^{2}}$ because of the quaternionic property. Next%
\begin{align}
A_{a1}^{b1} &  =%
{\displaystyle\sum}
a_{a1}^{c1}\left(  D_{c1}^{d1}b_{d1}^{b1}-b_{c1}^{d1}D_{d1}^{b1}\right)  \\
&  =\gamma^{\mu}%
{\displaystyle\sum}
X_{a}^{c}\left(  \partial_{\mu}Z_{c}^{b}\right)  \nonumber\\
&  =-\frac{i}{2}g_{2}W_{\mu}^{\alpha}\left(  \sigma^{\alpha}\right)  _{a}%
^{b}\nonumber
\end{align}
The reason we can write this as an $SU(2)$ gauge field is because it comes
from multiplying quaternions:%
\begin{align}
q_{1}\partial_{\mu}q_{2} &  =\left(
\begin{array}
[c]{cc}%
\alpha_{1} & \beta_{1}\\
-\overline{\beta}_{1} & \overline{\alpha}_{1}%
\end{array}
\right)  \left(
\begin{array}
[c]{cc}%
\partial_{\mu}\alpha_{2} & \partial_{\mu}\beta_{2}\\
-\partial_{\mu}\overline{\beta}_{2} & \overline{\alpha}_{1}%
\end{array}
\right)  \\
&  =\left(
\begin{array}
[c]{cc}%
\alpha_{1}\partial_{\mu}\alpha_{2}-\beta_{1}\partial_{\mu}\overline{\beta} &
\alpha_{1}\partial_{\mu}\beta_{2}+\beta_{1}\partial_{\mu}\overline{\alpha}%
_{2}\\
-\overline{\beta}_{1}\partial_{\mu}\alpha_{2}-\overline{\alpha}_{1}%
\partial_{\mu}\overline{\beta}_{2} & -\overline{\beta}_{1}\partial_{\mu}%
\beta_{2}+\overline{\alpha}_{1}\overline{\alpha}_{1}%
\end{array}
\right)  \nonumber
\end{align}
which is of the right form if we note that $A$ is Hermitian. The other
components for $A_{\alpha i}^{\beta j}$ give exactly the same results with the
replacements $k^{\nu}\rightarrow k^{u}$ and $k^{e}\rightarrow k^{d}$ and is
proportional to $\delta_{i}^{j}.$ For the $A_{A^{\prime}}^{B^{\prime}}$
elements we have
\begin{equation}
A_{A^{\prime}}^{B^{\prime}}=%
{\displaystyle\sum}
a_{A^{\prime}}^{C^{\prime}}\left(  D_{C^{\prime}}^{D^{\prime}}b_{D^{\prime}%
}^{B^{\prime}}-b_{C^{\prime}}^{D^{\prime}}D_{D^{\prime}}^{B^{\prime}}\right)
\end{equation}
In terms of components we have
\begin{align}
A_{\alpha^{\prime}1^{\prime}}^{\beta^{\prime}1^{\prime}} &  =%
{\displaystyle\sum}
a_{\alpha^{\prime}1^{\prime}}^{\gamma^{\prime}1^{\prime}}\gamma^{\mu}%
\partial_{\mu}b_{\gamma^{\prime}1^{\prime}}^{\beta^{\prime}1^{\prime}}\\
&  =\delta_{\alpha^{\prime}}^{\beta^{\prime}}%
{\displaystyle\sum}
Y_{1^{\prime}}^{1^{\prime}}\gamma^{\mu}\partial_{\mu}W_{1^{\prime}}%
^{1^{\prime}}\nonumber\\
&  =\delta_{\alpha^{\prime}}^{\beta^{\prime}}\left(  -\frac{i}{2}g_{1}%
\gamma^{\mu}B_{\mu}\right)  \nonumber
\end{align}
because $Y_{1}^{1}=X_{\overset{.}{1}}^{\overset{.}{1}}$ and $W_{1^{\prime}%
}^{1^{\prime}}=Z_{\overset{.}{1}}^{\overset{.}{1}}.$ Next
\begin{align}
A_{\alpha^{\prime}i^{\prime}}^{\beta^{\prime}j^{\prime}} &  =%
{\displaystyle\sum}
a_{\alpha^{\prime}i^{\prime}}^{\gamma^{\prime}k^{\prime}}\left(
D_{\gamma^{\prime}k^{\prime}}^{\delta^{\prime}l^{\prime}}b_{\delta^{\prime
}l^{^{\prime}\prime}}^{\beta^{\prime}j^{\prime}}-b_{\gamma^{\prime}k^{\prime}%
}^{\delta^{\prime}l^{\prime}}D_{\delta^{\prime}l^{\prime}}^{\beta^{\prime
}j^{\prime}}\right)  \\
&  =\delta_{\alpha^{\prime}}^{\beta^{\prime}}%
{\displaystyle\sum}
Y_{i^{\prime}}^{k^{\prime}}\left(  \gamma^{\mu}\partial_{\mu}W_{k^{\prime}%
}^{j^{\prime}}\right)  \nonumber\\
&  \equiv\delta_{\alpha^{\prime}}^{\beta^{\prime}}\gamma^{\mu}\left(  V_{\mu
}\right)  _{i^{\prime}}^{j^{\prime}}\nonumber
\end{align}
We shall require that the field $A$ is unimodular
\begin{equation}
\text{Tr}\left(  A\right)  =0.
\end{equation}
This condition turns out to be equivalent to the cancelation of all chiral
anomalies. In this respect, it is important to understand the connection
between chiral anomalies and the unimodularity conditions and we refer to \cite{chams1} and \cite{lizzi}. In fact we shall
discuss below in \S \ref{conclu} the meaning of this unimodularity condition.  This condition
implies that
\begin{equation}
\left(  A_{\mu}\right)  _{\alpha I}^{\alpha I}+\left(  A_{\mu}\right)
_{\alpha^{\prime}I^{\prime}}^{\alpha^{\prime}I^{\prime}}=0.
\end{equation}
Thus
\begin{equation}
-\frac{i}{2}g_{1}B_{\mu}+\left(  V_{\mu}\right)  _{i^{\prime}}^{i^{\prime}}=0
\end{equation}
and we can write
\begin{equation}
\left(  V_{\mu}\right)  _{i^{\prime}}^{j^{\prime}}=\frac{i}{6}g_{1}B_{\mu
}\delta_{i^{\prime}}^{j^{\prime}}+\frac{i}{2}g_{3}V_{\mu}^{m}\left(
\lambda^{m}\right)  _{i^{\prime}}^{j^{\prime}}%
\end{equation}
where $\left(  \lambda^{m}\right)  _{i}^{j}$ are the $8$ Gell-Mann matrices.

The mixed components are
\begin{align}
A_{\alpha I}^{\beta^{\prime}J^{\prime}}  &  =%
{\displaystyle\sum}
a_{\alpha I}^{\gamma K}\left(  D_{\gamma K}^{\delta^{\prime}L^{\prime}%
}b_{\delta^{\prime}L^{\prime}}^{\beta^{\prime}J^{\prime}}-b_{\gamma K}^{\delta
L}D_{\delta L}^{\beta^{\prime}J^{\prime}}\right) \\
&  =%
{\displaystyle\sum}
\left(  a_{\alpha I}^{\overset{.}{1}1}b_{\overset{.}{1^{\prime}}1^{\prime}%
}^{\beta^{\prime}J^{\prime}}-a_{\alpha I}^{\overset{.}{1}1}b_{\overset{.}{1}%
1}^{\overset{.}{1}1}\delta_{\overset{.}{1^{\prime}}}^{\beta^{\prime}}%
\delta_{1^{\prime}}^{J^{\prime}}\right)  D_{\overset{.}{1}1}^{\overset
{.}{1^{\prime}}1^{\prime}}\nonumber\\
&  =\delta_{\alpha}^{\overset{.}{1}}\delta_{I}^{1}\delta_{\overset
{.}{1^{\prime}}}^{\beta^{\prime}}\delta_{1^{\prime}}^{J^{\prime}}%
{\displaystyle\sum}
X_{\overset{.}{1}}^{\overset{.}{1}}\left(  W_{1^{\prime}}^{1^{\prime}%
}-Z_{\overset{.}{1}}^{\overset{.}{1}}\right) \nonumber\\
&  =0\nonumber
\end{align}
Thus whatever field would be placed in the mixed component of the Dirac
operator would stay unperturbed.

Evaluating the matrix $JAJ$ we now have%
\begin{align}
\left(  JAJ^{-1}\right)  _{\overset{.}{1}1}^{\overset{.}{1}1}  &  =\frac{i}%
{2}g_{1}\gamma^{\mu}B_{\mu}\\
\left(  JAJ^{-1}\right)  _{\overset{.}{2}1}^{\overset{.}{2}1}  &  =\frac{i}%
{2}g_{1}\gamma^{\mu}B_{\mu}\\
\left(  JAJ^{-1}\right)  _{a1}^{b1}  &  =\frac{i}{2}g_{1}\gamma^{\mu}B_{\mu
}\delta_{a}^{b}\\
\left(  JAJ^{-1}\right)  _{\alpha i}^{\beta j}  &  =\left(  -\frac{i}{6}%
g_{1}\gamma^{\mu}B_{\mu}\delta_{i}^{j}-\frac{i}{2}g_{3}\gamma^{\mu}V_{\mu}%
^{m}\left(  \lambda^{m}\right)  _{i}^{j}\right)  \delta_{\alpha}^{\beta}%
\end{align}
Adding $D+A+JAJ^{-1}$ gives the Dirac operator including inner fluctuations.
All the components are listed in the appendix A. It is important to note that
we have obtained all the correct representations of the fermions, with the
correct quantum numbers, including all hypercharges. We stress that the
unimodularity condition is essential for obtaining the correct hypercharge assignments.

At this point we can give more details about the cases which were ignored when
we restricted our choice of the algebra to the ``Symplectic-Unitary". These are
the five possibilities%
\begin{align}
&  M_{4}\left(  \mathbb{C}\right)  \oplus M_{4}\left(  \mathbb{C}\right)
\text{ }\\
&  \text{ }M_{4}\left(  \mathbb{R}\right)  \text{ }\oplus M_{4}\left(
\mathbb{C}\right) \\
&  \text{ }M_{2}\left(  \mathbb{H}\right)  \oplus\text{ }M_{2}\left(
\mathbb{H}\right) \\
&  \text{ }M_{4}\left(  \mathbb{R}\right)  \text{ }\oplus M_{2}\left(
\mathbb{H}\right) \\
&  M_{4}\left(  \mathbb{R}\right)  \oplus\text{ }M_{4}\left(  \mathbb{R}%
\right)
\end{align}
The last three cases can be discarded immediately. This can be seen as
follows. If grading is imposed on $M_{4}\left(  \mathbb{R}\right)  $ then this
will break the algebra into $M_{2}\left(  \mathbb{R}\right)  \oplus
M_{2}\left(  \mathbb{R}\right)  $ corresponding to the leptonic group
$SO(2)\times SO(2)$ which cannot accommodate the weak symmetry $SU\left(
2\right)  .$ If grading is imposed on $M_{2}\left(  \mathbb{H}\right)  $ this
will break the algebra into $\mathbb{H}\oplus$ $\mathbb{H}$ corresponding to
the group $SU(2)\times SU(2)$ and in this case the color group $SU\left(
3\right)  $ could not be accommodated. The same reason also hold for the
second case with the difference that $M_{4}\left(  \mathbb{C}\right)  $ will
break into $M_{2}\left(  \mathbb{C}\right)  \oplus M_{2}\left(  \mathbb{C}%
\right)  .$ Thus we are left only with the first possibility. This case must
be analyzed. When grading is imposed on the first algebra $M_{4}\left(
\mathbb{C}\right)  $ it will break into $M_{2}\left(  \mathbb{C}\right)
\oplus M_{2}\left(  \mathbb{C}\right)  .$ The condition that there is non
trivial mixing between fermions and conjugate fermions, would then break the
algebra $M_{2}\left(  \mathbb{C}\right)  \oplus M_{2}\left(  \mathbb{C}%
\right)  \oplus M_{4}\left(  \mathbb{C}\right)  $ into
\begin{equation}
\mathbb{C}\oplus\mathbb{C}^{^{\prime}}\oplus M_{2}\left(  \mathbb{C}\right)
\oplus\mathbb{C}\oplus M_{3}\left(  \mathbb{C}\right)
\end{equation}
where two of the algebras $\mathbb{C}$ must be identified to satisfy the first
order condition. This identification then implies that the first component of
the spinor $\psi_{A}$ will become neutral with respect to all gauge fields.
Working the components of the gauge field $A=%
{\displaystyle\sum}
a\left[  D,b\right]  $ as was carried out in the $M_{2}\left(  \mathbb{H}%
\right)  \oplus M_{4}\left(  \mathbb{C}\right)  $ case shows that we will get
two complex Higgs fields instead of one, because in this case the different
components will not be related by the quaternionic conditions. If in addition
the unimodularity condition is imposed restricting the unitary action of the
algebra to be $SU\left(  \mathcal{A}\right)  $ the gauge group becomes
\begin{equation}
U(1)^{3}\times SU(2)\times SU\left(  3\right)
\end{equation}
and there are two additional $U(1)$ gauge fields to those of the standard
model. In normal situations it is possible to take one of the $U(1)$ to be the
hypercharge and the other $U(1)$ to be the $B-L$, however, the last $U(1)$ if
not truncated will be anomalous. However, in this case, because the neutrino
will be neutral with respect to the two additional $U(1)$ gauge fields implies
that these $U(1)$ are anomalous. Thus this case can only be discarded after
analyzing the model, and showing that it is inconsistent at the quantum level.
If the unimodularity condition is not imposed, then the gauge group becomes
\begin{equation}
U(1)^{4}\times SU(2)\times SU\left(  3\right)
\end{equation}
which will also be anomalous. It remains to be seen whether in these cases a
Green-Schwarz mechanism can be employed to cancel one of the $U(1)$ anomalies.
From this analysis, it should be clear that the only compelling case to
consider is when the "Symplectic-Unitary" symmetry is imposed together with
the unimodularity condition.

\bigskip

\section{Spectral Action}

\medskip

The relevant Dirac operator is $D_{A}$
which includes both inner and outer automorphisms. The fermionic part of the
action is simple and of the Dirac type. Since our considerations are
Euclidean, one cannot impose the Majorana condition
\begin{equation}
J\,\psi=\psi
\end{equation}
as this could only be done in the Minkowski case. The appropriate action turns
out to be given by
\begin{equation}
\left\langle J\,\psi,D_{A}\psi\right\rangle
\end{equation}
which is an antisymmetric bilinear form. To show this we have%
\begin{align}
\left\langle J\,\zeta^{\prime},D_{A}\zeta\right\rangle  &  =-\left\langle
J\,\zeta^{\prime},J^{2}D_{A}\zeta\right\rangle \\
&  =-\left\langle JD_{A}\zeta,\zeta^{\prime}\right\rangle =-\left\langle
D_{A}J\,\zeta,\zeta^{\prime}\right\rangle \\
&  =-\left\langle J\,\zeta,D_{A}\zeta^{\prime}\right\rangle
\end{align}
where $\zeta,$ $\zeta^{\prime}\in\mathcal{H}$ are commuting sections, and
where we have used $J\,^{2}=-1,$ the unitarity of $J$%
\begin{equation}
\left\langle J\,\zeta^{\prime},J\zeta\right\rangle =\left\langle \zeta
,\zeta^{\prime}\right\rangle
\end{equation}
and the hermiticity of $D_{A}.$ Because of the anticommutativity of the
Grassmann variables $\psi$ the expression $\left\langle J\,\psi,D_{A}%
\psi\right\rangle $ is nonzero. Moreover one can impose the chirality
condition because
\begin{equation}
\gamma JD=JD\,\gamma
\end{equation}
The path integral
\begin{equation}%
{\displaystyle\int}
\exp\left(  -\frac{1}{2}\left\langle J\,\psi,D_{A}\psi\right\rangle \right)
D\,\psi=\mathrm{Pf}\left(  D_{A}\right)
\end{equation}
where the Pfaffian is the square root of the determinant. Thus it is possible
to integrate only the chiral fermions $\psi$ and the correct degrees of
freedom are obtained because of the appearance of the Pfaffian.

All details of the standard model as well as its unification with gravity are
achieved by postulating the action
\begin{equation}
\frac{1}{2}\left\langle J\,\psi,D_{A}\psi\right\rangle +\mathrm{Trace}%
\,f\left(  D_{A}/\Lambda\right)
\end{equation}
where $\Lambda$\ is some scale to be determined, and the trace is taken over
all eigenvalues below the scale $\Lambda.$ We restrict the function $f$ \ to
be even and positive. It can be shown, using heat kernel methods that this
trace can be expressed in terms of the geometrical Seeley deWitt coefficients,%
\begin{equation}
\mathrm{Trace}\,f\left(  D_{A}/\Lambda\right)  =%
{\displaystyle\sum\limits_{n=0}^{\infty}}
F_{4-n}\Lambda^{4-n}a_{n}%
\end{equation}
where the function $F$ is defined by $F(u)=f\,(v)$ where $u=v^{2},$ thus
$F(D^{2})=f\,(D)$. We define
\begin{equation}
f_{k}=%
{\displaystyle\int_{0}^{\infty}}
f\left(  v\right)  v^{k-1}dv,\qquad k>0
\end{equation}
then
\begin{align}
F_{4}  &  =%
{\displaystyle\int_{0}^{\infty}}
F(u)udu=2%
{\displaystyle\int_{0}^{\infty}}
f(v)v^{3}dv=2f_{4}\\
F_{2}  &  =%
{\displaystyle\int_{0}^{\infty}}
F(u)du=2%
{\displaystyle\int_{0}^{\infty}}
f(v)vdv=2f_{2}\\
F_{0}  &  =F(0)=f\,(0)=f_{0}\\
F_{-2n}  &  =\left(  -1\right)  ^{n}F^{\left(  n\right)  }\left(  0\right)
=\left[  \left(  -1\right)  ^{n}\left(  \frac{1}{2v}\frac{d}{dv}\right)
^{n}f\right]  \left(  0\right)  \qquad n\geq1
\end{align}
The $a_{n}$ are the Seeley deWitt coefficients, and fortunately are given by
general formulas for any second order elliptic differential operator. These
formulas, derived by Gilkey, can be conveniently used in our case. The first
step is to expand $D^{2}$ into the form
\begin{equation}
D^{2}=-\left(  g^{\mu\nu}\partial_{\mu}\partial_{\nu}+\mathcal{A}^{\mu
}\partial_{\mu}+B\right)
\end{equation}
and from this extract the connection $\omega_{\mu}$
\begin{equation}
D^{2}=-\left(  g^{\mu\nu}\nabla_{\mu}\nabla_{\nu}+E\right)
\end{equation}
where
\begin{equation}
\nabla_{\mu}=\partial_{\mu}+\mathbb{\omega}_{\mu}.
\end{equation}
This gives
\begin{align}
\mathbb{\omega}_{\mu}  &  =\frac{1}{2}g_{\mu\nu}\left(  \mathcal{A}^{\nu
}+\Gamma^{\nu}\right) \\
E  &  =B-g^{\mu\nu}\left(  \partial_{\mu}\mathbb{\omega}_{\nu}+\mathbb{\omega
}_{\mu}\mathbb{\omega}_{\nu}-\Gamma_{\mu\nu}^{\rho}\mathbb{\omega}_{\rho
}\right) \\
\Omega_{\mu\nu}  &  =\partial_{\mu}\mathbb{\omega}_{\nu}-\partial_{\nu
}\mathbb{\omega}_{\mu}+\left[  \mathbb{\omega}_{\mu},\mathbb{\omega}_{\nu
}\right]
\end{align}
where $\Gamma^{\nu}=g^{\rho\sigma}\Gamma_{\rho\sigma}^{\nu}$ and $\Gamma
_{\mu\nu}^{\rho}$ is the Christoffel connection of the metric $g_{\mu\nu}.$
The first few Seeley-deWitt coefficients $a_{n}$ for manifolds without
boundary, are given by%
\begin{equation}
a_{0}=\frac{1}{16\pi^{2}}%
{\displaystyle\int}
d^{4}x\sqrt{g}\text{\textrm{Tr}}\left(  1\right)
\end{equation}%
\begin{equation}
a_{2}=\frac{1}{16\pi^{2}}%
{\displaystyle\int}
d^{4}x\sqrt{g}\text{\textrm{Tr}}\left(  E+\frac{1}{6}R\right)
\end{equation}%
\begin{align}
a_{4}  &  =\frac{1}{16\pi^{2}}\frac{1}{360}%
{\displaystyle\int\limits_{M}}
d^{4}x\sqrt{g}\,\mathrm{Tr}\left(  12R_{;\mu}^{\;\;\mu}+5R^{2}-2R_{\mu\nu
}R^{\mu\nu}\right. \\
&  \quad\left.  +\,2R_{\mu\nu\rho\sigma}R^{\mu\nu\rho\sigma}+60RE+180E^{2}%
+60E_{;\mu}^{\quad\mu}+30\,\Omega_{\mu\nu}\Omega^{\mu\nu}\right) \nonumber
\end{align}
while the odd ones all vanish for manifolds without boundary
\begin{equation}
a_{2n+1}=0.
\end{equation}
We will deal with higher order terms such as $a_{6}$ later. Using these
formulas, it is simple and straightforward to compute the spectral action.
Having listed all the matrix components of the Dirac operator $D_{M}^{N}$ \ we
now proceed to evaluate the matrix $D^{2}$%
\begin{align}
\left(  D^{2}\right)  _{A}^{B}  &  =D_{A}^{C}D_{C}^{B}+D_{A}^{C^{\prime}%
}D_{C^{\prime}}^{B}\\
\left(  D^{2}\right)  _{A}^{B^{\prime}}  &  =D_{A}^{C}D_{C}^{B^{\prime}}%
+D_{A}^{C^{\prime}}D_{C^{\prime}}^{B^{\prime}}\\
\left(  D^{2}\right)  _{A^{\prime}}^{B}  &  =D_{A^{\prime}}^{C}D_{C}%
^{B}+D_{A^{\prime}}^{C^{\prime}}D_{C^{\prime}}^{B}\\
\left(  D^{2}\right)  _{A^{\prime}}^{B^{\prime}}  &  =D_{A^{\prime}}^{C}%
D_{C}^{B^{\prime}}+D_{A^{\prime}}^{C^{\prime}}D_{C^{\prime}}^{B^{\prime}}%
\end{align}
We can use the properties
\begin{equation}
D_{A^{\prime}}^{B^{\prime}}=\overline{D}_{A}^{B},\qquad D_{A^{\prime}}%
^{B}=\overline{D}_{A}^{B^{\prime}},\qquad D_{A}^{B^{\prime}}=\overline
{D}_{A^{\prime}}^{B}%
\end{equation}
and thus it will not be necessary to compute all the traces by taking
advantage of the fact that some of the traces will be related to each other by
complex conjugation. As an example we calculate the first few component of
$D^{2}$
\begin{align}
\left(  D^{2}\right)  _{\overset{.}{1}1}^{\overset{.}{1}1}  &  =D_{\overset
{.}{1}1}^{\overset{.}{1}1}D_{\overset{.}{1}1}^{\overset{.}{1}1}+D_{\overset
{.}{1}1}^{a1}D_{a1}^{\overset{.}{1}1}+k^{\ast\nu_{R}}k^{\nu_{R}}\sigma^{2}\\
&  =\gamma^{\mu}D_{\mu}\gamma^{\nu}D_{\nu}\otimes1_{3}+k^{\nu\ast}k^{\nu}%
H_{a}\overline{H}^{a}+k^{\ast\nu_{R}}k^{\nu_{R}}\sigma^{2}\nonumber
\end{align}%
\begin{align}
\left(  D^{2}\right)  _{\overset{.}{1}1}^{a1}  &  =D_{\overset{.}{1}%
1}^{\overset{.}{1}1}D_{\overset{.}{1}1}^{a1}+D_{\overset{.}{1}1}^{b1}%
D_{b1}^{a1}\\
&  =\gamma^{\mu}D_{\mu}\gamma_{5}k^{\ast\nu}\epsilon^{ab}H_{b}+\gamma
_{5}k^{\ast\nu}\epsilon^{bc}H_{c}\gamma^{\mu}\left(  \left(  D_{\mu}+\frac
{i}{2}g_{1}B_{\mu}\right)  \delta_{b}^{a}-\frac{i}{2}g_{2}W_{\mu}^{\alpha
}\left(  \sigma^{\alpha}\right)  _{b}^{a}\right) \nonumber\\
&  =\gamma^{\mu}\gamma_{5}k^{\ast\nu}\epsilon^{ab}\nabla_{\mu}H_{b}\nonumber
\end{align}
where%
\begin{equation}
\nabla_{\mu}H_{a}=\left(  \left(  \partial_{\mu}-\frac{i}{2}g_{1}B_{\mu
}\right)  \delta_{b}^{a}-\frac{i}{2}g_{2}W_{\mu}^{\alpha}\left(
\sigma^{\alpha}\right)  _{b}^{a}\right)  H_{b}%
\end{equation}
We list all the components of $D^{2}$ in Appendix B. \ Using the form for
$D^{2}$
\begin{equation}
D^{2}=-\left(  g^{\mu\nu}\partial_{\mu}\partial_{\nu}+\mathcal{A}^{\mu
}\partial_{\mu}+B\right)
\end{equation}
we can read then $\mathbb{\omega}_{\mu}$ and $E$
\begin{align}
\mathbb{\omega}_{\mu}  &  =\frac{1}{2}g_{\mu\nu}\left(  \mathcal{A}^{\nu
}+\Gamma^{\nu}\right) \\
E  &  =B-g^{\mu\nu}\left(  \partial_{\mu}\mathbb{\omega}_{\nu}+\mathbb{\omega
}_{\mu}\mathbb{\omega}_{\nu}-\Gamma_{\mu\nu}^{\rho}\mathbb{\omega}_{\rho
}\right)
\end{align}
The matrix elements $\left(  \mathbb{\omega}_{\mu}\right)  _{M}^{N}$ and
$\left(  E\right)  _{M}^{N}$ are listed in Appendix C. It is now possible to
summarize the results
\begin{equation}
-\frac{1}{2}\text{Tr }\left(  E\right)  =4\left[  12R+2a\overline
{H}H+c\,\sigma^{2}\right]
\end{equation}%
\begin{align}
\frac{1}{2}\text{tr }\left(  E^{2}\right)   &  =4\left[  5g_{1}^{2}B_{\mu\nu
}^{2}+3g_{2}^{2}\left(  W_{\mu\nu}^{\alpha}\right)  ^{2}+3g_{3}^{2}\left(
V_{\mu\nu}^{m}\right)  ^{2}+3R^{2}+aR\overline{H}H\right. \\
&  \left.  +\frac{1}{2}cR\sigma^{2}+2b\left(  \overline{H}H\right)
^{2}+2a\left\vert \nabla_{\mu}H_{a}\right\vert ^{2}+4e\overline{H}%
H\,\sigma^{2}+c\left(  \partial_{\mu}\sigma\right)  ^{2}+d\,\sigma^{4}\right]
\nonumber
\end{align}%
\begin{equation}
\frac{1}{2}\text{Tr }\left(  \Omega_{\mu\nu}^{2}\right)  _{M}^{M}=4\left[
-6R_{\mu\nu\rho\sigma}^{2}-10g_{1}^{2}B_{\mu\nu}^{2}-6g_{2}^{2}\left(
W_{\mu\nu}^{\alpha}\right)  ^{2}-6g_{3}^{2}\left(  V_{\mu\nu}^{m}\right)
^{2}\right]
\end{equation}
where
\begin{align}
a  &  =\text{\textrm{tr}}\left(  k^{\ast\nu}k^{\nu}+k^{\ast e}k^{e}+3\left(
k^{\ast u}k^{u}+k^{\ast d}k^{d}\right)  \right) \\
b  &  =\text{\textrm{tr}}\left(  \left(  k^{\ast\nu}k^{\nu}\right)
^{2}+\left(  k^{\ast e}k^{e}\right)  ^{2}+3\left(  \left(  k^{\ast u}%
k^{u}\right)  ^{2}+\left(  k^{\ast d}k^{d}\right)  ^{2}\right)  \right) \\
c  &  =\text{\textrm{tr}}\left(  k^{\ast\nu_{R}}k^{\nu_{R}}\right) \\
d  &  =\text{\textrm{tr}}\left(  \left(  k^{\ast\nu_{R}}k^{\nu_{R}}\right)
^{2}\right) \\
e  &  =\text{\textrm{tr}}\left(  k^{\ast\nu}k^{\nu}k^{\ast\nu_{R}}k^{\nu_{R}%
}\right)
\end{align}
The first two Seeley-deWitt coefficients are
\begin{align}
a_{0}  &  =\frac{1}{16\pi^{2}}%
{\displaystyle\int}
d^{4}x\sqrt{g}\text{\textrm{Tr}}\left(  1\right) \\
&  =\frac{24}{\pi^{2}}%
{\displaystyle\int}
d^{4}x\sqrt{g}\nonumber
\end{align}%
\begin{align}
a_{2}  &  =\frac{1}{16\pi^{2}}%
{\displaystyle\int}
d^{4}x\sqrt{g}\text{\textrm{Tr}}\left(  E+\frac{1}{6}R\right) \\
&  =-\frac{2}{\pi^{2}}%
{\displaystyle\int}
d^{4}x\sqrt{g}\left(  R+\frac{1}{2}a\overline{H}H+\frac{1}{4}c\,\sigma
^{2}\right) \nonumber
\end{align}%
\begin{align}
a_{4}  &  =\frac{1}{2\pi^{2}}%
{\displaystyle\int}
d^{4}x\sqrt{g}\left[  -\frac{3}{5}C_{\mu\nu\rho\sigma}^{\,2}+\frac{11}%
{30}R^{\ast}R^{\ast}+\frac{5}{3}g_{1}^{2}B_{\mu\nu}^{2}+g_{2}^{2}\left(
W_{\mu\nu}^{\alpha}\right)  ^{2}+g_{3}^{2}\left(  V_{\mu\nu}^{m}\right)
^{2}\right. \\
&  \qquad\qquad\qquad+\frac{1}{6}aR\overline{H}H+b\left(  \overline
{H}H\right)  ^{2}\sigma^{2}+a\left\vert \nabla_{\mu}H_{a}\right\vert
^{2}+2e\overline{H}H\,\sigma^{2}\nonumber\\
&  \qquad\qquad\qquad\left.  +\frac{1}{2}d\,\sigma^{4}+\frac{1}{12}%
cR\,\sigma^{2}+\frac{1}{2}c\left(  \partial_{\mu}\sigma\right)  ^{2}-\frac
{2}{5}R_{;\mu}^{\,\,;\mu}-\frac{a}{3}\left(  \overline{H}H\right)  _{;\mu
}^{\,\,;\mu}-\frac{c}{6}\left(  \sigma^{2}\right)  _{;\mu}^{\,\,;\mu}\right]
\nonumber
\end{align}
Thus the bosonic spectral action to second order is given by%
\begin{equation}
S=F_{4}\Lambda^{4}a_{0}+F_{2}\Lambda^{2}a_{2}+F_{0}a_{4}+F_{-2}\Lambda
^{-2}a_{6}+\cdots
\end{equation}
and
\begin{align}
S_{\mathrm{b}}  &  =\frac{24}{\pi^{2}}F_{4}\Lambda^{4}%
{\displaystyle\int}
d^{4}x\sqrt{g}\\
&  -\frac{2}{\pi^{2}}F_{2}\Lambda^{2}%
{\displaystyle\int}
d^{4}x\sqrt{g}\left(  R+\frac{1}{2}a\overline{H}H+\frac{1}{4}c\sigma
^{2}\right) \nonumber\\
&  +\frac{1}{2\pi^{2}}F_{0}%
{\displaystyle\int}
d^{4}x\sqrt{g}\left[  \frac{1}{30}\left(  -18C_{\mu\nu\rho\sigma}%
^{2}+11R^{\ast}R^{\ast}\right)  +\frac{5}{3}g_{1}^{2}B_{\mu\nu}^{2}+g_{2}%
^{2}\left(  W_{\mu\nu}^{\alpha}\right)  ^{2}+g_{3}^{2}\left(  V_{\mu\nu}%
^{m}\right)  ^{2}\right. \nonumber\\
&  \qquad\left.  +\frac{1}{6}aR\overline{H}H+b\left(  \overline{H}H\right)
^{2}+a\left\vert \nabla_{\mu}H_{a}\right\vert ^{2}+2e\overline{H}H\,\sigma
^{2}+\frac{1}{2}d\,\sigma^{4}+\frac{1}{12}cR\sigma^{2}+\frac{1}{2}c\left(
\partial_{\mu}\sigma\right)  ^{2}\right] \nonumber\\
&  +F_{-2}\Lambda^{-2}a_{6}+\cdots\nonumber
\end{align}
It is worth to also summarize the the fermionic action
\begin{align}
S_{\mathrm{f}}  &  =\nu_{R}^{\ast}\gamma^{\mu}D_{\mu}\nu_{R}\\
+  &  e_{R}^{\ast}\gamma^{\mu}\left(  D_{\mu}+ig_{1}B_{\mu}\right)
e_{R}\nonumber\\
+  &  l_{L}^{a\ast}\gamma^{\mu}\left(  \left(  D_{\mu}+\frac{i}{2}g_{1}B_{\mu
}\right)  \delta_{a}^{b}-\frac{i}{2}g_{2}W_{\mu}^{\alpha}\left(
\sigma^{\alpha}\right)  _{a}^{b}\right)  l_{_{b}L}\nonumber\\
+  &  u_{R\ }^{i\ast}\gamma^{\mu}\left(  \left(  D_{\mu}-\frac{2i}{3}%
g_{1}B_{\mu}\right)  \delta_{i}^{j}-\frac{i}{2}g_{3}V_{\mu}^{m}\left(
\lambda^{m}\right)  _{i}^{j}\right)  u_{jR}\nonumber\\
+  &  d_{R}^{i\ast}\gamma^{\mu}\left(  \left(  D_{\mu}+\frac{i}{3}g_{1}B_{\mu
}\right)  \delta_{i}^{j}-\frac{i}{2}g_{3}V_{\mu}^{m}\left(  \lambda
^{m}\right)  _{i}^{j}\right)  d_{jR}\nonumber\\
+  &  q_{L}^{ia\ast}\gamma^{\mu}\left(  \left(  D_{\mu}-\frac{i}{6}g_{1}%
B_{\mu}\right)  \delta_{a}^{b}\delta_{i}^{j}-\frac{i}{2}g_{2}W_{\mu}^{\alpha
}\left(  \sigma^{\alpha}\right)  _{a}^{b}\delta_{i}^{j}-\frac{i}{2}g_{3}%
V_{\mu}^{m}\left(  \lambda^{m}\right)  _{i}^{j}\delta_{a}^{b}\right)
q_{jbL}\nonumber\\
&  +\nu_{R}^{\ast}\gamma_{5}k^{\ast\nu}\epsilon^{ab}H_{b}l_{_{a}L}+e_{R}%
^{\ast}\gamma_{5}k^{\ast e}\overline{H}^{a}l_{_{a}L}\nonumber\\
&  +u_{R\ }^{i\ast}\gamma_{5}k^{\ast u}\epsilon^{ab}H_{b}\delta_{i}^{j}%
q_{jaL}+d_{R}^{i\ast}\gamma_{5}k^{\ast d}\overline{H}^{a}\delta_{i}^{j}%
q_{jaL}+\nu_{R}^{\ast}\gamma_{5}k^{\ast\nu_{R}}\sigma\left(  \nu_{R}^{\ast
}\right)  ^{c}+\mathrm{h.c}\nonumber
\end{align}
Our strategy is to use the spectral action as an effective action at a fixed
scale, of the order of the unification scale, and to impose the additional
relations between the independent parameters of the Standard Model coupled to
gravity as a boundary condition at that scale. One can then let these
parameters run down using the RG equations to their value at ordinary scale.
As a first example one has the unification of the three gauge couplings in the
form
\begin{equation}
g_{3}^{2}=g_{2}^{2}=\frac{5}{3}\,g_{1}^{2}%
\end{equation}
In applying the above strategy we have limited ourselves to the first three
terms in the expansion of the spectral action, the reason being that the
natural spectral functions $F(D^{2}/\Lambda^{2})$ used in the spectral action
are meant to count the number of eigenvalues of $D^{2}$ which are less than
$\Lambda^{2}$. These functions are \textquotedblleft cutoff" functions which
are completely flat near $0$ and thus have all their Taylor coefficients
$F^{(n)}(0)$ vanishing except $F(0)$. The question of whether we should
ignore the higher order terms will be  dealt with in Part II.

\section{QFT analysis: zeroth order}

As we have seen in the previous section, the spectral action gives a well
defined form for all interactions. We shall take the Wilsonian point of view
where the geometrical action is considered as an effective theory valid at
some scale $\Lambda,$ which is related to the low energy action by running all
masses and coupling constants as determined by the RG equations. \ The special
relations that exist between the different coupling constants are taken as
boundary conditions for the integration of the RG equations. To check the
validity of the model, we examine consequences of \ these boundary conditions.
We shall assume that the spectral action is determined by the cutoff function,
so that all higher order terms in the heat kernel expansion are truncated to
zero. In this case, the normalization of the kinetic terms imposes a relation
between the coupling constants {$g_{1}$, $g_{2}$, $g_{3}$} and the coefficient
$F_{0}$, of the form
\begin{equation}
{\frac{g_{3}^{2}\,F_{0}}{2\pi^{2}}=\frac{1}{4},\ \ \ \ \ \ g_{3}^{2}=g_{2}%
^{2}=\frac{5}{3}\,g_{1}^{2}\,.}%
\end{equation}
This gives that {$\sin^{2}\theta_{W}=\frac{3}{8}$} a value also obtained in
{$SU(5)$ }and {$SO(10)$} grand unified theories. The three momenta of the
function $F${$_{0},F_{2}$} and $F${$_{4}$} can be used to specify the initial
conditions on the gauge couplings, the {Newton constant} and the {cosmological
constant}. The fine structure constant {$\alpha_{em}$ }is thus given by
\begin{equation}
{\alpha_{em}=\,\sin(\theta_{w})^{2}\,\alpha_{2}\,,\quad\alpha_{i}=\frac
{g_{i}^{2}}{4\pi}}%
\end{equation}
Its infrared value is {$\sim1/137.036$} but it is running as a function of the
energy and increases to the value {$\alpha_{em}(M_{Z})=1/128.09$ }already, at
the energy {$M_{Z}\sim91.188$ Gev.}

Assuming the {\textquotedblleft big desert\textquotedblright} hypothesis, the
running of the three couplings $\alpha_{i}$ is known. With 1-loop corrections
only, it is given by
\begin{equation}
{\beta_{g_{i}}=(4\pi)^{-2}\,b_{i}\,g_{i}^{3},\ \ \ {\hbox{ with }}%
\ \ b=(\frac{41}{6},-\frac{19}{6},-7),}%
\end{equation}
so that {%
\begin{align}
\alpha_{1}^{-1}(\Lambda)  &  =\,\alpha_{1}^{-1}(M_{Z})-\frac{41}{12\pi}%
\,\log\,\frac{\Lambda}{M_{Z}}\\
\alpha_{2}^{-1}(\Lambda)  &  =\,\alpha_{2}^{-1}(M_{Z})+\frac{19}{12\pi}%
\,\log\,\frac{\Lambda}{M_{Z}}\\
\alpha_{3}^{-1}(\Lambda)  &  =\,\alpha_{3}^{-1}(M_{Z})+\frac{42}{12\pi}%
\,\log\,\frac{\Lambda}{M_{Z}}%
\end{align}
} where $M_{Z}$ is the mass of the $Z^{0}$ vector boson.

\smallskip

It is known that the predicted unification of the coupling constants does not
hold exactly. In fact, if one considers the actual experimental values {%
\begin{equation}
g_{1}(M_{Z})=0.3575,\ \ \ g_{2}(M_{Z})=0.6514,\ \ \ g_{3}(M_{Z})=1.221,
\end{equation}
} one obtains the values
\begin{equation}
\alpha_{1}(M_{Z})=0.0101,\ \ \ \alpha_{2}(M_{Z})=0.0337,\ \ \ \alpha_{3}%
(M_{Z})=0.1186.
\end{equation}
and one knows that the graphs of the running of the three constants
$\alpha_{i}$ do not meet exactly, hence do not specify a unique unification
energy. The discrepancy comes mostly from the running of the $\alpha_{1}$
coupling as we should expect unification of the gauge couplings with the
Newton coupling near the Planck energy.

We first note that the relations between the gauge coupling constants, and the
RG equations are carried for the interactions obtained by  assuming that the
spectral function is a cut-off function, and thus suppressing all higher
order terms. In Part II we shall show that if the spectral function $F\left(  D^{2}\right)  $
  deviates by small perturbations from the cut-off function, higher order interactions can lead to small corrections which
alter the running of each of the gauge coupling constants.  In other words, we shall investigate the
possibility that all couplings are unified at $\Lambda\ $ provided that the
function $F$ \ is chosen appropriately, and higher order corrections from the
spectral action are included.

A distinctive feature of the spectral action is that the Higgs coupling is
proportional to the gauge couplings. \ This implies a restriction on its mass.
To see this consider the equation
\begin{equation}
\frac{d\lambda}{dt}=\lambda\gamma+\frac{1}{8\pi^{2}}(12\lambda^{2}+B)
\end{equation}
 where
\begin{align}
\gamma &  =\,\frac{1}{16\pi^{2}}(12y_{t}^{2}-9g_{2}^{2}-3g_{1}^{2})\\
B  &  =\,\frac{3}{16}(3g_{2}^{4}+2g_{1}^{2}\,g_{2}^{2}+g_{1}^{4})-3\,y_{t}%
^{4}\,.
\end{align}
The Higgs mass is then given by
\begin{equation}
m_{H}^{2}=\,8\lambda\,\frac{M^{2}}{g^{2}}\,,\quad m_{H}=\sqrt{2\lambda}%
\,\frac{2M}{g}%
\end{equation}
One can solve this equation numerically, provided the boundary condition for
$\lambda$ is given. This depends on the value of the gauge coupling at
unification, and where the unification scale is taken. If for example the
boundary value {$\lambda_{0}=0.356$} is taken at {$\Lambda=10^{17}$ Gev, this
}gives {$\lambda(M_{Z})\sim0.241$} and a Higgs mass of the order of {$170$ Gev
which is disfavored by experiment (and was even ruled out for some period).
This answer is sensitive to the value of the unification scale, and since we
expect that it can have substantial consequences to let the spectral
function deviate from the cutoff function, we should include the higher order
corrections to the spectral action in our analysis of the Higgs mass. A
reliable value for the mass of the Higgs depends on the form of the spectral
function, which in turn determines the unification scale.}

On the other hand, the mass of the top quark is governed by the top quark
Yukawa coupling $k^{t}$ through the equation
\begin{equation}
m_{\mathrm{top}}(t)=\frac{1}{\sqrt{2}}\frac{2M}{g}\,k^{t}=\frac{1}{\sqrt{2}%
}\,v\,k^{t},
\end{equation}
 where {$v=\frac{2M}{g}$ }is the vacuum expectation value of the Higgs field.
\ All fermions get their masses by coupling to the Higgs through interactions
of the form {
\begin{equation}
kH\overline{\psi}\psi
\end{equation}
} After normalizing the kinetic energy of the Higgs field through the
redefinition $H\rightarrow${$\frac{\pi}{\sqrt{aF_{0}}}H,$} the mass term
becomes
\begin{equation}
\frac{\pi}{\sqrt{F_{0}}}\frac{k}{\sqrt{a}}H\overline{\psi}\psi
\end{equation}
and we notice that $%
{\displaystyle\sum\limits_{i}}
\left(  \frac{k_{i}}{\sqrt{a}}\right)  ^{2}=1.$ This gives a relation among
the fermions masses and the W- mass
\begin{equation}
{\sum_{\mathrm{generations}}m_{e}^{2}+m_{\nu}^{2}+3m_{d}^{2}+3m_{u}^{2}%
=8M_{W}^{2}.}%
\end{equation}
If the value of $g$ at a unification scale of {$10^{17}$ Gev} is taken to be
{$\sim0.517$ and }neglecting the $\tau$ neutrino Yukawa coupling, we get {%
\begin{equation}
k^{t}=\,\frac{2}{\sqrt{3}}\,g\sim0.597\,.
\end{equation}
} The numerical integration of the differential equation gives a top quark
mass of the order of $\,179$ Gev, and the agreement with experiment becomes
quite good if one takes into account the Yukawa coupling for neutrinos as
explained in details in \cite{mc2}. This indicates that the top quark mass is
less sensitive than the Higgs mass to the unification scale ambiguities. This
could be related to the fact that the fermionic action is much simpler than
the bosonic one which is only determined by an infinite expansion whose
reliability depends on the convergence of the higher order terms.

\section{Parity violating terms}

It is possible to add to the spectral action terms that will violate parity
such as the gravitational term $\epsilon^{\mu\nu\rho\sigma}R_{\mu\nu
ab}R_{\rho\sigma}^{\quad ab}$ and the non-abelian $\theta$ term $\epsilon
^{\mu\nu\rho\sigma}V_{\mu\nu}^{m}V_{\rho\sigma}^{m}.$ These  arise by
allowing for the spectral action to include the term%
\begin{equation}
\text{Tr}\left(  \gamma G\,\left(\frac{D^{2}}{\Lambda^{2}}\right)  \right)
\end{equation}
where $G$ is a function not necessarily equal to the function $F,$ and
\begin{equation}
\gamma=\gamma_{5}\otimes\gamma_{F}%
\end{equation}
is the total grading. In this case it is easy to see that there are no
contributions coming from $a_{0}$ and $a_{2}$ and the first new term occurs in $a_{4}$ where there are only two contributions:%
\begin{equation}
\frac{1}{16\pi^{2}}\frac{1}{12}\text{Tr}\left(  \gamma_{5}\gamma_{F}%
\,\Omega_{\mu\nu}^{2}\right)  =\frac{1}{16\pi^{2}}\epsilon^{\mu\nu\rho\sigma
}R_{\mu\nu ab}R_{\rho\sigma}^{\quad ab}(24-24)=0
\end{equation}
and
\begin{align}
&  \frac{1}{16\pi^{2}}\frac{1}{2}\text{Tr}\left(  \gamma_{5}\gamma_{F}%
E^{2}\right) \\
&  =-\frac{4}{16\pi^{2}}\epsilon^{\mu\nu\rho\sigma}\left(  \left(  1-\left(
\frac{1}{2}\right)  ^{2}\left(  2\right)  +\left(  \frac{2}{3}\right)
^{2}\left(  3\right)  +\left(  \frac{1}{3}\right)  ^{2}\left(  3\right)
-\left(  \frac{1}{6}\right)  ^{2}\left(  3\right)  \left(  2\right)  \right)
3g_{1}^{2}B_{\mu\nu}B_{\rho\sigma}\right. \nonumber\\
&  \left.  \left(  -\left(  \frac{1}{2}\right)  ^{2}\left(  2\right)  -\left(
\frac{1}{2}\right)  ^{2}\left(  2\right)  \left(  3\right)  \right)
3g_{2}^{2}W_{\mu\nu}^{\alpha}W_{\rho\sigma}^{\alpha}+\left(  \left(  \frac
{1}{2}\right)  ^{2}\left(  2\right)  \left(  1+1-2\right)  \right)  3V_{\mu
\nu}^{m}V_{\rho\sigma}^{m}\right) \nonumber\\
&  =-\frac{3}{4\pi^{2}}\epsilon^{\mu\nu\rho\sigma}\left(  2g_{1}^{2}B_{\mu\nu
}B_{\rho\sigma}-2g_{2}^{2}W_{\mu\nu}^{\alpha}W_{\rho\sigma}^{\alpha}\right)
\nonumber
\end{align}
Thus the additional terms to the spectral action, up to orders $\frac
{1}{\Lambda^{2}},$ are
\begin{equation}
\frac{3G_{0}}{8\pi^{2}}\epsilon^{\mu\nu\rho\sigma}\left(  2g_{1}^{2}B_{\mu\nu
}B_{\rho\sigma}-2g_{2}^{2}W_{\mu\nu}^{\alpha}W_{\rho\sigma}^{\alpha}\right)
\end{equation}
where $G_{0}=G\left(  0\right)  .$ The $B_{\mu\nu}B_{\rho\sigma}$ is a surface
term, while $W_{\mu\nu}^{\alpha}W_{\rho\sigma}^{\alpha}$ is topological, and
both violate PC invariance. The surprising thing is the vanishing of both the
gravitational PC violating term $\epsilon^{\mu\nu\rho\sigma}R_{\mu\nu
ab}R_{\rho\sigma}^{\quad ab}$ and the $\theta$ \ QCD term $\epsilon^{\mu
\nu\rho\sigma}V_{\mu\nu}^{m}V_{\rho\sigma}^{m}.$ In this way the $\theta$
parameter is naturally zero, and can only be generated by the higher order
interactions. The reason behind the vanishing of both terms is that in these
two sectors there is a left-right symmetry graded with the matrix $\gamma_{F}$
giving an exact cancelation between the left-handed sectors and the
right-handed ones. In other words the trace of $\gamma_{F}$ vanishes and this
implies that the index of the full Dirac operator, using the total grading,
vanishes. There is one more condition to solve the strong CP problem which is
to have the following condition on the mass matrices of the up quark and down
quark
\begin{equation}
\det k^{u}\det k^{d}=\operatorname{real}.
\end{equation}
At present, it is not clear what condition must be imposed on the quarks Dirac
operator, in order to obtain such relation. If this condition can be imposed
naturally, then it will be possible to show that (\cite{Mohapatra})
\begin{equation}
\theta_{QT}+\theta_{QCD}=0
\end{equation}
at the tree level, and loop corrections can only change this by orders of less
than $10^{-9}.$

\section{Dilaton Interactions}

The scale $\Lambda$ appears as a free parameter in the spectral action. It is
more natural if it can arise as the vev of a dynamical field. We thus
introduce the dilaton field $\phi$ and replace the operator $D^{2}$ in the
spectral action by
\begin{equation}
P=e^{-\phi}D^{2}e^{-\phi}%
\end{equation}
A shift in the dilaton field $\phi\rightarrow\phi+\ln\Lambda$ transforms
$P\rightarrow\frac{1}{\Lambda^{2}}P.$ The interactions of the dilaton can be
determined by observing that geometrical constructs $\omega_{\mu}$ and $E$
\ that appeared in the heat kernel expansion for $D^{2}$ are related to
$\Omega_{\mu}$ and $\mathcal{E}$ of $P$ \ by
\begin{align}
\Omega_{\mu}  &  =\omega_{\mu}-2\partial_{\mu}\phi\\
\mathcal{E}  &  =e^{-2\phi}\left(  E+g^{\mu\nu}\nabla_{\mu}^{g}\nabla_{\nu
}^{g}\phi+\partial_{\mu}\phi\partial_{\nu}\phi\right)
\end{align}
where the covariant derivative $\nabla_{\mu}^{g}$ is with respect to the
metric $g_{\mu\nu}.$ We have shown that the first four terms in the spectral
action are independent of the dilaton field when expressed in the Einstein
frame with the metric
\begin{equation}
G_{\mu\nu}=g_{\mu\nu}e^{2\phi}%
\end{equation}
and in terms of a rescale Higgs field, except for one term which is the
dilaton kinetic energy. The Higgs fields are rescaled according to
\begin{equation}
H^{\prime}=He^{-\phi}%
\end{equation}
and the fermions according to
\begin{equation}
\psi^{\prime}=\psi e^{-\frac{3}{2}\phi}%
\end{equation}
From the relations between $\mathcal{E}$ and $E$ it should be clear that the
full potential of the theory can only get a scaling factor. This factor is
absorbed when the rescaled fields are used. In other words, the potential is
independent of the dilaton. Thus at the classical level, the vev of the
dilaton is undetermined. This situation changes when quantum radiative
corrections are taken into account. By taking the corrections to be at the
Planck scale, and assuming that there are also non-perturbative effects, one
finds that the vev of the dilaton is of order one in Planck units. It is
interesting to note that this model is exactly what became to be known as the
Randall-Sundrum model, although it was obtained in the noncommutative
formulation of the standard model long before that. In this picture the Higgs
fields $H$ gets a vev of the order of the Planck scale, however, the physical
field $H^{\prime}$ has its vev suppressed through the dilaton coupling
$e^{-\phi}$. Thus if $\left\langle \phi\right\rangle \sim40$ in Planck units,
then $e^{-\phi}\sim10^{-19}$. Thus the problem of explaining the very low mass
scale of fermion masses reduces to explaining the origin of a dilaton vev of
the order of $10^{2}.$

\section{Conclusions and Outlook}

We summarize the main assumptions made in determining the noncommutative space:

\begin{enumerate}
\item Space-time is a product of a continuous four-dimensional manifold times
a finite space.

\item One of the algebras $M_{4}\left(  \mathbb{C}\right)  $ is subject to
symplectic symmetry reducing it to $M_{2}\left(  \mathbb{H}\right)  .$

\item The commutator of the Dirac operator with the center of the algebra is
non trivial $\left[  D,Z\left(  \mathcal{A}\right)  \right]  $ $\neq0.$

\item The unitary algebra $U\left(  \mathcal{A}\right)  $ is restricted to
$SU\left(  \mathcal{A}\right)  .$
\end{enumerate}

\emph{These give rise to the following predictions:}

\begin{enumerate}
\item The number of fundamental fermions is $16.$

\item The algebra of the finite space is $\mathbb{C}\oplus\mathbb{H}\oplus
M_{3}\left(  \mathbb{C}\right)  .$

\item The correct representations of the fermions with respect to $SU(3)\times
SU(2)\times U(1)$.

\item Higgs doublet and spontaneous symmetry breaking mechanism. This is
highly non-trivial especially that the mass term of the Higgs field comes with
the correct negative sign.

\item Mass of the top quark compatible with experiment.

\item See-saw mechanism to give very light left-handed neutrinos.
\end{enumerate}

We give here a brief outline of open directions.

\subsection{The variant of the Einstein-Yang Mills system}\medskip \hfill\label{conclu}

Before the reduction to the subgroup $U(1)\times SU(2)\times SU(3)$ (coming from the order one condition
and the hypothesis of finite distance between the two copies of the four-dimensional manifold $M$) the model
one gets is the product of $M$ with the finite space whose algebra  is $A_F=M_{2}\left(  \mathbb{H}%
\right)  \oplus M_{4}\left(  \mathbb{C}\right)$. This model is thus very closely related to the
Einstein-Yang Mills system in which one simply replaces the algebra $C^\infty(M)$ of functions by the algebra $C^\infty(M)\otimes M_n(\mathbb{C})=M_n(C^\infty(M))$.

There are a number of reasons to take seriously the variant of the
Einstein-Yang Mills system  obtained with the algebra $A_F=M_{2}\left(  \mathbb{H}%
\right)  \oplus M_{4}\left(  \mathbb{C}\right)$.

\begin{enumerate}
  \item The first one is that, as was shown recently in \cite{Broek}, the usual Einstein-Yang Mills system  is closely related to supersymmetry
as suggested in \cite{cc2} and in particular the fermions are in the adjoint representation. While the corresponding $SU(n)$ model are far away from realistic models, the situation changes for the above variant with $A_F=M_{2}\left(  \mathbb{H}%
\right)  \oplus M_{4}\left(  \mathbb{C}\right)$ since the gauge group coming from the even part of the graded algebra is the $SU(2)_L\times SU(2)_R\times SU(4)$ of the Patti-Salam model which is much more realistic already.
  \item The second reason is that the $SU(4)$ which appears naturally from inner automorphisms of the algebra $A_F$ is a conceptual explanation for the ``unimodularity condition" which is an odd ingredient when taken at the level of the reduction to the subgroup. The point here is that it is only at the level of this algebra $A_F$ that the unimodularity condition does acquire a conceptual meaning instead of being an ad-hoc prescription.
  \item The third reason is that the conceptual description of the Hilbert space of Fermions for one generation, {\it i.e.\/} of the irreducible representation of $(A_F,J)$, is as the space of maps ${\rm Hom}(E,F)\oplus {\rm Hom}(F,E)$  where $E$ is a two dimensional vector space over the quaternions $\mathbb{H}$ and $F$ a $4$-dimensional vector space over $\mathbb{C}$. It is hard to miss the hint to twistors since the latter involve the relation between the corresponding projective spaces namely $\mathbb{P}^1\mathbb{H}$ and $\mathbb{P}^3\mathbb{C}$.
\end{enumerate}

While it is natural to try to extend the results of \cite{Broek} to the above variant by assuming that the Dirac operator of the finite space is $0$ as for the Einstein-Yang Mills system, it is also quite desirable to come up with a dynamical mechanism for the reduction to the subgroup $U(1)\times SU(2)\times SU(3)$ coming from the order one condition
and the finite distance between the two copies of the four-dimensional manifold $M$. At the moment the reduction is imposed by a mathematical requirement which is non-dynamical and the corresponding symmetry breaking should in fact come from additional terms in the action showing a preference for the physically desirable ``finite distance" condition.

\subsection{Role of $M_4(\mathbb{C})$}\medskip \hfill\label{mfourc}

Let us first ignore the fact that we have two simple components  $M_{2}\left(  \mathbb{H}%
\right)  \oplus M_{4}\left(  \mathbb{C}\right)$ and explain briefly in what sense one obtains a simpler
presentation by replacing the algebra $C^\infty(M)$ of functions by the algebra $C^\infty(M)\otimes M_4(\mathbb{C})=M_4(C^\infty(M))$.

We start by the two dimensional case, and give a very simple presentation of the algebra $C^\infty(S^2)\otimes M_2(\mathbb{C})=M_2(C^\infty(S^2))$. The algebra is generated by a symbol $e$ and the scalar matrices
$m\in M_2(\mathbb{C})$. Elements of the algebra are sums of words of the form
\begin{equation}
w=e\,m_1\,e\,m_2\,e\cdots m_k\,e\,,\ \  m_j\in M_2(\mathbb{C})
\end{equation}
One multiplies them according to the following rules.
The algebraic rules are the usual ones for $M_2(\mathbb{C})$ and
                  one has the additional relations
\begin{equation}
e=e^*=e^2\, ,\ \ \left\langle e-\frac{1}{2}\right\rangle=0\,.
\end{equation}
Here the trace $X\mapsto \left\langle X\right\rangle$ with values in the commutant of $M_2(\mathbb{C})$ is \begin{equation}
\left\langle X\right\rangle=e_{11}Xe_{11}+e_{21}Xe_{12}+e_{12}Xe_{21}+e_{22}Xe_{22}
\end{equation}
with the standard  notation for the $4$ matrix units
$$
e_{11}=\left(
                    \begin{array}{cc}
                      1 & 0 \\
                      0 & 0 \\
                    \end{array}
                  \right)\,, \
                  e_{12}=\left(
                    \begin{array}{cc}
                      0 & 1 \\
                      0 & 0 \\
                    \end{array}
                  \right)\,, \
                  e_{21}=\left(
                    \begin{array}{cc}
                      0 & 0 \\
                      1 & 0 \\
                    \end{array}
                  \right)\,, \
                  e_{22}=\left(
                    \begin{array}{cc}
                      0 & 0 \\
                      0 & 1 \\
                    \end{array}
                  \right)
                  $$
 The point then is that one obtains in this
way a dense subalgebra of $C^\infty(S^2)\otimes M_2(\mathbb{C})=M_2(C^\infty(S^2))$. This follows since once $e$ is expressed in matrix form with coefficients in the commutant of $M_2(\mathbb{C})$ it takes the form
\begin{equation}
    e=\left(
        \begin{array}{cc}
          \frac{1}{2}+t & z \\
          z^* &\frac{1}{2}-t\\
        \end{array}
      \right)
\end{equation}
and the equation $e^2=e$ implies that $t,z,z^*$  commute pairwise and fulfill the relation
\begin{equation}
zz^*+t^2=\frac 14
\end{equation}
Moreover the orientability condition which fixes the volume form of the metric and guarantees that the metric is non-degenerate takes the simple form
\begin{equation}
    \left\langle \left(  e-\frac{1}{2}\right)  \left[  D,e\right]  ^{2}%
\right\rangle=\gamma
\end{equation}
where $\gamma$ is the chirality operator satisfying
\begin{equation}
\gamma^{2}=\gamma,\qquad\gamma=\gamma^{\ast},\qquad\gamma e=e\gamma,\qquad
D\gamma=-\gamma D
\end{equation}
In dimension $4$ one has a similar description of the commutative solution given by the $4$-sphere
(with a not necessarily round metric having the prescribed volume form).
 The algebra $M_2(\mathbb{C})$ is replaced by
$4\times4$ matrices and as above the algebra is generated by
$M_{4}\left(  \mathbb{C}\right)  $ and a projection $e=e^{2}=e^{\ast}$ of the form
\begin{equation}
e=\left(
\begin{array}
[c]{cccc}%
\frac{1}{2}+t & 0 & \alpha & \beta\\
0 & \frac{1}{2}+t & -\beta^{\ast} & \alpha^{\ast}\\
\alpha^{\ast} & -\beta & \frac{1}{2}-t & 0\\
\beta^{\ast} & \alpha & 0 & \frac{1}{2}-t
\end{array}
\right)
\end{equation}
where $t,\alpha,\alpha^{\ast},\beta$ and $\beta^{\ast}$ all commute and
satisfy the relation%
\[
t^{2}+\left\vert \alpha\right\vert ^{2}+\left\vert \beta\right\vert ^{2}%
=\frac{1}{4}%
\]
One can then check that $\mathcal{A}=C\left(  S^{4}\right)  .$ The
differential constraints
\begin{equation}
    \left\langle \left(  e-\frac{1}{2}\right)  \left[  D,e\right]  ^{4}%
\right\rangle=\gamma
\end{equation}
are then satisfied by any Riemannian structure with a
given volume form on $S^{4}.$

The really new feature which appears in dimension $4$ is that the equations admit non-trivial noncommutative solutions. This fact was discovered in \cite{CLa} and the problem of classification of solutions has been solved in three dimensions in \cite{CDV1}, \cite{CDV2}, \cite{CDV3}, while the $4$-dimensional case is still under investigation.

The simplest noncommutative solution is obtained (\cite{CLa}) as a deformation by considering the
algebra to be generated by $M_{4}\left(  \mathbb{C}\right)  $ and $e$ where
\begin{equation}
e=\left(
\begin{array}
[c]{cc}%
q_{11} & q_{12}\\
q_{21} & q_{22}%
\end{array}
\right)
\end{equation}
where each $q$ is a $2\times2$ matrix of the form
\begin{equation}
q=\left(
\begin{array}
[c]{cc}%
\alpha & \beta\\
-\lambda\beta & \alpha^{\ast}%
\end{array}
\right)
\end{equation}
In this case the projection constraints imply that
\begin{equation}
e=\left(
\begin{array}
[c]{cccc}%
\frac{1}{2}+t & 0 & \alpha & \beta\\
0 & \frac{1}{2}+t & -\lambda\beta^{\ast} & \alpha^{\ast}\\
\alpha^{\ast} & -\overline{\lambda}\beta & \frac{1}{2}-t & 0\\
\beta^{\ast} & \alpha & 0 & \frac{1}{2}-t
\end{array}
\right)
\end{equation}
satisfying%
\begin{equation}
\alpha\alpha^{\ast}=\alpha^{\ast}\alpha,\quad\beta\beta^{\ast}=\beta^{\ast
}\beta,\quad\alpha\beta=\lambda\beta\alpha,\quad\alpha^{\ast}\beta
=\overline{\lambda}\beta\alpha
\end{equation}
giving rise to deformed $S^{4}.$

Assuming that the unification scale is not far away from the Planck scale, it
is natural to modify the basic assumption we made that space-time is a product
of a continuous four dimensional manifold times a finite space. This leads
us to investigate the postulate that at very high energies, the structure of
space time becomes noncommutative in a nontrivial way, which will change in an
intrinsic way the particle spectrum. On the other hand, the encouraging
results we obtained about the almost unique prediction of the spectrum of the
standard model for the gauge group and particle representations, can be taken
as a guide that the true geometry should reproduce at lower energies, the
product structure we assumed. The starting point is to look for a
noncommutative space whose KO dimension is ten (mod 8) and whose metric
dimension as dictated by the growth of eigenvalues of the Dirac operator is
four. \ A good starting point would be to mesh in a smooth manner the
four-dimensional manifold with the finite space $M_{2}\left(  \mathbb{H}%
\right)  \oplus M_{4}\left(  \mathbb{C}\right)  .$
The next step is to define the noncommutative space by marrying the concept of
generating a manifold as instantonic solution of a set of equations, and to
blend these with the finite space.

\subsection{Generations}\medskip \hfill\label{generations}

In this short section we shall speculate on a possible relation between the fundamental group of space-time and the three generations of fermions. Our starting point is the intimate relation in topology:

\centerline{Manifold $\leftrightarrow$  Poincar\'e duality in $KO$-homology}

and the coincidence of the basic ingredients of cycles in $KO$-homology, namely spectral triples $$
({\mathcal A},{\mathcal H},D)\,,\ \   \ ds = D^{-1}\,, \ J , \ \gamma
$$
$$
J^2=\varepsilon\,, \ DJ=\varepsilon^{\prime}JD,\quad
J\,\gamma=\varepsilon^{\prime\prime }\gamma J,\quad D\gamma=-\gamma D
$$

with the ingredients of the quantum theory:
\begin{itemize}
  \item ${\mathcal H}$: one particle Euclidean Fermions
  \item $D$: inverse propagator
  \item $J$: charge conjugation
  \item $\gamma$: chirality
\end{itemize}

The point about the fundamental group that we wish to make here is that the above equivalence between ``manifolds" $M$ and spaces which fulfill Poincar\'e duality in $KO$-homology is only fully encoded by the fundamental cycle in $KO$-homology if one also takes into account the fundamental group $\Gamma=\pi_1(M)$. More specifically, in the non-simply connected case when $\pi_1(M)$ is non-trivial, the natural datum is not the Dirac operator $D$  on the manifold $M$ but rather the Dirac operator $\tilde D$ on the universal cover $\tilde M$ of $M$. Even though these operators look alike locally the action of the group $\Gamma=\pi_1(M)$ on the $L^2$ spinors on the universal cover $\tilde M$ breaks this space into ``sectors" and there is a (might be superficial) resemblance between this decomposition into sectors and the decomposition of the Hilbert space of Fermions as a sum of Hilbert spaces corresponding to generations. We are fully aware of the subtleties inherent to the mixing of generations from the CKM matrix \cite{ckm} but there is room in the geometric formalism, with basic examples coming for instance from non-Galois coverings, to investigate the possibility of a geometric origin for the multiplicity of generations.

\bigskip

\subsection{Unification of couplings}\medskip \hfill\label{unif}

 The one loop RG equations for the running of the gauge couplings and
Newton constant do not meet exactly at one point which is expected to be at
the Planck scale. The error, however, is within few percent. Higher order
corrections will change the running of all coupling constants and we shall see in Part II that rather surprisingly, if one no longer assumes that the function $f(D/\Lambda)$ is flat at $0$ as any cut-off function, then the contributions of the higher order terms alter the simple unification rule involving the $\frac 35$. This allows one to improve on the above issue under the assumption that the Yukawa coupling of the tau neutrino is of the same order as the Yukawa coupling of the top quark, an hypothesis that already appeared naturally when dealing with the prediction on the top quark mass (\cite{mc2}).

\subsection{Mass of the Higgs}\medskip \hfill\label{Higgs}

 The mass of the Higgs field in the zeroth order approximation of the
spectral action is around $170$ Gev. This however, depends on the value of the
gauge couplings at the unification scale. Higher order corrections will
definitely change this predicted value, but since the prediction comes from the value at unification of the coupling constant of the quartic term in the Higgs potential, the finiteness of this coupling implies the same qualitative results as the hypothesis of the ``big desert" and the assumption that the Standard Model is still valid at this very high scale. Of course this hinges on the naturalness problem whose only accepted resolution involves supersymmetry. It is rather striking however that the spectral action naturally contains a quadratic mass term in the Higgs field which has the correct sign and size to allow one to do fine tuning. In any case we consider that the experimental determination of the Higgs mass will give a precious indication

\subsection{New particles}\medskip \hfill\label{New}

 The reduction of the gauge group to $U(1)\times SU(2)\times SU(3)$ was obtained above from the order one condition using the hypothesis that the two layers of space-time corresponding to the two-dimensional center of the algebra $A_F$ of the finite space are at a finite distance apart. This reduces the natural gauge group $SU(2)_L\times SU(2)_R\times SU(4)$ given by the even part of the algebra $A_F=M_2(\mathbb{H})\oplus M_4(\mathbb{C})$ to the Standard Model gauge group. The justification that we have given, starting in \cite{mc2}, for this reduction, is based on the order one condition for the Dirac operator and is imposed as a mathematical condition. It is desirable to improve this point by finding a dynamical mechanism that effects the same symmetry breaking  from $SU(2)_L\times SU(2)_R\times SU(4)$ to $U(1)\times SU(2)\times SU(3)$. Such a mechanism should generate mass terms for the broken part of the gauge sector. This will thus correspond to new particles not present in the Standard Model but well motivated from the above considerations. What is missing at the mathematical level is to understand how the order one condition can be imposed at the dynamical level, and also how the inner fluctuations of the metric behave if one no longer assumes the order one condition.

\subsection{Quantum Level}\medskip \hfill\label{quant}

 So far we have used the renormalization group in a very straightforward manner
   starting from the simple idea that the spectral action holds at the unification scale and using the values of the couplings as boundary conditions.  The compatibility between the values at low energy (obtained by integration over the fluctuations in the intermediate scales) and observation is a basic test of the general idea but in case this test is passed, one needs to go much further and develop a theory that takes over at higher scales.
Since the model we developed contains both gravity and the Standard Model it is clear that  this problem is the problem
of quantizing gravity. We refer the reader to  \cite{Wulk} for interesting suggestions concerning the role of the ghost fields. One challenging problem at this point is to compute the bosonic propagator for the inner fluctuations of the metric using the spectral action and functional derivatives of tracial functions. One may hope that the techniques developed in the context of renormalization of QFT on noncommutative spaces will be useful in the building of the quantum theory of the spectral action. In  \cite{CoMM}
 an analogy was developed between the phase transitions which occur in the number theoretic context and a scenario of spontaneous symmetry breaking involving the full gravitational sector. If substantiated, this could show how geometry would emerge from the computation of the KMS states of an operator theoretic system, closely related to a matrix model with basic variable the Dirac operator $D$. It is worthwhile to note, at this point, that, at the conceptual level, the spectral action is closely related to an entropy since it can be written as the logarithm of a number of states in the second quantized Fermionic Hilbert space.

 There is another very interesting mathematical problem which is suggested by the quantum theory. While we have a simple prescription for the inner fluctuations of the metric, the formulas for modifying the ``outer" part of the metric are surely more subtle, but we want to point out that
 \begin{enumerate}
   \item A change of the Weyl factor in the metric is given by a beautifully simple formula for the Dirac operator which extends to the noncommutative case \cite{cc3}, \cite{comosc}.
   \item There is a simple and efficient analogue in noncommutative geometry for the modification of the conformal structure encoded by a Beltrami differential (\cite{Co-book}).
    \end{enumerate}
    Finally there are interesting  developments on cosmology \cite{marco}, \cite{Sakel} which open a new line of investigations where  the KMS condition should play a leading role in the analysis of phase transitions following the model developed in \cite{Sher} for the case of the electroweak transition.

\newpage

\section{Appendix A: components of the Dirac operator}

We summarize our results by listing all matrix entries of the full Dirac
operator $\left(  D_{A}\right)  _{M}^{N}$, but will omit the index $A$ of
$D_{A}$ in what follows:%
\begin{align*}
\left(  D\right)  _{\overset{.}{1}1}^{\overset{.}{1}1}  &  =\gamma^{\mu
}\otimes D_{\mu}\otimes1_{3},\quad D_{\mu}=\partial_{\mu}+\frac{1}{4}%
\omega_{\mu}^{cd}\left(  e\right)  \gamma_{cd},\quad1_{3}=\text{generations}\\
\left(  D\right)  _{\overset{.}{1}1}^{a1}  &  =\gamma_{5}\otimes k^{\ast\nu
}\otimes\epsilon^{ab}H_{b}\qquad k^{\nu}=3\times3\text{ neutrino mixing
matrix}\\
\left(  D\right)  _{\overset{.}{2}1}^{\overset{.}{2}1}  &  =\gamma^{\mu
}\otimes\left(  D_{\mu}+ig_{1}B_{\mu}\right)  \otimes1_{3}\\
\left(  D\right)  _{\overset{.}{2}1}^{a1}  &  =\gamma_{5}\otimes k^{\ast
e}\otimes\overline{H}^{a}\\
\left(  D\right)  _{a1}^{\overset{.}{1}1}  &  =\gamma_{5}\otimes k^{\nu
}\otimes\epsilon_{ab}\overline{H}^{b}\\
\left(  D\right)  _{a1}^{\overset{.}{2}1}  &  =\gamma_{5}\otimes k^{e}\otimes
H_{a}\\
\left(  D\right)  _{a1}^{b1}  &  =\gamma^{\mu}\otimes\left(  \left(  D_{\mu
}+\frac{i}{2}g_{1}B_{\mu}\right)  \delta_{a}^{b}-\frac{i}{2}g_{2}W_{\mu
}^{\alpha}\left(  \sigma^{\alpha}\right)  _{a}^{b}\right)  \otimes1_{3},\text{
\qquad}\sigma^{\alpha}=\text{Pauli}\\
\left(  D\right)  _{\overset{.}{1}i}^{\overset{.}{1}j}  &  =\gamma^{\mu
}\otimes\left(  \left(  D_{\mu}-\frac{2i}{3}g_{1}B_{\mu}\right)  \delta
_{i}^{j}-\frac{i}{2}g_{3}V_{\mu}^{m}\left(  \lambda^{m}\right)  _{i}%
^{j}\right)  \otimes1_{3},\qquad\lambda^{i}=\text{Gell-Mann}\\
\left(  D\right)  _{\overset{.}{1}i}^{aj}  &  =\gamma_{5}\otimes k^{\ast
u}\otimes\epsilon^{ab}H_{b}\delta_{i}^{j}\\
\left(  D\right)  _{\overset{.}{2}i}^{\overset{.}{2}j}  &  =\gamma^{\mu
}\otimes\left(  \left(  D_{\mu}+\frac{i}{3}g_{1}B_{\mu}\right)  \delta_{i}%
^{j}-\frac{i}{2}g_{3}V_{\mu}^{m}\left(  \lambda^{m}\right)  _{i}^{j}\right)
\otimes1_{3}\\
\left(  D\right)  _{\overset{.}{2}i}^{aj}  &  =\gamma_{5}\otimes k^{\ast
d}\otimes\overline{H}^{a}\delta_{i}^{j}\\
\left(  D\right)  _{ai}^{bj}  &  =\gamma^{\mu}\otimes\left(  \left(  D_{\mu
}-\frac{i}{6}g_{1}B_{\mu}\right)  \delta_{a}^{b}\delta_{i}^{j}-\frac{i}%
{2}g_{2}W_{\mu}^{\alpha}\left(  \sigma^{\alpha}\right)  _{a}^{b}\delta_{i}%
^{j}-\frac{i}{2}g_{3}V_{\mu}^{m}\left(  \lambda^{m}\right)  _{i}^{j}\delta
_{a}^{b}\right)  \otimes1_{3}\\
\left(  D\right)  _{ai}^{\overset{.}{1}j}  &  =\gamma_{5}\otimes k^{u}%
\otimes\epsilon_{ab}\overline{H}^{b}\delta_{i}^{j}\\
\left(  D\right)  _{ai}^{\overset{.}{2}j}  &  =\gamma_{5}\otimes k^{d}\otimes
H_{a}\delta_{i}^{j}\\
\left(  D\right)  _{\overset{.}{1}1}^{\overset{.}{1^{\prime}}1^{\prime}}  &
=\gamma_{5}\otimes k^{\ast\nu_{R}}\sigma\qquad\text{generate scale }%
M_{R}\text{ by }\sigma\rightarrow M_{R}\\
\left(  D\right)  _{\overset{.}{1^{\prime}}1^{\prime}}^{\overset{.}{1}1}  &
=\gamma_{5}\otimes k^{\nu_{R}}\sigma\\
D_{A^{\prime}}^{B^{\prime}}  &  =\overline{D}_{A}^{B},\qquad D_{A^{\prime}%
}^{B}=\overline{D}_{A}^{B^{\prime}},\qquad D_{A}^{B^{\prime}}=\overline
{D}_{A^{\prime}}^{B}%
\end{align*}
\newpage

\section{Appendix B: components of the square of the Dirac operator}

Next we list all the components of the matrix $\left(  D^{2}\right)  _{M}%
^{N}$
\begin{align}
\left(  D^{2}\right)  _{\overset{.}{1}1}^{\overset{.}{1}1} &  =D_{\overset
{.}{1}1}^{\overset{.}{1}1}D_{\overset{.}{1}1}^{\overset{.}{1}1}+D_{\overset
{.}{1}1}^{a1}D_{a1}^{\overset{.}{1}1}+k^{\ast\nu_{R}}k^{\nu_{R}}\sigma^{2}\\
&  =\gamma^{\mu}D_{\mu}\gamma^{\nu}D_{\nu}\otimes1_{3}+k^{\nu\ast}k^{\nu}%
H_{a}\overline{H}^{a}+k^{\ast\nu_{R}}k^{\nu_{R}}\sigma^{2}\nonumber
\end{align}%
\begin{align}
\left(  D^{2}\right)  _{\overset{.}{1}1}^{a1} &  =D_{\overset{.}{1}%
1}^{\overset{.}{1}1}D_{\overset{.}{1}1}^{a1}+D_{\overset{.}{1}1}^{b1}%
D_{b1}^{a1}\\
&  =\gamma^{\mu}D_{\mu}\gamma_{5}k^{\ast\nu}\epsilon^{ab}H_{b}+\gamma
_{5}k^{\ast\nu}\epsilon^{bc}H_{c}\gamma^{\mu}\left(  \left(  D_{\mu}+\frac
{i}{2}g_{1}B_{\mu}\right)  \delta_{b}^{a}-\frac{i}{2}g_{2}W_{\mu}^{\alpha
}\left(  \sigma^{\alpha}\right)  _{b}^{a}\right)  \nonumber\\
&  =\gamma^{\mu}\gamma_{5}k^{\ast\nu}\epsilon^{ab}\nabla_{\mu}H_{b}\nonumber
\end{align}
where%
\begin{equation}
\nabla_{\mu}H_{a}=\left(  \left(  \partial_{\mu}-\frac{i}{2}g_{1}B_{\mu
}\right)  \delta_{b}^{a}-\frac{i}{2}g_{2}W_{\mu}^{\alpha}\left(
\sigma^{\alpha}\right)  _{b}^{a}\right)  H_{b}%
\end{equation}
and we have used the identity $\epsilon^{ab}\epsilon_{cd}\left(
\sigma^{\alpha}\right)  _{b}^{d}=-\left(  \sigma^{\alpha}\right)  _{c}^{a}.$
Next
\begin{align}
\left(  D^{2}\right)  _{\overset{.}{2}1}^{\overset{.}{2}1} &  =D_{\overset
{.}{2}1}^{\overset{.}{2}1}D_{\overset{.}{2}1}^{\overset{.}{2}1}+D_{\overset
{.}{2}1}^{a1}D_{a1}^{\overset{.}{2}1}\\
&  =\gamma^{\mu}\left(  D_{\mu}+ig_{1}B_{\mu}\right)  \gamma^{\nu}\left(
D_{\nu}+ig_{1}B_{\nu}\right)  \otimes1_{3}+k^{e\ast}k^{e}\overline
{H}H\nonumber
\end{align}
where we have denoted $\overline{H}H=\overline{H}^{a}H_{a}=H_{a}\overline
{H}^{a}.$
\begin{align}
\left(  D^{2}\right)  _{\overset{.}{2}1}^{a1} &  =D_{\overset{.}{2}%
1}^{\overset{.}{2}1}D_{\overset{.}{2}1}^{a1}+D_{\overset{.}{2}1}^{b1}%
D_{b1}^{a1}\\
&  =\gamma^{\mu}\left(  D_{\mu}+ig_{1}B_{\mu}\right)  \gamma_{5}k^{\ast
e}\overline{H}^{a}+\gamma_{5}k^{\ast e}\overline{H}^{b}\gamma^{\mu}\left(
\left(  D_{\mu}+\frac{i}{2}g_{1}B_{\mu}\right)  \delta_{b}^{a}-\frac{i}%
{2}g_{2}W_{\mu}^{\alpha}\left(  \sigma^{\alpha}\right)  _{b}^{a}\right)
\nonumber\\
&  =\gamma^{\mu}\gamma_{5}k^{\ast e}\nabla_{\mu}\overline{H}^{a}\nonumber
\end{align}%
\begin{align}
\left(  D^{2}\right)  _{a1}^{b1} &  =D_{a1}^{\overset{.}{1}1}D_{\overset{.}%
{1}1}^{b1}+D_{a1}^{\overset{.}{2}1}D_{\overset{.}{2}1}^{b1}+D_{a1}^{c1}%
D_{c1}^{b1}\\
&  =k^{e}k^{\ast e}H_{a}\overline{H}^{b}+k^{\nu}k^{\ast\nu}\epsilon
_{ac}\epsilon^{bd}\overline{H}^{c}H_{d}\nonumber\\
&  +\gamma^{\mu}\left(  \left(  D_{\mu}+\frac{i}{2}g_{1}B_{\mu}\right)
\delta_{a}^{c}-\frac{i}{2}g_{2}W_{\mu}^{\alpha}\left(  \sigma^{\alpha}\right)
_{a}^{c}\right)  \gamma^{\nu}\left(  \left(  D_{\nu}+\frac{i}{2}g_{1}B_{\nu
}\right)  \delta_{c}^{b}-\frac{i}{2}g_{2}W_{\nu}^{\alpha}\left(
\sigma^{\alpha}\right)  _{c}^{b}\right)  \otimes1_{3}\nonumber
\end{align}%
\begin{align}
\left(  D^{2}\right)  _{\overset{.}{1}i}^{\overset{.}{1}j} &  =D_{\overset
{.}{1}i}^{\overset{.}{1}k}D_{\overset{.}{1}k}^{\overset{.}{1}j}+D_{\overset
{.}{1}1}^{ak}D_{ak}^{\overset{.}{1}j}=\\
&  \gamma^{\mu}\left(  \left(  D_{\mu}-\frac{2i}{3}g_{1}B_{\mu}\right)
\delta_{i}^{k}-\frac{i}{2}g_{3}V_{\mu}^{m}\left(  \lambda^{m}\right)  _{i}%
^{k}\right)  \gamma^{\nu}\left(  \left(  D_{\nu}-\frac{2i}{3}g_{1}B_{\nu
}\right)  \delta_{i}^{k}-\frac{i}{2}g_{3}V_{\nu}^{m}\left(  \lambda
^{m}\right)  _{i}^{k}\right)  \otimes1_{3}\nonumber\\
&  +k^{u\ast}k^{u}\overline{H}H\,\delta_{i}^{j}\nonumber
\end{align}%
\begin{align}
\left(  D^{2}\right)  _{\overset{.}{2}i}^{aj} &  =D_{\overset{.}{2}i}%
^{bk}D_{bk}^{aj}+D_{\overset{.}{2}i}^{\overset{.}{2}k}D_{\overset{.}{2}k}%
^{aj}\\
&  =\gamma_{5}k^{\ast d}\overline{H}^{b}\delta_{i}^{k}\gamma^{\mu}\left(
\left(  D_{\mu}-\frac{i}{6}g_{1}B_{\mu}\right)  \delta_{b}^{a}\delta_{k}%
^{j}-\frac{i}{2}g_{2}W_{\mu}^{\alpha}\left(  \sigma^{\alpha}\right)  _{b}%
^{a}\delta_{k}^{j}-\frac{i}{2}g_{3}V_{\mu}^{m}\left(  \lambda^{m}\right)
_{k}^{j}\delta_{b}^{a}\right)  \nonumber\\
&  +\gamma^{\mu}\left(  \left(  D_{\mu}+\frac{i}{3}g_{1}B_{\mu}\right)
\delta_{i}^{k}-\frac{i}{2}g_{3}V_{\mu}^{m}\left(  \lambda^{m}\right)  _{i}%
^{k}\right)  \gamma_{5}k^{\ast d}H_{a}\delta_{k}^{j}\nonumber\\
&  =\gamma^{\mu}\gamma_{5}k^{\ast d}\nabla_{\mu}\overline{H}^{a}\delta_{i}%
^{j}\nonumber
\end{align}%
\begin{align}
\left(  D^{2}\right)  _{\overset{.}{1}i}^{aj} &  =D_{\overset{.}{1}i}%
^{bk}D_{bk}^{aj}+D_{\overset{.}{1}i}^{\overset{.}{1}k}D_{\overset{.}{1}k}%
^{aj}\\
&  =\gamma_{5}k^{\ast u}\epsilon^{bc}H_{c}\delta_{i}^{k}\gamma^{\mu}\left(
\left(  D_{\mu}-\frac{i}{6}g_{1}B_{\mu}\right)  \delta_{b}^{a}\delta_{k}%
^{j}-\frac{i}{2}g_{2}W_{\mu}^{\alpha}\left(  \sigma^{\alpha}\right)  _{b}%
^{a}\delta_{k}^{j}-\frac{i}{2}g_{3}V_{\mu}^{m}\left(  \lambda^{m}\right)
_{k}^{j}\delta_{b}^{a}\right)  \nonumber\\
&  +\gamma^{\mu}\left(  \left(  D_{\mu}-\frac{2i}{3}g_{1}B_{\mu}\right)
\delta_{i}^{k}-\frac{i}{2}g_{3}V_{\mu}^{m}\left(  \lambda^{m}\right)  _{i}%
^{k}\right)  \gamma_{5}k^{\ast u}\epsilon^{ab}H_{b}\delta_{k}^{j}\nonumber\\
&  =\gamma^{\mu}\gamma_{5}k^{\ast u}\epsilon^{ab}\nabla_{\mu}H_{b}\delta
_{i}^{j}\nonumber
\end{align}%
\begin{align}
\left(  D^{2}\right)  _{\overset{.}{2}i}^{\overset{.}{2}j} &  =D_{\overset
{.}{2}i}^{\overset{.}{2}k}D_{\overset{.}{2}k}^{\overset{.}{2}j}+D_{\overset
{.}{2}i}^{ak}D_{ak}^{\overset{.}{2}j}\\
&  =\gamma^{\mu}\left(  \left(  D_{\mu}+\frac{i}{3}g_{1}B_{\mu}\right)
\delta_{i}^{k}-\frac{i}{2}g_{3}V_{\mu}^{m}\left(  \lambda^{m}\right)  _{i}%
^{k}\right)  \gamma^{\nu}\left(  \left(  D_{\nu}+\frac{i}{3}g_{1}B_{\nu
}\right)  \delta_{k}^{j}-\frac{i}{2}g_{3}V_{\nu}^{m}\left(  \lambda
^{m}\right)  _{k}^{j}\right)  \nonumber\\
&  +k^{d\ast}k^{d}\,\overline{H}H\,\delta_{i}^{j}\nonumber
\end{align}%
\begin{align}
\left(  D^{2}\right)  _{ai}^{\overset{.}{1}j} &  =D_{ai}^{\overset{.}{1}%
k}D_{\overset{.}{1}k}^{\overset{.}{1}j}+D_{ai}^{ck}D_{ck}^{\overset{.}{1}j}\\
&  =\gamma_{5}k^{u}\epsilon_{ab}\overline{H}^{b}\delta_{i}^{k}\gamma^{\mu
}\left(  \left(  D_{\mu}-\frac{2i}{3}g_{1}B_{\mu}\right)  \delta_{k}^{j}%
-\frac{i}{2}g_{3}V_{\mu}^{m}\left(  \lambda^{m}\right)  _{k}^{j}\right)
\nonumber\\
&  +\gamma^{\mu}\left(  \left(  D_{\mu}-\frac{i}{6}g_{1}B_{\mu}\right)
\delta_{a}^{c}\delta_{i}^{k}-\frac{i}{2}g_{2}W_{\mu}^{\alpha}\left(
\sigma^{\alpha}\right)  _{a}^{c}\delta_{i}^{k}-\frac{i}{2}g_{3}V_{\mu}%
^{m}\left(  \lambda^{m}\right)  _{i}^{k}\delta_{a}^{c}\right)  \gamma_{5}%
k^{u}\epsilon_{cb}\overline{H}^{b}\delta_{k}^{j}\nonumber\\
&  =\gamma^{\mu}\gamma_{5}k^{u}\epsilon_{ab}\nabla_{\mu}\overline{H}^{b}%
\delta_{i}^{j}\nonumber
\end{align}%
\begin{align}
\left(  D^{2}\right)  _{ai}^{\overset{.}{2}j} &  =D_{ai}^{\overset{.}{2}%
k}D_{\overset{.}{2}k}^{\overset{.}{2}j}+D_{ai}^{ck}D_{ck}^{\overset{.}{2}j}\\
&  =\gamma_{5}k^{d}H_{a}\delta_{i}^{k}\gamma^{\mu}\left(  \left(  D_{\mu
}+\frac{i}{3}g_{1}B_{\mu}\right)  \delta_{k}^{j}-\frac{i}{2}g_{3}V_{\mu}%
^{m}\left(  \lambda^{m}\right)  _{k}^{j}\right)  \nonumber\\
&  +\gamma^{\mu}\left(  \left(  D_{\mu}-\frac{i}{6}g_{1}B_{\mu}\right)
\delta_{a}^{c}\delta_{i}^{k}-\frac{i}{2}g_{2}W_{\mu}^{\alpha}\left(
\sigma^{\alpha}\right)  _{a}^{c}\delta_{i}^{k}-\frac{i}{2}g_{3}V_{\mu}%
^{m}\left(  \lambda^{m}\right)  _{i}^{k}\delta_{a}^{c}\right)  \gamma_{5}%
k^{d}H_{c}\delta_{k}^{j}\nonumber\\
&  =\gamma^{\mu}\gamma_{5}k^{d}\nabla_{\mu}H_{a}\delta_{i}^{j}\nonumber
\end{align}
Finally
\begin{align}
\left(  D^{2}\right)  _{ai}^{bj} &  =D_{ai}^{\overset{.}{1}k}D_{\overset{.}%
{1}k}^{bj}+D_{ai}^{\overset{.}{2}k}D_{\overset{.}{2}k}^{bj}+D_{ai}^{ck}%
D_{ck}^{bj}\\
&  =k^{u}k^{\ast u}\overline{H}^{b}H_{a}\delta_{i}^{j}+k^{d}k^{\ast d}%
\epsilon_{ac}\epsilon^{bd}\overline{H}^{c}H_{d}\delta_{i}^{j}\nonumber\\
&  +\left[  \gamma^{\mu}\left(  \left(  D_{\mu}-\frac{i}{6}g_{1}B_{\mu
}\right)  \delta_{a}^{c}\delta_{i}^{k}-\frac{i}{2}g_{2}W_{\mu}^{\alpha}\left(
\sigma^{\alpha}\right)  _{a}^{c}\delta_{i}^{k}-\frac{i}{2}g_{3}V_{\mu}%
^{m}\left(  \lambda^{m}\right)  _{i}^{k}\delta_{a}^{c}\right)  \right.
\nonumber\\
&  \left.  \gamma^{\nu}\left(  \left(  D_{\nu}-\frac{i}{6}g_{1}B_{\nu}\right)
\delta_{c}^{b}\delta_{k}^{j}-\frac{i}{2}g_{2}W_{\nu}^{\alpha}\left(
\sigma^{\alpha}\right)  _{c}^{a}\delta_{k}^{j}-\frac{i}{2}g_{3}V_{\nu}%
^{m}\left(  \lambda^{m}\right)  _{k}^{j}\delta_{c}^{a}\right)  \right]
\nonumber
\end{align}
There are also terms in the off-diagonal part
\begin{align}
\left(  D^{2}\right)  _{\overset{.}{1}1}^{a^{\prime}1^{\prime}} &
=D_{\overset{.}{1}1}^{\overset{.}{1}^{^{\prime}}1^{^{\prime}}}D_{\overset
{.}{1}^{^{\prime}}1^{^{\prime}}}^{a^{\prime}1^{\prime}}\\
&  =k^{\ast\nu_{R}}\overline{k^{\ast\nu}}\epsilon^{ab}\overline{H}_{b}%
\sigma\nonumber\\
\left(  D^{2}\right)  _{a^{\prime}1^{\prime}}^{\overset{.}{1}1} &
=D_{a^{\prime}1^{\prime}}^{\overset{.}{1}^{^{\prime}}1^{^{\prime}}}%
D_{\overset{.}{1}^{^{\prime}}1^{^{\prime}}}^{\overset{.}{1}1}\\
&  =\overline{k^{\nu}}k^{\nu_{R}}\epsilon_{ab}H^{b}\sigma\nonumber\\
\left(  D^{2}\right)  _{a1}^{\overset{.}{1}^{^{\prime}}1^{^{\prime}}} &
=D_{a1}^{\overset{.}{1}1}D_{\overset{.}{1}1}^{\overset{.}{1}^{^{\prime}%
}1^{^{\prime}}}\\
&  =k^{\nu}k^{\ast\nu_{R}}\epsilon_{ab}\overline{H}^{b}\sigma\nonumber\\
\left(  D^{2}\right)  _{\overset{.}{1}^{^{\prime}}1^{^{\prime}}}^{a1} &
=D_{\overset{.}{1}^{^{\prime}}1^{^{\prime}}}^{\overset{.}{1}1}D_{\overset
{.}{1}1}^{a1}\\
&  =k^{\nu_{R}}k^{\ast\nu}\epsilon^{ab}H_{b}\sigma\nonumber
\end{align}
\newpage

\section{Appendix C: connection $\mathbb{\omega}_{\mu},$ curvature
$\Omega_{\mu\nu}$ and invariant $E$}

We list here the entries of the matrices $\left(  \mathbb{\omega}_{\mu
}\right)  _{M}^{N}$ , $\left(  E\right)  _{M}^{N}$ which are defined in terms
of the operator $D^{2\text{ }}$and the curvature $\left(  \Omega_{\mu\nu
}\right)  _{M}^{N}$ where
\begin{align*}
\left(  \mathbb{\omega}_{\mu}\right)  _{\overset{.}{1}1}^{\overset{.}{1}1}  &
=\frac{1}{4}\omega_{\mu}^{cd}\left(  e\right)  \gamma_{cd}\otimes1_{3}\\
\left(  \mathbb{\omega}_{\mu}\right)  _{\overset{.}{2}1}^{\overset{.}{2}1}  &
=\left(  \frac{1}{4}\omega_{\mu}^{cd}\left(  e\right)  \gamma_{cd}%
+ig_{1}B_{\mu}\right)  \otimes1_{3}\\
\left(  \mathbb{\omega}_{\mu}\right)  _{a1}^{b1}  &  =\left(  \left(  \frac
{1}{4}\omega_{\mu}^{cd}\left(  e\right)  \gamma_{cd}+\frac{i}{2}g_{1}B_{\mu
}\right)  \delta_{a}^{b}-\frac{i}{2}g_{2}W_{\mu}^{\alpha}\left(
\sigma^{\alpha}\right)  _{a}^{b}\right)  \otimes1_{3}\\
\left(  \mathbb{\omega}_{\mu}\right)  _{\overset{.}{1}i}^{\overset{.}{1}j}  &
=\left(  \left(  \frac{1}{4}\omega_{\mu}^{cd}\left(  e\right)  \gamma
_{cd}-\frac{2i}{3}g_{1}B_{\mu}\right)  \delta_{i}^{j}-\frac{i}{2}g_{3}V_{\mu
}^{m}\left(  \lambda^{m}\right)  _{i}^{j}\right)  \otimes1_{3}\\
\left(  \mathbb{\omega}_{\mu}\right)  _{\overset{.}{2}i}^{\overset{.}{2}j}  &
=\left(  \left(  \frac{1}{4}\omega_{\mu}^{cd}\left(  e\right)  \gamma
_{cd}+\frac{i}{3}g_{1}B_{\mu}\right)  \delta_{i}^{j}-\frac{i}{2}g_{3}V_{\mu
}^{m}\left(  \lambda^{i}\right)  _{i}^{j}\right)  \otimes1_{3}\\
\left(  \mathbb{\omega}_{\mu}\right)  _{ai}^{bj}  &  =\left(  \left(  \frac
{1}{4}\omega_{\mu}^{cd}\left(  e\right)  \gamma_{cd}-\frac{i}{6}g_{1}B_{\mu
}\right)  \delta_{a}^{b}\delta_{i}^{j}-\frac{i}{2}g_{2}W_{\mu}^{\alpha}\left(
\sigma^{\alpha}\right)  _{a}^{b}\delta_{i}^{j}-\frac{i}{2}g_{3}V_{\mu}%
^{m}\left(  \lambda^{m}\right)  _{i}^{j}\delta_{a}^{b}\right)  \otimes1_{3}\\
\left(  \mathbb{\omega}_{\mu}\right)  _{A}^{B^{\prime}}  &  =0=\left(
\mathbb{\omega}_{\mu}\right)  _{A^{\prime}}^{B},\qquad\left(  \mathbb{\omega
}_{\mu}\right)  _{A^{\prime}}^{B^{\prime}}=\left(  \overline{\mathbb{\omega}%
}_{\mu}\right)  _{A}^{B}%
\end{align*}
\nolinebreak

The components of the curvature $\ \Omega_{\mu\nu}=\partial_{\mu}\omega_{\nu
}-\partial_{\nu}\omega_{\mu}+\left[  \omega_{\mu},\omega_{\nu}\right]  $ are
given by\nolinebreak\ \nolinebreak%
\begin{align*}
\left(  \Omega_{\mu\nu}\right)  _{\overset{.}{1}1}^{\overset{.}{1}1}  &
=\frac{1}{4}R_{\mu\nu}^{cd}\gamma_{cd}\otimes1_{3}\\
\left(  \Omega_{\mu\nu}\right)  _{\overset{.}{2}1}^{\overset{.}{2}1}  &
=\left(  \frac{1}{4}R_{\mu\nu}^{cd}\gamma_{cd}+ig_{1}B_{\mu\nu}\right)
\otimes1_{3}\\
\left(  \Omega_{\mu\nu}\right)  _{a1}^{b1}  &  =\left(  \left(  \frac{1}%
{4}R_{\mu\nu}^{cd}\gamma_{cd}+\frac{i}{2}g_{1}B_{\mu\nu}\right)  \delta
_{a}^{b}-\frac{i}{2}g_{2}W_{\mu\nu}^{\alpha}\left(  \sigma^{\alpha}\right)
_{a}^{b}\right)  \otimes1_{3}\\
\left(  \Omega_{\mu\nu}\right)  _{\overset{.}{1}i}^{\overset{.}{1}j}  &
=\left(  \left(  \frac{1}{4}R_{\mu\nu}^{cd}\gamma_{cd}-\frac{2i}{3}g_{1}%
B_{\mu\nu}\right)  \delta_{i}^{j}-\frac{i}{2}g_{3}V_{\mu\nu}^{m}\left(
\lambda^{m}\right)  _{i}^{j}\right)  \otimes1_{3}\\
\left(  \Omega_{\mu\nu}\right)  _{\overset{.}{2}i}^{\overset{.}{2}j}  &
=\left(  \left(  \frac{1}{4}R_{\mu\nu}^{cd}\gamma_{cd}+\frac{i}{3}g_{1}%
B_{\mu\nu}\right)  \delta_{i}^{j}-\frac{i}{2}g_{3}V_{\mu\nu}^{m}\left(
\lambda^{m}\right)  _{i}^{j}\right)  \otimes1_{3}\\
\left(  \Omega_{\mu\nu}\right)  _{ai}^{bj}  &  =\left(  \left(  \frac{1}%
{4}R_{\mu\nu}^{cd}\gamma_{cd}-\frac{i}{6}g_{1}B_{\mu\nu}\right)  \delta
_{a}^{b}\delta_{i}^{j}-\frac{i}{2}g_{2}W_{\mu\nu}^{\alpha}\left(
\sigma^{\alpha}\right)  _{a}^{b}\delta_{i}^{j}-\frac{i}{2}g_{3}V_{\mu\nu}%
^{m}\left(  \lambda^{m}\right)  _{i}^{j}\delta_{a}^{b}\right)  \otimes1_{3}\\
\left(  \Omega_{\mu\nu}\right)  _{A}^{B^{\prime}}  &  =0=\left(  \Omega
_{\mu\nu}\right)  _{A^{\prime}}^{B},\qquad\left(  \Omega_{\mu\nu}\right)
_{A^{\prime}}^{B^{\prime}}=\left(  \overline{\Omega}_{\mu\nu}\right)  _{A}^{B}%
\end{align*}
Finally%
\begin{align*}
-\left(  E\right)  _{\overset{.}{1}1}^{\overset{.}{1}1}  &  =\frac{1}%
{4}R\otimes1_{3}+\left(  k^{\ast\nu}k^{\nu}\overline{H}H+k^{\ast\nu_{R}}%
k^{\nu_{R}}\sigma^{2}\right) \\
-\left(  E\right)  _{\overset{.}{1}1}^{a1}  &  =\gamma^{\mu}\gamma_{5}\otimes
k^{\ast\nu}\otimes\epsilon^{ab}\nabla_{\mu}H_{b}\\
-\left(  E\right)  _{\overset{.}{2}1}^{\overset{.}{2}1}  &  =\left(  \frac
{1}{4}R+\frac{1}{2}\gamma^{\mu\nu}\left(  ig_{1}B_{\mu\nu}\right)  \right)
\otimes1_{3}+\left(  k^{\ast e}k^{e}\overline{H}H\right) \\
-\left(  E\right)  _{\overset{.}{2}1}^{a1}  &  =\gamma^{\mu}\gamma_{5}\otimes
k^{\ast e}\otimes\nabla_{\mu}\overline{H}^{a}\\
-\left(  E\right)  _{a1}^{\overset{.}{1}1}  &  =\left(  \gamma^{\mu}\gamma
_{5}\otimes k^{\nu}\otimes\epsilon_{ab}\nabla_{\mu}\overline{H}^{b}\right) \\
-\left(  E\right)  _{a1}^{\overset{.}{2}1}  &  =\gamma^{\mu}\gamma_{5}\otimes
k^{e}\otimes\nabla_{\mu}H_{a}\\
-\left(  E\right)  _{a1}^{b1}  &  =\left(  \frac{1}{4}R\delta_{a}^{b}+\frac
{1}{2}\gamma^{\mu\nu}\left(  \frac{i}{2}g_{1}B_{\mu\nu}\delta_{a}^{b}-\frac
{i}{2}g_{2}W_{\mu\nu}^{\alpha}\left(  \sigma^{\alpha}\right)  _{a}^{b}\right)
\right)  \otimes1_{3}\\
&  +\left(  k^{e}k^{\ast e}H_{a}\overline{H}^{b}+k^{\nu}k^{\ast\nu}%
\epsilon_{ac}\epsilon^{bd}\overline{H}^{c}H_{d}\right) \\
-\left(  E\right)  _{\overset{.}{1}i}^{\overset{.}{1}j}  &  =\left(  \frac
{1}{4}R\delta_{i}^{j}+\frac{1}{2}\gamma^{\mu\nu}\left(  -\frac{2i}{3}%
g_{1}B_{\mu\nu}\delta_{i}^{j}-\frac{i}{2}g_{3}V_{\mu\nu}^{m}\left(
\lambda^{m}\right)  _{i}^{j}\right)  \right)  \otimes1_{3}\\
&  +\left(  k^{\ast u}k^{u}H_{a}\overline{H}^{a}\right)  \delta_{i}^{j}\\
-\left(  E\right)  _{\overset{.}{1}i}^{aj}  &  =\gamma^{\mu}\gamma_{5}\otimes
k^{\ast u}\otimes\epsilon^{ab}\nabla_{\mu}H_{b}\delta_{i}^{j}\\
-\left(  E\right)  _{\overset{.}{2}i}^{\overset{.}{2}j}  &  =\left(  \frac
{1}{4}R\delta_{i}^{j}+\frac{1}{2}\gamma^{\mu\nu}\left(  \frac{i}{3}g_{1}%
B_{\mu\nu}\delta_{i}^{j}-\frac{i}{2}g_{3}V_{\mu\nu}^{m}\left(  \lambda
^{m}\right)  _{i}^{j}\right)  \right)  \otimes1_{3}\\
&  +\left(  k^{\ast d}k^{d}\overline{H}H\right)  \delta_{i}^{j}\\
-\left(  E\right)  _{\overset{.}{2}i}^{aj}  &  =\gamma^{\mu}\gamma_{5}\otimes
k^{\ast d}\otimes\nabla_{\mu}\overline{H}^{a}\delta_{i}^{j}\\
-\left(  E\right)  _{ai}^{\overset{.}{1}j}  &  =\gamma^{\mu}\gamma_{5}\otimes
k^{u}\otimes\epsilon_{ab}\nabla_{\mu}\overline{H}^{b}\delta_{i}^{j}\\
-\left(  E\right)  _{ai}^{\overset{.}{2}j}  &  =\gamma^{\mu}\gamma_{5}\otimes
k^{d}\otimes\nabla_{\mu}H_{a}\delta_{i}^{j}\\
-\left(  E\right)  _{ai}^{bj}  &  =\left(  \frac{1}{4}R\delta_{a}^{b}%
\delta_{i}^{j}+\frac{1}{2}\gamma^{\mu\nu}\left(  -\frac{i}{6}g_{1}B_{\mu\nu
}\delta_{a}^{b}\delta_{i}^{j}-\frac{i}{2}g_{2}W_{\mu\nu}^{\alpha}\left(
\sigma^{\alpha}\right)  _{a}^{b}\delta_{i}^{j}-\frac{i}{2}g_{3}V_{\mu\nu}%
^{m}\left(  \lambda^{m}\right)  _{i}^{j}\delta_{a}^{b}\right)  \right)
\otimes1_{3}\\
&  +\left(  k^{e}k^{\ast e}H_{a}\overline{H}^{b}+k^{\nu}k^{\ast\nu}%
\epsilon_{ac}\epsilon^{bd}\overline{H}^{c}H_{d}\right)  \delta_{i}^{j}\\
-\left(  E\right)  _{\overset{.}{1}1}^{a^{\prime}1^{\prime}}  &  =k^{\ast
\nu_{R}}\overline{k^{\ast\nu}}\otimes\epsilon^{ab}\overline{H}_{b}\sigma\\
-\left(  E\right)  _{a^{\prime}1^{\prime}}^{\overset{.}{1}1}  &
=\overline{k^{\nu}}k^{\nu_{R}}\otimes\epsilon_{ab}H^{b}\sigma\\
-\left(  E\right)  _{a1}^{\overset{.}{1}^{^{\prime}}1^{^{\prime}}}  &
=k^{\nu}k^{\ast\nu_{R}}\otimes\epsilon_{ab}\overline{H}^{b}\sigma\\
-\left(  E\right)  _{\overset{.}{1}^{^{\prime}}1^{^{\prime}}}^{a1}  &
=k^{\ast\nu_{R}}\overline{k^{\ast\nu}}\otimes\epsilon^{ab}\overline{H}%
_{b}\sigma\\
-\left(  E\right)  _{\overset{.}{1}1}^{\overset{.}{1}^{^{\prime}}1^{^{\prime}%
}}  &  =\gamma^{\mu}\gamma_{5}\otimes k^{\ast\nu_{R}}\otimes\partial_{\mu
}\sigma\\
-\left(  E\right)  _{\overset{.}{1}^{^{\prime}}1^{^{\prime}}}^{\overset{.}%
{1}1}  &  =\gamma^{\mu}\gamma_{5}\otimes k^{\nu_{R}}\otimes\partial_{\mu
}\sigma
\end{align*}

We can easily compute the trace of $E$%
\begin{equation}
\text{\textrm{Tr} }\left(  E\right)  =\text{\textrm{tr} }\left(  E_{A}%
^{A}+E_{A^{\prime}}^{A^{\prime}}\right)  =\text{\textrm{tr} }\left(  E_{A}%
^{A}+\overline{E}_{A}^{A}\right)
\end{equation}
Thus
\begin{align}
-\text{\textrm{tr} }\left(  E\right)  _{\overset{.}{1}1}^{\overset{.}{1}1}  &
=\text{\textrm{tr}}\left(  \frac{1}{4}R\otimes1_{3}+\left(  k^{\ast\nu}k^{\nu
}\overline{H}H+k^{\ast\nu_{R}}k^{\nu_{R}}\sigma^{2}\right)  \right)  \text{
}\\
&  =4\left[  \frac{3}{4}R+k^{\ast\nu}k^{\nu}\overline{H}H+k^{\ast\nu_{R}%
}k^{\nu_{R}}\sigma^{2}\right] \nonumber
\end{align}%
\begin{align}
-\text{\textrm{tr} }\left(  E\right)  _{\overset{.}{2}1}^{\overset{.}{2}1}  &
=\text{\textrm{tr}}\left(  \left(  \frac{1}{4}R+\frac{1}{2}\gamma^{\mu\nu
}\left(  ig_{1}B_{\mu\nu}\right)  \right)  1_{3}+\left(  k^{\ast e}%
k^{e}\overline{H}H\right)  \right) \\
&  =4\left[  \frac{3}{4}R+k^{\ast e}k^{e}\overline{H}H\right] \nonumber
\end{align}%
\begin{equation}
-\text{\textrm{tr} }\left(  E\right)  _{a1}^{a1}=4\left[  \frac{3}{4}R\left(
2\right)  +\left(  k^{e}k^{\ast e}+k^{\nu}k^{\ast\nu}\right)  \overline
{H}H\right]
\end{equation}%
\begin{equation}
-\text{\textrm{tr} }\left(  E\right)  _{\overset{.}{1}i}^{\overset{.}{1}%
i}=4\left[  \frac{3}{4}\left(  3\right)  R+3k^{\ast u}k^{u}\overline
{H}H\right]
\end{equation}%
\begin{equation}
-\text{\textrm{tr} }\left(  E\right)  _{\overset{.}{1}i}^{\overset{.}{1}%
i}=4\left[  \frac{3}{4}\left(  3\right)  R+3k^{\ast d}k^{d}\overline
{H}H\right]
\end{equation}%
\begin{equation}
-\text{\textrm{tr} }\left(  E\right)  _{ai}^{ai}=4\left[  \frac{3}{4}R\left(
2\right)  \left(  3\right)  +3\left(  k^{u}k^{\ast u}+k^{d}k^{\ast d}\right)
\overline{H}H\right]
\end{equation}
Collecting all terms we get
\begin{align}
-\frac{1}{2}\text{\textrm{Tr} }\left(  E\right)   &  =4\left[  \frac{3}%
{4}R\left(  1+1+2+3+3+6\right)  \right. \\
&  \left.  +2\left(  k^{\ast\nu}k^{\nu}+k^{\ast e}k^{e}+3\left(  k^{\ast
u}k^{u}+k^{\ast d}k^{d}\right)  \right)  \overline{H}H+k^{\ast\nu_{R}}%
k^{\nu_{R}}\,\sigma^{2}\right] \nonumber\\
&  =4\left[  12R+2a\overline{H}H+c\,\sigma^{2}\right] \nonumber
\end{align}

\begin{align}
\text{\textrm{Tr}}\left(  \Omega_{\mu\nu}^{2}\right)  _{ai}^{ai}  &  =\\
&  \text{\textrm{Tr}}\left\{  \left(  \left(  \left(  \frac{1}{4}R_{\mu\nu
}^{cd}\gamma_{cd}-\frac{i}{6}g_{1}B_{\mu\nu}\right)  \delta_{a}^{b}\delta
_{m}^{n}-\frac{i}{2}g_{2}W_{\mu\nu}^{\alpha}\left(  \sigma^{\alpha}\right)
_{a}^{b}\delta_{m}^{n}-\frac{i}{2}g_{2}V_{\mu\nu}^{m}\left(  \lambda
^{m}\right)  _{i}^{j}\delta_{a}^{b}\right)  \otimes1_{3}\right)  ^{2}\right\}
\nonumber\\
&  =4\left[  -\frac{1}{8}R_{\mu\nu\rho\sigma}^{2}\left(  3\right)  \left(
2\right)  \left(  3\right)  -\frac{1}{36}g_{1}^{2}B_{\mu\nu}^{2}\left(
3\right)  \left(  2\right)  \left(  3\right)  \right. \nonumber\\
&  \qquad\left.  -\frac{1}{4}g_{2}^{2}\left(  W_{\mu\nu}^{\alpha}\right)
^{2}\left(  3\right)  \left(  2\right)  \left(  3\right)  -\frac{1}{4}%
g_{3}^{2}\left(  V_{\mu\nu}^{m}\right)  ^{2}\left(  3\right)  \left(
2\right)  \left(  2\right)  \right] \nonumber
\end{align}
Collecting these terms we have
\begin{align}
\frac{1}{2}\text{\textrm{Tr} }\left(  \Omega_{\mu\nu}^{2}\right)  _{M}^{M}  &
=4\left[  -\frac{3}{8}R_{\mu\nu\rho\sigma}^{2}\left(  16\right)  -3g_{1}%
^{2}B_{\mu\nu}^{2}\left(  1+\frac{1}{2}+\frac{4}{3}+\frac{1}{3}+\frac{1}%
{6}\right)  \right. \\
&  \left.  -3g_{2}^{2}\left(  W_{\mu\nu}^{\alpha}\right)  ^{2}\left(  \frac
{1}{2}+\frac{3}{2}\right)  -3g_{3}^{2}\left(  V_{\mu\nu}^{m}\right)
^{2}\left(  \frac{1}{2}+\frac{1}{2}+1\right)  \right] \nonumber\\
&  =4\left[  -6R_{\mu\nu\rho\sigma}^{2}-10g_{1}^{2}B_{\mu\nu}^{2}-6g_{2}%
^{2}\left(  W_{\mu\nu}^{\alpha}\right)  ^{2}-6g_{3}^{2}\left(  V_{\mu\nu}%
^{m}\right)  ^{2}\right] \nonumber
\end{align}

\section{Appendix D: components and traces of $E^{2}$ and $\Omega^{2}$}

Next we compute \nolinebreak$\left(  E^{2}\right)  _{A}^{B}=E_{A}^{C}E_{C}%
^{B}+E_{A}^{C^{\prime}}E_{C^{\prime}}^{B}:\nolinebreak$%
\begin{align}
\left(  E^{2}\right)  _{\overset{.}{1}1}^{\overset{.}{1}1} &  =E_{\overset
{.}{1}1}^{\overset{.}{1}1}E_{\overset{.}{1}1}^{\overset{.}{1}1}+E_{\overset
{.}{1}1}^{a1}E_{a1}^{\overset{.}{1}1}+E_{\overset{.}{1}1}^{a^{\prime}%
1^{\prime}}E_{a^{\prime}1^{\prime}}^{\overset{.}{1}1}+E_{\overset{.}{1}%
1}^{\overset{.}{1}^{^{\prime}}1^{^{\prime}}}E_{\overset{.}{1}^{^{\prime}%
}1^{^{\prime}}}^{\overset{.}{1}1}\\
\left(  E^{2}\right)  _{\overset{.}{1}1}^{\overset{.}{2}1} &  =E_{\overset
{.}{1}1}^{a1}E_{a1}^{\overset{.}{2}1}\\
\left(  E^{2}\right)  _{\overset{.}{2}1}^{\overset{.}{1}1} &  =E_{\overset
{.}{2}1}^{a1}E_{a1}^{\overset{.}{1}1}\\
\left(  E^{2}\right)  _{\overset{.}{2}1}^{\overset{.}{2}1} &  =E_{\overset
{.}{2}1}^{\overset{.}{2}1}E_{\overset{.}{2}1}^{\overset{.}{2}1}+E_{\overset
{.}{2}1}^{a1}E_{a1}^{\overset{.}{2}1}\\
\left(  E^{2}\right)  _{\overset{.}{1}1}^{a1} &  =E_{\overset{.}{1}%
1}^{\overset{.}{1}1}E_{\overset{.}{1}1}^{a1}+E_{\overset{.}{1}1}^{b1}%
E_{b1}^{a1}+E_{\overset{.}{1}1}^{\overset{.}{1}^{^{\prime}}1^{^{\prime}}%
}E_{\overset{.}{1}^{^{\prime}}1^{^{\prime}}}^{a1}\\
\left(  E^{2}\right)  _{\overset{.}{2}1}^{a1} &  =E_{\overset{.}{2}%
1}^{\overset{.}{2}1}E_{\overset{.}{2}1}^{a1}+E_{\overset{.}{2}1}^{b1}%
E_{b1}^{a1}\\
\left(  E^{2}\right)  _{a1}^{\overset{.}{1}1} &  =E_{a1}^{\overset{.}{1}%
1}E_{\overset{.}{1}1}^{\overset{.}{1}1}+E_{a1}^{b1}E_{b1}^{\overset{.}{1}%
1}+E_{a1}^{\overset{.}{1}^{^{\prime}}1^{^{\prime}}}E_{\overset{.}{1}%
^{^{\prime}}1^{^{\prime}}}^{\overset{.}{1}1}\\
\left(  E^{2}\right)  _{a1}^{\overset{.}{2}1} &  =E_{a1}^{\overset{.}{2}%
1}E_{\overset{.}{2}1}^{\overset{.}{2}1}+E_{a1}^{b1}E_{b1}^{\overset{.}{2}1}\\
\left(  E^{2}\right)  _{a1}^{b1} &  =E_{a1}^{\overset{.}{1}1}E_{\overset{.}%
{1}1}^{b1}+E_{a1}^{\overset{.}{2}1}E_{\overset{.}{2}1}^{b1}+E_{a1}^{c1}%
E_{c1}^{b1}+E_{a1}^{\overset{.}{1}^{^{\prime}}1^{^{\prime}}}E_{\overset{.}%
{1}^{^{\prime}}1^{^{\prime}}}^{b1}\\
\left(  E^{2}\right)  _{\overset{.}{1}i}^{\overset{.}{1}j} &  =E_{\overset
{.}{1}i}^{\overset{.}{1}k}E_{\overset{.}{1}k}^{\overset{.}{1}j}+E_{\overset
{.}{1}i}^{ak}E_{ak}^{\overset{.}{1}j}\\
\left(  E^{2}\right)  _{\overset{.}{1}i}^{\overset{.}{2}j} &  =E_{\overset
{.}{1}i}^{\overset{.}{1}k}E_{\overset{.}{1}k}^{\overset{.}{2}j}+E_{\overset
{.}{1}i}^{ak}E_{ak}^{\overset{.}{2}j}\\
\left(  E^{2}\right)  _{\overset{.}{1}i}^{aj} &  =E_{\overset{.}{1}%
i}^{\overset{.}{1}k}E_{\overset{.}{1}k}^{aj}+E_{\overset{.}{1}i}^{bk}%
E_{bk}^{aj}\\
\left(  E^{2}\right)  _{\overset{.}{2}i}^{\overset{.}{2}j} &  =E_{\overset
{.}{2}i}^{\overset{.}{2}k}E_{\overset{.}{2}k}^{\overset{.}{2}j}+E_{\overset
{.}{2}i}^{ak}E_{ak}^{\overset{.}{2}j}\\
\left(  E^{2}\right)  _{\overset{.}{2}i}^{aj} &  =E_{\overset{.}{2}%
i}^{\overset{.}{2}k}E_{\overset{.}{2}k}^{aj}+E_{\overset{.}{2}i}^{bk}%
E_{bk}^{aj}%
\end{align}%
\begin{align}
\left(  E^{2}\right)  _{ai}^{\overset{.}{1}j} &  =E_{ai}^{\overset{.}{1}%
k}E_{\overset{.}{1}k}^{\overset{.}{1}j}+E_{ai}^{bk}E_{bk}^{\overset{.}{1}j}\\
\left(  E^{2}\right)  _{ai}^{\overset{.}{2}j} &  =E_{ai}^{\overset{.}{2}%
k}E_{\overset{.}{2}k}^{\overset{.}{2}j}+E_{ai}^{bk}E_{bk}^{\overset{.}{2}j}\\
\left(  E^{2}\right)  _{ai}^{bj} &  =E_{ai}^{\overset{.}{1}k}E_{\overset{.}%
{1}k}^{bj}+E_{ai}^{\overset{.}{2}k}E_{\overset{.}{2}k}^{bj}+E_{ai}^{ck}%
E_{ck}^{bj}\\
\left(  E^{2}\right)  _{a^{\prime}1^{\prime}}^{\overset{.}{1}1} &
=E_{a^{\prime}1^{\prime}}^{\overset{.}{1}1}E_{\overset{.}{1}1}^{\overset{.}%
{1}1}+E_{a^{\prime}1^{\prime}}^{b^{\prime}1^{\prime}}E_{b^{\prime}1^{\prime}%
}^{\overset{.}{1}1}+E_{a^{\prime}1^{\prime}}^{\overset{.}{1}^{^{\prime}%
}1^{^{\prime}}}E_{\overset{.}{1}^{^{\prime}}1^{^{\prime}}}^{\overset{.}{1}1}\\
\left(  E^{2}\right)  _{a^{\prime}1^{\prime}}^{b1} &  =E_{a^{\prime}1^{\prime
}}^{\overset{.}{1}1}E_{\overset{.}{1}1}^{b1}+E_{a^{\prime}1^{\prime}%
}^{\overset{.}{1}^{^{\prime}}1^{^{\prime}}}E_{\overset{.}{1}^{^{\prime}%
}1^{^{\prime}}}^{b1}\\
\left(  E^{2}\right)  _{\overset{.}{1}^{^{\prime}}1^{^{\prime}}}^{a1} &
=E_{\overset{.}{1}^{^{\prime}}1^{^{\prime}}}^{b1}E_{b1}^{a1}+E_{\overset{.}%
{1}^{^{\prime}}1^{^{\prime}}}^{\overset{.}{1}^{^{\prime}}1^{^{\prime}}%
}E_{\overset{.}{1}^{^{\prime}}1^{^{\prime}}}^{a1}+E_{\overset{.}{1}^{^{\prime
}}1^{^{\prime}}}^{\overset{.}{1}1}E_{\overset{.}{1}1}^{a1}\\
\left(  E^{2}\right)  _{\overset{.}{1}1}^{a^{\prime}1^{\prime}} &
=E_{\overset{.}{1}1}^{b^{\prime}1^{\prime}}E_{b^{\prime}1^{\prime}}%
^{a^{\prime}1^{\prime}}+E_{\overset{.}{1}1}^{\overset{.}{1}1}E_{\overset{.}%
{1}1}^{a^{\prime}1^{\prime}}+E_{\overset{.}{1}1}^{\overset{.}{1}^{^{\prime}%
}1^{^{\prime}}}E_{\overset{.}{1}^{^{\prime}}1^{^{\prime}}}^{a^{\prime
}1^{\prime}}\\
\left(  E^{2}\right)  _{a1}^{\overset{.}{1}^{^{\prime}}1^{^{\prime}}} &
=E_{a1}^{b1}E_{b1}^{\overset{.}{1}^{^{\prime}}1^{^{\prime}}}+E_{a1}%
^{\overset{.}{1}^{^{\prime}}1^{^{\prime}}}E_{\overset{.}{1}^{^{\prime}%
}1^{^{\prime}}}^{\overset{.}{1}^{^{\prime}}1^{^{\prime}}}+E_{a1}^{\overset
{.}{1}1}E_{\overset{.}{1}1}^{\overset{.}{1}^{^{\prime}}1^{^{\prime}}}\\
\left(  E^{2}\right)  _{\overset{.}{1}^{^{\prime}}1^{^{\prime}}}^{\overset
{.}{1}1} &  =E_{\overset{.}{1}^{^{\prime}}1^{^{\prime}}}^{\overset{.}{1}%
1}E_{\overset{.}{1}1}^{\overset{.}{1}1}+E_{\overset{.}{1}^{^{\prime}%
}1^{^{\prime}}}^{a1}E_{a1}^{\overset{.}{1}1}+E_{\overset{.}{1}^{^{\prime}%
}1^{^{\prime}}}^{a^{\prime}1^{\prime}}E_{a^{\prime}1^{\prime}}^{\overset{.}%
{1}1}+E_{\overset{.}{1}^{^{\prime}}1^{^{\prime}}}^{\overset{.}{1}^{^{\prime}%
}1^{^{\prime}}}E_{\overset{.}{1}^{^{\prime}}1^{^{\prime}}}^{\overset{.}{1}1}%
\end{align}
We list the various traces for $E^{2}$
\begin{align}
\text{\textrm{tr} }\left(  E^{2}\right)  _{\overset{.}{1}1}^{\overset{.}{1}1}
&  =\text{\textrm{tr}}\left\{  \left(  \frac{1}{4}R.1_{3}+\left(  k^{\ast\nu
}k^{\nu}\overline{H}H+k^{\ast\nu_{R}}k^{\nu_{R}}\,\sigma^{2}\right)  \right)
^{2}+\gamma^{\mu}\gamma_{5}k^{\ast\nu_{R}}\partial_{\mu}\sigma\gamma^{\nu
}\gamma_{5}k^{\nu_{R}}\partial_{\nu}\sigma\right.  \\
&  \quad\left.  +\gamma^{\mu}\gamma_{5}k^{\ast\nu}\epsilon^{ab}\nabla_{\mu
}H_{b}\gamma^{\nu}\gamma_{5}k^{\nu}\epsilon_{ac}\nabla_{\nu}\overline{H}%
^{c}+k^{\ast\nu_{R}}\overline{k^{\ast\nu}}\epsilon^{ab}\overline{H}%
_{b}\overline{k^{\nu}}k^{\nu_{R}}\epsilon_{ac}H^{c}\right\}  \nonumber\\
&  =4\left[  \frac{1}{16}R^{2}\left(  3\right)  +\left(  k^{\ast\nu}k^{\nu
}\right)  ^{2}\left(  \overline{H}H\right)  ^{2}+3k^{\ast\nu}k^{\nu}k^{\ast
\nu_{R}}k^{\nu_{R}}\overline{H}H\,\sigma^{2}+k^{\ast\nu_{R}}k^{\nu_{R}}\left(
\partial_{\mu}\sigma\right)  ^{2}\right.  \nonumber\\
&  \quad\left.  +\left(  k^{\ast\nu_{R}}k^{\nu_{R}}\right)  ^{2}\sigma
^{4}+k^{\ast\nu}k^{\nu}\left\vert \nabla_{\mu}H_{a}\right\vert ^{2}+\frac
{1}{2}R\left(  k^{\ast\nu}k^{\nu}\overline{H}H+k^{\ast\nu_{R}}k^{\nu_{R}%
}\sigma^{2}\right)  \right]  \nonumber
\end{align}
where we have used \textrm{tr}$\left(  \gamma^{\mu}\gamma_{5}\gamma^{\nu
}\gamma_{5}\right)  =4g^{\mu\nu}.$%
\begin{align}
\text{\textrm{tr }}\left(  E^{2}\right)  _{\overset{.}{2}1}^{\overset{.}{2}1}
&  =\text{\textrm{tr}}\left\{  \gamma^{\mu}\gamma_{5}k^{\ast e}\nabla_{\mu
}\overline{H}^{a}\gamma^{\nu}\gamma_{5}\otimes k^{e}\nabla_{\nu}H_{a}\right.
\\
&  \left.  +\left(  \left(  \frac{1}{4}R+\frac{1}{2}\gamma^{\mu\nu}\left(
ig_{1}B_{\mu\nu}\right)  \right)  1_{3}+\left(  k^{\ast e}k^{e}\overline
{H}H\right)  \right)  ^{2}\right\}  \nonumber\\
&  =4\left[  \left(  \frac{1}{4}\left(  -2\right)  \left(  -g_{1}^{2}B_{\mu
\nu}^{2}\right)  +\frac{1}{16}R^{2}\right)  3+\frac{1}{2}Rk^{\ast e}%
k^{e}\overline{H}H\right.  \nonumber\\
&  \left.  +\left(  k^{\ast e}k^{e}\right)  ^{2}\left(  \overline{H}H\right)
^{2}+k^{\ast e}k^{e}\left\vert \nabla_{\mu}H_{a}\right\vert ^{2}\right]
\nonumber
\end{align}
where we have used \textrm{tr}$\left(  \gamma^{\mu\nu}\gamma_{\kappa\lambda
}\right)  =-4\left(  \delta_{\kappa}^{\mu}\delta_{\lambda}^{\nu}%
-\delta_{\lambda}^{\mu}\delta_{\kappa}^{\nu}\right)  .$
\begin{align}
&  \text{\textrm{tr} }\left(  E^{2}\right)  _{a1}^{a1}\\
&  =\text{\textrm{tr}}\left\{  \left(  \left(  \frac{R}{4}\delta_{a}^{b}%
+\frac{1}{2}\gamma^{\mu\nu}\left(  \frac{i}{2}g_{1}B_{\mu\nu}\delta_{a}%
^{b}-\frac{i}{2}g_{2}W_{\mu\nu}^{\alpha}\left(  \sigma^{\alpha}\right)
_{a}^{b}\right)  \right)  +\left(  k^{e}k^{\ast e}H_{a}\overline{H}^{b}%
+k^{\nu}k^{\ast\nu}\epsilon_{ac}\epsilon^{bd}\overline{H}^{c}H_{d}\right)
\right)  ^{2}\right.  \nonumber\\
&  \left.  +\gamma^{\mu}\gamma_{5}k^{\ast\nu}\epsilon^{ab}\nabla_{\mu}%
H_{b}\gamma^{\nu}\gamma_{5}k^{\nu}\epsilon_{ab}\nabla_{\nu}\overline{H}%
^{b}+\gamma^{\mu}\gamma_{5}k^{\ast e}\nabla_{\mu}\overline{H}^{a}\gamma^{\nu
}\gamma_{5}k^{e}\nabla_{\nu}H_{a}\right\}  \nonumber\\
&  =4\left[  \frac{1}{4}\left(  -2\right)  \left(  -\frac{1}{4}g_{1}^{2}%
B_{\mu\nu}^{2}\left(  2\right)  \left(  3\right)  -\frac{1}{4}g_{2}^{2}\left(
W_{\mu\nu}^{\alpha}\right)  ^{2}\left(  2\right)  \left(  3\right)  \right)
+\frac{1}{16}R^{2}\left(  2\right)  \left(  3\right)  \right.  \nonumber\\
&  +\frac{1}{2}R\left(  k^{\ast\nu}k^{\nu}+k^{\ast e}k^{e}\right)
\overline{H}H+\left(  \left(  k^{\ast\nu}k^{\nu}\right)  ^{2}+\left(  k^{\ast
e}k^{e}\right)  ^{2}\right)  \left(  \overline{H}H\right)  ^{2}\nonumber\\
&  \left.  +\left(  k^{\ast\nu}k^{\nu}+k^{\ast e}k^{e}\right)  \left\vert
\nabla_{\mu}H_{a}\right\vert ^{2}+k^{\ast\nu}k^{\nu}k^{\ast\nu_{R}}k^{\nu_{R}%
}\overline{H}H\,\sigma^{2}\right]  \nonumber
\end{align}
where the factor $\left(  2\right)  =\delta_{a}^{a}$ and \textrm{tr}$\left(
\sigma^{\alpha}\sigma^{\beta}\right)  =2\delta^{\alpha\beta}$ and the factor
$\left(  3\right)  =$\textrm{tr} $1_{3}$ of the $3$ generations. Next
\begin{align}
&  \text{\textrm{tr} }\left(  E^{2}\right)  _{\overset{.}{1}i}^{\overset{.}%
{1}i}\\
&  =\text{\textrm{tr}}\left\{  \left(  \left(  \frac{R}{4}\delta_{i}^{j}%
+\frac{1}{2}\gamma^{\mu\nu}\left(  -\frac{2i}{3}g_{1}B_{\mu\nu}\delta_{i}%
^{j}-\frac{i}{2}g_{3}V_{\mu\nu}^{m}\left(  \lambda^{m}\right)  _{i}%
^{j}\right)  \right)  1_{3}+\left(  k^{\ast u}k^{u}\overline{H}H\right)
\delta_{i}^{j}\right)  ^{2}\right.  \nonumber\\
&  \left.  +\gamma^{\mu}\gamma_{5}k^{\ast u}\epsilon^{ab}\nabla_{\mu}%
H_{b}\delta_{i}^{j}\gamma^{\nu}\gamma_{5}k^{u}\epsilon_{ab}\nabla_{\nu
}\overline{H}^{b}\delta_{j}^{i}\right\}  \nonumber\\
&  =4\left[  \frac{1}{4}\left(  -2\right)  \left(  -\frac{4}{9}g_{1}^{2}%
B_{\mu\nu}^{2}\left(  3\right)  \left(  3\right)  -\frac{1}{4}g_{3}^{2}\left(
V_{\mu\nu}^{m}\right)  ^{2}\left(  2\right)  \left(  3\right)  \right)
+\frac{1}{16}R^{2}\left(  3\right)  \left(  3\right)  \right.  \nonumber\\
&  \left.  +\left(  k^{\ast u}k^{u}\right)  ^{2}\left(  \overline{H}H\right)
^{2}\left(  3\right)  +\frac{1}{2}R\left(  3\right)  \left(  k^{\ast u}%
k^{u}\right)  \left(  \overline{H}H\right)  +\left(  3\right)  \left(  k^{\ast
u}k^{u}\right)  \left\vert \nabla_{\mu}H_{a}\right\vert ^{2}\right]  \nonumber
\end{align}
where $\left(  3\right)  =\delta_{i}^{i}$ and \textrm{tr}$\left(  \lambda
^{m}\lambda^{n}\right)  =2\delta^{mn}.$
\begin{align}
\text{\textrm{tr} }\left(  E^{2}\right)  _{\overset{.}{2}i}^{\overset{.}{2}i}
&  =\text{\textrm{tr}}\left\{  \left(  \left(  \frac{R}{4}R\delta_{i}%
^{j}+\frac{1}{2}\gamma^{\mu\nu}\left(  \frac{i}{3}g_{1}B_{\mu\nu}\delta
_{i}^{j}-\frac{i}{2}g_{3}V_{\mu\nu}^{m}\left(  \lambda^{m}\right)  _{i}%
^{j}\right)  \right)  1_{3}+\left(  k^{\ast d}k^{d}\overline{H}H\right)
\delta_{i}^{j}\right)  ^{2}\right.  \\
&  \left.  +\gamma^{\mu}\gamma_{5}k^{\ast d}\nabla_{\mu}\overline{H}_{a}%
\delta_{i}^{j}\gamma^{\nu}\gamma_{5}k^{d}\nabla_{\nu}H^{a}\delta_{j}%
^{i}\right\}  \nonumber\\
&  =4\left[  \frac{1}{4}\left(  -2\right)  \left(  -\frac{1}{9}g_{1}^{2}%
B_{\mu\nu}^{2}\left(  3\right)  \left(  3\right)  -\frac{1}{4}g_{3}^{2}\left(
V_{\mu\nu}^{m}\right)  ^{2}\left(  2\right)  \left(  3\right)  \right)
+\frac{1}{16}R^{2}\left(  3\right)  \left(  3\right)  \right.  \nonumber\\
&  \left.  +\left(  k^{\ast d}k^{d}\right)  ^{2}\left(  \overline{H}H\right)
^{2}\left(  3\right)  +\frac{1}{2}R\left(  3\right)  \left(  k^{\ast d}%
k^{d}\right)  \left(  \overline{H}H\right)  +\left(  3\right)  \left(  k^{\ast
d}k^{d}\right)  \left\vert \nabla_{\mu}H_{a}\right\vert ^{2}\right]  \nonumber
\end{align}
Finally
\begin{align}
&  \text{\textrm{tr} }\left(  E^{2}\right)  _{ai}^{ai}\\
&  =\text{\textrm{tr}}\left\{  \left(  \left(  \frac{R}{4}\delta_{a}^{b}%
\delta_{i}^{j}+\frac{1}{2}\gamma^{\mu\nu}\left(  -\frac{i}{6}g_{1}B_{\mu\nu
}\delta_{a}^{b}\delta_{i}^{j}-\frac{i}{2}g_{2}W_{\mu\nu}^{\alpha}\left(
\sigma^{\alpha}\right)  _{a}^{b}\delta_{i}^{j}-\frac{i}{2}g_{3}V_{\mu\nu}%
^{m}\left(  \lambda^{m}\right)  _{i}^{j}\delta_{a}^{b}\right)  \right)
\right.  \right.  \nonumber\\
&  \left.  \left(  k^{e}k^{\ast e}H_{a}\overline{H}^{b}+k^{\nu}k^{\ast\nu
}\epsilon_{ac}\epsilon^{bd}\overline{H}^{c}H_{d}\right)  \delta_{i}%
^{j}\right)  ^{2}\nonumber\\
&  \left.  +\gamma^{\mu}\gamma_{5}k^{u}\epsilon_{ab}\nabla_{\mu}\overline
{H}^{b}\delta_{i}^{j}\gamma^{\mu}\gamma_{5}k^{\ast u}\epsilon^{ac}\nabla_{\mu
}H_{c}\delta_{j}^{i}+\gamma^{\mu}\gamma_{5}k^{d}\nabla_{\mu}H_{a}\delta
_{i}^{j}\gamma^{\nu}\gamma_{5}k^{\ast d}\nabla_{\nu}\overline{H}^{a}\delta
_{j}^{i}\right\}  \nonumber\\
&  =4\left[  \frac{1}{4}\left(  -2\right)  \left(  -\frac{1}{36}g_{1}%
^{2}B_{\mu\nu}^{2}\left(  3\right)  \left(  2\right)  \left(  3\right)
-\frac{1}{4}g_{2}^{2}\left(  W_{\mu\nu}^{\alpha}\right)  ^{2}\left(  3\right)
\left(  2\right)  \left(  3\right)  -\frac{1}{4}g_{3}^{2}\left(  V_{\mu\nu
}^{m}\right)  ^{2}\left(  2\right)  \left(  3\right)  \left(  3\right)
\right)  \right.  \nonumber\\
&  +\frac{1}{16}R^{2}\left(  3\right)  \left(  2\right)  \left(  3\right)
+\frac{1}{2}R\left(  3\right)  \left(  k^{\ast u}k^{u}+k^{\ast d}k^{d}\right)
\left(  \overline{H}H\right)  +3\left(  k^{\ast u}k^{u}+k^{\ast d}%
k^{d}\right)  \left\vert \nabla_{\mu}H_{a}\right\vert ^{2}\nonumber\\
&  \left.  +3\left(  \left(  k^{\ast u}k^{u}\right)  ^{2}+\left(  k^{\ast
d}k^{d}\right)  ^{2}\right)  \left(  \overline{H}H\right)  ^{2}\right]
\nonumber
\end{align}
Collecting all terms
\begin{align}
\frac{1}{2}\text{\textrm{tr} }\left(  E^{2}\right)   &  =4\left[  g_{1}%
^{2}B_{\mu\nu}^{2}\left(  \frac{3}{2}+2+\frac{1}{2}+\frac{1}{4}+\frac{3}%
{4}\right)  \right.  \\
&  +g_{2}^{2}\left(  W_{\mu\nu}^{\alpha}\right)  ^{2}\left(  \frac{9}{4}%
+\frac{3}{4}\right)  +g_{3}^{2}\left(  V_{\mu\nu}^{m}\right)  ^{2}\left(
\frac{3}{4}+\frac{3}{4}+\frac{3}{2}\right)  \nonumber\\
&  +\frac{1}{16}R^{2}\left(  3+3+9+9+18+6\right)  +\frac{1}{2}R\sigma
^{2}k^{\ast\nu_{R}}k^{\nu_{R}}+\left(  k^{\ast\nu_{R}}k^{\nu_{R}}\right)
^{2}\sigma^{4}\nonumber\\
&  +R\overline{H}H\left(  k^{\ast\nu}k^{\nu}+k^{\ast e}k^{e}+3\left(  k^{\ast
u}k^{u}+k^{\ast d}k^{d}\right)  \right)  \nonumber\\
&  +2\left(  \overline{H}H\right)  ^{2}\left(  \left(  k^{\ast\nu}k^{\nu
}\right)  ^{2}+\left(  k^{\ast e}k^{e}\right)  ^{2}+3\left(  \left(  k^{\ast
u}k^{u}\right)  ^{2}+\left(  k^{\ast d}k^{d}\right)  ^{2}\right)  \right)
\nonumber\\
&  \left.  +2\left\vert \nabla_{\mu}H_{a}\right\vert ^{2}\left(  k^{\ast\nu
}k^{\nu}+k^{\ast e}k^{e}+3\left(  k^{\ast u}k^{u}+k^{\ast d}k^{d}\right)
\right)  +4k^{\ast\nu}k^{\nu}k^{\ast\nu_{R}}k^{\nu_{R}}\overline{H}%
H\,\sigma^{2}\right]  \nonumber\\
&  =4\left[  5g_{1}^{2}B_{\mu\nu}^{2}+3g_{2}^{2}\left(  W_{\mu\nu}^{\alpha
}\right)  ^{2}+3g_{3}^{2}\left(  V_{\mu\nu}^{m}\right)  ^{2}+3R^{2}%
+aR\overline{H}H\right.  \nonumber\\
&  \left.  +\frac{1}{2}cR\sigma^{2}+2b\left(  \overline{H}H\right)
^{2}+2a\left\vert \nabla_{\mu}H_{a}\right\vert ^{2}+4e\overline{H}%
H\,\sigma^{2}+c\left(  \partial_{\mu}\sigma\right)  ^{2}+d\,\sigma^{4}\right]
\nonumber
\end{align}
where
\begin{align}
a &  =\text{\textrm{tr}}\left(  k^{\ast\nu}k^{\nu}+k^{\ast e}k^{e}+3\left(
k^{\ast u}k^{u}+k^{\ast d}k^{d}\right)  \right)  \\
b &  =\text{\textrm{tr}}\left(  \left(  k^{\ast\nu}k^{\nu}\right)
^{2}+\left(  k^{\ast e}k^{e}\right)  ^{2}+3\left(  \left(  k^{\ast u}%
k^{u}\right)  ^{2}+\left(  k^{\ast d}k^{d}\right)  ^{2}\right)  \right)  \\
c &  =\text{\textrm{tr}}\left(  k^{\ast\nu_{R}}k^{\nu_{R}}\right)  \\
d &  =\text{\textrm{tr}}\left(  \left(  k^{\ast\nu_{R}}k^{\nu_{R}}\right)
^{2}\right)  \\
e &  =\text{\textrm{tr}}\left(  k^{\ast\nu}k^{\nu}k^{\ast\nu_{R}}k^{\nu_{R}%
}\right)
\end{align}
Next
\begin{align}
\text{\textrm{Tr }}\left(  \Omega_{\mu\nu}^{2}\right)  _{M}^{M} &
=2\text{\textrm{Tr} }\left(  \Omega_{\mu\nu}^{2}\right)  _{A}^{A}\\
&  =2\text{\textrm{Tr}}\left\{  \left(  \Omega_{\mu\nu}^{2}\right)
_{\overset{.}{1}1}^{\overset{.}{1}1}+\left(  \Omega_{\mu\nu}^{2}\right)
_{\overset{.}{2}1}^{\overset{.}{2}1}+\left(  \Omega_{\mu\nu}^{2}\right)
_{a1}^{a1}+\left(  \Omega_{\mu\nu}^{2}\right)  _{\overset{.}{1}i}^{\overset
{.}{1}i}+\left(  \Omega_{\mu\nu}^{2}\right)  _{\overset{.}{2}i}^{\overset
{.}{2}i}+\left(  \Omega_{\mu\nu}^{2}\right)  _{ai}^{ai}\right\}  \nonumber
\end{align}%
\begin{align}
\text{\textrm{Tr}}\left(  \Omega_{\mu\nu}^{2}\right)  _{\overset{.}{1}%
1}^{\overset{.}{1}1} &  =\text{\textrm{Tr}}\left\{  \left(  \frac{1}{4}%
R_{\mu\nu}^{cd}\gamma_{cd}\otimes1_{3}\right)  ^{2}\right\}  \\
&  =4\left[  -\frac{1}{8}R_{\mu\nu\rho\sigma}^{2}\left(  3\right)  \right]
\nonumber
\end{align}%
\begin{align}
\text{\textrm{Tr}}\left(  \Omega_{\mu\nu}^{2}\right)  _{\overset{.}{2}%
1}^{\overset{.}{2}1} &  =\text{\textrm{Tr}}\left\{  \left(  \left(  \frac
{1}{4}R_{\mu\nu}^{cd}\gamma_{cd}+ig_{1}B_{\mu\nu}\right)  \otimes1_{3}\right)
^{2}\right\}  \\
&  =4\left[  -\frac{1}{8}R_{\mu\nu\rho\sigma}^{2}\left(  3\right)  -g_{1}%
^{2}B_{\mu\nu}^{2}\left(  3\right)  \right]  \nonumber
\end{align}%
\begin{align}
\text{\textrm{Tr}}\left(  \Omega_{\mu\nu}^{2}\right)  _{a1}^{a1} &
=\text{\textrm{Tr}}\left\{  \left(  \left(  \left(  \frac{1}{4}R_{\mu\nu}%
^{cd}\gamma_{cd}-\frac{i}{2}g_{1}B_{\mu\nu}\right)  \delta_{a}^{b}-\frac{i}%
{2}g_{2}W_{\mu\nu}^{\alpha}\left(  \sigma^{\alpha}\right)  _{a}^{b}\right)
\otimes1_{3}\right)  ^{2}\right\}  \\
&  =4\left[  -\frac{1}{8}R_{\mu\nu\rho\sigma}^{2}\left(  3\right)  \left(
2\right)  -\frac{1}{4}g_{1}^{2}B_{\mu\nu}^{2}\left(  3\right)  \left(
2\right)  -\frac{1}{4}g_{2}^{2}\left(  W_{\mu\nu}^{\alpha}\right)  ^{2}\left(
3\right)  \left(  2\right)  \right]  \nonumber
\end{align}%
\begin{align}
\text{\textrm{Tr}}\left(  \Omega_{\mu\nu}^{2}\right)  _{\overset{.}{1}%
i}^{\overset{.}{1}i} &  =\text{\textrm{Tr}}\left\{  \left(  \left(  \left(
\frac{1}{4}R_{\mu\nu}^{cd}\gamma_{cd}-\frac{2i}{3}g_{1}B_{\mu\nu}\right)
\delta_{i}^{j}-\frac{i}{2}g_{2}V_{\mu\nu}^{m}\left(  \lambda^{m}\right)
_{i}^{j}\right)  \otimes1_{3}\right)  ^{2}\right\}  \\
&  =4\left[  -\frac{1}{8}R_{\mu\nu\rho\sigma}^{2}\left(  3\right)  \left(
3\right)  -\frac{4}{9}g_{1}^{2}B_{\mu\nu}^{2}\left(  3\right)  \left(
3\right)  -\frac{1}{4}g_{3}^{2}\left(  V_{\mu\nu}^{m}\right)  ^{2}\left(
3\right)  \left(  2\right)  \right]  \nonumber
\end{align}%
\begin{align}
\text{\textrm{Tr}}\left(  \Omega_{\mu\nu}^{2}\right)  _{\overset{.}{2}%
i}^{\overset{.}{2}i} &  =\text{\textrm{Tr}}\left\{  \left(  \left(  \left(
\frac{1}{4}R_{\mu\nu}^{cd}\gamma_{cd}+\frac{i}{3}g_{1}B_{\mu\nu}\right)
\delta_{i}^{j}-\frac{i}{2}g_{2}V_{\mu\nu}^{m}\left(  \lambda^{m}\right)
_{i}^{j}\right)  \otimes1_{3}\right)  ^{2}\right\}  \\
&  =4\left[  -\frac{1}{8}R_{\mu\nu\rho\sigma}^{2}\left(  3\right)  \left(
3\right)  -\frac{1}{9}g_{1}^{2}B_{\mu\nu}^{2}\left(  3\right)  \left(
3\right)  -\frac{1}{4}g_{3}^{2}\left(  V_{\mu\nu}^{m}\right)  ^{2}\left(
3\right)  \left(  2\right)  \right]  \nonumber
\end{align}%
\begin{align}
\text{\textrm{Tr}}\left(  \Omega_{\mu\nu}^{2}\right)  _{ai}^{ai} &  =\\
&  \text{\textrm{Tr}}\left\{  \left(  \left(  \left(  \frac{1}{4}R_{\mu\nu
}^{cd}\gamma_{cd}-\frac{i}{6}g_{1}B_{\mu\nu}\right)  \delta_{a}^{b}\delta
_{m}^{n}-\frac{i}{2}g_{2}W_{\mu\nu}^{\alpha}\left(  \sigma^{\alpha}\right)
_{a}^{b}\delta_{m}^{n}-\frac{i}{2}g_{2}V_{\mu\nu}^{m}\left(  \lambda
^{m}\right)  _{i}^{j}\delta_{a}^{b}\right)  \otimes1_{3}\right)  ^{2}\right\}
\nonumber\\
&  =4\left[  -\frac{1}{8}R_{\mu\nu\rho\sigma}^{2}\left(  3\right)  \left(
2\right)  \left(  3\right)  -\frac{1}{36}g_{1}^{2}B_{\mu\nu}^{2}\left(
3\right)  \left(  2\right)  \left(  3\right)  \right.  \nonumber\\
&  \qquad\left.  -\frac{1}{4}g_{2}^{2}\left(  W_{\mu\nu}^{\alpha}\right)
^{2}\left(  3\right)  \left(  2\right)  \left(  3\right)  -\frac{1}{4}%
g_{3}^{2}\left(  V_{\mu\nu}^{m}\right)  ^{2}\left(  3\right)  \left(
2\right)  \left(  2\right)  \right]  \nonumber
\end{align}
Collecting these terms we have
\begin{align}
\frac{1}{2}\text{\textrm{Tr} }\left(  \Omega_{\mu\nu}^{2}\right)  _{M}^{M} &
=4\left[  -\frac{3}{8}R_{\mu\nu\rho\sigma}^{2}\left(  16\right)  -3g_{1}%
^{2}B_{\mu\nu}^{2}\left(  1+\frac{1}{2}+\frac{4}{3}+\frac{1}{3}+\frac{1}%
{6}\right)  \right.  \\
&  \left.  -3g_{2}^{2}\left(  W_{\mu\nu}^{\alpha}\right)  ^{2}\left(  \frac
{1}{2}+\frac{3}{2}\right)  -3g_{3}^{2}\left(  V_{\mu\nu}^{m}\right)
^{2}\left(  \frac{1}{2}+\frac{1}{2}+1\right)  \right]  \nonumber\\
&  =4\left[  -6R_{\mu\nu\rho\sigma}^{2}-10g_{1}^{2}B_{\mu\nu}^{2}-6g_{2}%
^{2}\left(  W_{\mu\nu}^{\alpha}\right)  ^{2}-6g_{3}^{2}\left(  V_{\mu\nu}%
^{m}\right)  ^{2}\right]  \nonumber
\end{align}
We also have
\begin{align}
\frac{1}{6}\text{\textrm{Tr}}\left(  E+\frac{1}{5}R\right)  _{;\mu}^{\,\,;\mu}
&  =\frac{4}{6}\left[  -24R-4a\overline{H}H-2c\sigma^{2}+\frac{96}{5}R\right]
_{;\mu}^{\,\,;\mu}\\
&  =-4\left[  \frac{4}{5}R+\frac{2}{3}a\overline{H}H+\frac{1}{3}c\sigma
^{2}\right]  _{;\mu}^{\,\,;\mu}\nonumber
\end{align}
The first two Seely-de Witt coefficients are, first for $a_{0}$
\begin{align}
a_{0} &  =\frac{1}{16\pi^{2}}%
{\displaystyle\int}
d^{4}x\sqrt{g}\text{Tr}\left(  1\right)  \\
&  =\frac{1}{16\pi^{2}}\left(  4\right)  \left(  32\right)  \left(  3\right)
{\displaystyle\int}
d^{4}x\sqrt{g}\nonumber\\
&  =\frac{24}{\pi^{2}}%
{\displaystyle\int}
d^{4}x\sqrt{g}\nonumber
\end{align}
then for $a_{2}:$%
\begin{align}
a_{2} &  =\frac{1}{16\pi^{2}}%
{\displaystyle\int}
d^{4}x\sqrt{g}\text{\textrm{Tr}}\left(  E+\frac{1}{6}R\right)  \\
&  =\frac{1}{16\pi^{2}}%
{\displaystyle\int}
d^{4}x\sqrt{g}\left(  \left(  R(-96+64\right)  -16a\overline{H}H-8c\,\sigma
^{2}\right)  \nonumber\\
&  =-\frac{2}{\pi^{2}}%
{\displaystyle\int}
d^{4}x\sqrt{g}\left(  R+\frac{1}{2}a\overline{H}H+\frac{1}{4}c\,\sigma
^{2}\right)  \nonumber
\end{align}
With this information we can now compute the Seeley-de Witt coefficient
$a_{4}$%
\begin{equation}
a_{4}=\frac{1}{16\pi^{2}}%
{\displaystyle\int}
d^{4}x\sqrt{g}\text{\textrm{Tr}}\left(  \frac{1}{360}\left(  5R^{2}-2R_{\mu
\nu}^{2}+2R_{\mu\nu\rho\sigma}^{2}\right)  1+\frac{1}{2}\left(  E^{2}+\frac
{1}{3}RE+\frac{1}{6}\Omega_{\mu\nu}^{2}\right)  \right)
\end{equation}
and where we\ have omitted the surface terms. Thus%
\begin{align}
\frac{1}{2}\text{\textrm{Tr}}\left(  E^{2}+\frac{1}{3}RE+\frac{1}{6}%
\Omega_{\mu\nu}^{2}\right)   &  =4\left[  5g_{1}^{2}B_{\mu\nu}^{2}+3g_{2}%
^{2}\left(  W_{\mu\nu}^{\alpha}\right)  ^{2}+3g_{3}^{2}\left(  V_{\mu\nu}%
^{m}\right)  ^{2}+3R^{2}+aR\overline{H}H\right.  \\
&  +\frac{1}{2}cR\sigma^{2}+2b\left(  \overline{H}H\right)  ^{2}+2a\left\vert
\nabla_{\mu}H_{a}\right\vert ^{2}+4e\overline{H}H\sigma^{2}+d\,\sigma
^{4}+c\left(  \partial_{\mu}\sigma\right)  ^{2}\nonumber\\
&  -\frac{1}{3}R\left(  12R+2a\overline{H}H+c\,\sigma^{2}\right)  \nonumber\\
&  \left.  -R_{\mu\nu\rho\sigma}^{2}-\frac{5}{3}g_{1}^{2}B_{\mu\nu}^{2}%
-g_{2}^{2}\left(  W_{\mu\nu}^{\alpha}\right)  ^{2}-g_{3}^{2}\left(  V_{\mu\nu
}^{m}\right)  ^{2}\right]  \nonumber\\
&  =4\left[  -R_{\mu\nu\rho\sigma}^{2}-R^{2}+\frac{10}{3}g_{1}^{2}B_{\mu\nu
}^{2}+2g_{2}^{2}\left(  W_{\mu\nu}^{\alpha}\right)  ^{2}+2g_{3}^{2}\left(
V_{\mu\nu}^{m}\right)  ^{2}+\frac{1}{3}aR\overline{H}H\right.  \nonumber\\
&  \left.  +2b\left(  \overline{H}H\right)  ^{2}+2a\left\vert \nabla_{\mu
}H_{a}\right\vert ^{2}+4eH_{a}\overline{H}^{a}\sigma^{2}+d\,\sigma
^{4}+c\left(  \partial_{\mu}\sigma\right)  ^{2}+\frac{1}{6}cR\sigma
^{2}\right]  \nonumber
\end{align}
Thus
\begin{align}
a_{4} &  =\frac{1}{2\pi^{2}}%
{\displaystyle\int}
d^{4}x\sqrt{g}\left[  \frac{1}{30}\left(  5R^{2}-8R_{\mu\nu}^{2}-7R_{\mu
\nu\rho\sigma}^{2}\right)  +\frac{5}{3}g_{1}^{2}B_{\mu\nu}^{2}+g_{2}%
^{2}\left(  W_{\mu\nu}^{\alpha}\right)  ^{2}+g_{3}^{2}\left(  V_{\mu\nu}%
^{m}\right)  ^{2}\right.  \\
&  \qquad\qquad\qquad\qquad+\frac{1}{6}aR\overline{H}H+b\left(  \overline
{H}H\right)  ^{2}\sigma^{2}+a\left\vert \nabla_{\mu}H_{a}\right\vert
^{2}+2e\overline{H}H\,\sigma^{2}+\frac{1}{2}d\,\sigma^{4}\nonumber\\
&  \qquad\qquad\qquad\qquad\left.  +\frac{1}{12}cR\,\sigma^{2}+\frac{1}%
{2}c\left(  \partial_{\mu}\sigma\right)  ^{2}-\frac{2}{5}R_{;\mu}^{\,\,;\mu
}-\frac{a}{3}\left(  \overline{H}H\right)  _{;\mu}^{\,\,;\mu}-\frac{c}%
{6}\left(  \sigma^{2}\right)  _{;\mu}^{\,\,;\mu}\right]  \nonumber
\end{align}
Using the identities%
\begin{align}
R_{\mu\nu\rho\sigma}^{2} &  =2C_{\mu\nu\rho\sigma}^{2}+\frac{1}{3}%
R^{2}-R^{\ast}R^{\ast}\\
R_{\mu\nu}^{2} &  =\frac{1}{2}C_{\mu\nu\rho\sigma}^{2}+\frac{1}{3}R^{2}%
-\frac{1}{2}R^{\ast}R^{\ast}%
\end{align}
where\ $R^{\ast}R^{\ast}=\frac{1}{4}\epsilon^{\mu\nu\rho\sigma}\epsilon
_{\alpha\beta\gamma\delta}R_{\mu\nu}^{\quad\alpha\beta}R_{\rho\sigma}%
^{\quad\gamma\delta}.$
\begin{align}
\frac{1}{30}\left(  5R^{2}-8R_{\mu\nu}^{2}-7R_{\mu\nu\rho\sigma}^{2}\right)
&  =R^{2}\frac{1}{30}\left(  5-\frac{8}{3}-\frac{7}{3}\right)  +\frac{1}%
{30}C_{\mu\nu\rho\sigma}^{2}\left(  -4-14\right)  +\frac{1}{30}R^{\ast}%
R^{\ast}\left(  4+7\right)  \\
&  =-\frac{3}{5}C_{\mu\nu\rho\sigma}^{2}+\frac{11}{30}R^{\ast}R^{\ast
}\nonumber
\end{align}

\section{Appendix E: a concrete example}

\bigskip

We start with a two dimensional example, and on a flat two torus the Dirac
operator with coefficient in a trivial bundle $V$ of dimension $2$. We let
$\sigma_{j}$ be the Pauli matrices acting in $V$ and use the gauge potential
$W_{\mu}=\sigma_{\mu}$. Thus the Dirac is
\begin{equation}
D=\gamma^{\mu}\otimes(D_{\mu}+ig\sigma_{\mu})
\end{equation}
We use the notation $\nabla_{\mu}=D_{\mu}+ig\sigma_{\mu}$ for the covariant
derivative. One has $D_{\mu}=ip_{\mu}$ where the $p_{\mu}$ are the momenta.
The square of $D$ gives two terms
\begin{equation}
D^{2}=1_{S}\otimes\Delta_{2}-E\,,\ \Delta_{2}=(p_{\mu}+g\sigma_{\mu}%
)^{2}\,,\ E=-\gamma_{1}\gamma_{2}\otimes(\nabla_{1}\nabla_{2}-\nabla_{2}%
\nabla_{1})
\end{equation}
We begin by computing the eigenvalues of $\Delta_{2}$. It is given by the
$2\times2$ matrix
\begin{equation}
\left(
\begin{array}
[c]{cc}%
p_{2}^{2}+p_{1}^{2}+2g^{2} & -2ip_{2}g+2p_{1}g\\
2ip_{2}g+2p_{1}g & p_{2}^{2}+p_{1}^{2}+2g^{2}%
\end{array}
\right)
\end{equation}
whose eigenvalues are
\begin{equation}
\left\{  p_{2}^{2}+p_{1}^{2}+2g^{2}-2\sqrt{p_{2}^{2}g^{2}+p_{1}^{2}g^{2}%
},p_{2}^{2}+p_{1}^{2}+2g^{2}+2\sqrt{p_{2}^{2}g^{2}+p_{1}^{2}g^{2}}\right\}
\end{equation}
We compute the asymptotic expansion using the limit of flat space. Thus the
trace of $e^{-t\Delta}$ corresponds to the integral (up to an overall $2\pi$)
\begin{equation}
I=\int_{0}^{\infty}e^{-t(\rho^{2}-2g\rho+2g^{2})}\rho d\rho+\int_{0}^{\infty
}e^{-t(\rho^{2}+2g\rho+2g^{2})}\rho d\rho
\end{equation}
We take $g>0$ and compute the integrals as follows.
\begin{equation}
I_{+}=\int_{0}^{\infty}e^{-t(\rho^{2}-2g\rho+2g^{2})}\rho d\rho=e^{-tg^{2}%
}\int_{-g}^{\infty}e^{-tv^{2}}(v+g)dv
\end{equation}%
\[
=e^{-tg^{2}}\int_{-g}^{\infty}e^{-tv^{2}}vdv+ge^{-tg^{2}}\int_{-g}^{\infty
}e^{-tv^{2}}dv
\]
The first integral is the same (since $e^{-tv^{2}}v$ is odd) as
\begin{equation}
e^{-tg^{2}}\int_{g}^{\infty}e^{-tv^{2}}vdv=\frac{e^{-2tg^{2}}}{2t}%
\end{equation}
The second integral is expressed using the error function
\begin{equation}
\mathrm{Erf}(u)=\frac{2}{\sqrt{\pi}}\int_{0}^{u}e^{-v^{2}}dv
\end{equation}
One has $\mathrm{Erf}(\infty)=1$ and the second integral is
\begin{equation}
ge^{-tg^{2}}\int_{-g}^{\infty}e^{-tv^{2}}dv=ge^{-tg^{2}}\frac{\sqrt{\pi}%
}{2\sqrt{t}}\left(  1+\mathrm{Erf}(g\sqrt{t})\right)
\end{equation}
Thus one has
\begin{equation}
I_{1}=\frac{e^{-2tg^{2}}\left(  1+e^{tg^{2}}\sqrt{\pi}\sqrt{t}g\left(
1+\text{Erf}(g\sqrt{t})\right)  \right)  }{2t}%
\end{equation}
Next one has
\begin{equation}
I_{2}=\int_{0}^{\infty}e^{-t(\rho^{2}+2g\rho+2g^{2})}\rho d\rho=e^{-tg^{2}%
}\int_{g}^{\infty}e^{-tv^{2}}(v-g)dv
\end{equation}%
\[
=e^{-tg^{2}}\int_{g}^{\infty}e^{-tv^{2}}vdv-ge^{-tg^{2}}\int_{g}^{\infty
}e^{-tv^{2}}dv
\]%
\[
=\frac{e^{-2tg^{2}}}{2t}-ge^{-tg^{2}}\frac{\sqrt{\pi}}{2\sqrt{t}}\left(
1-\mathrm{Erf}(g\sqrt{t})\right)
\]
Thus one has
\begin{equation}
I_{2}=\frac{e^{-2tg^{2}}\left(  1-e^{tg^{2}}\sqrt{\pi}\sqrt{t}g\left(
1-\mathrm{Erf}(g\sqrt{t})\right)  \right)  }{2t}%
\end{equation}
which shows that $I_{2}(g)=I_{1}(-g)$ since $\mathrm{Erf}$ is an odd function.
One thus gets
\begin{equation}
I=I_{1}+I_{2}=\frac{e^{-2tg^{2}}}{t}+ge^{-tg^{2}}\frac{\sqrt{\pi}}{\sqrt{t}%
}\mathrm{Erf}(g\sqrt{t})
\end{equation}
One has the Taylor expansion
\begin{equation}
\frac{\sqrt{\pi}}{2}\mathrm{Erf}(u)=\sum_{0}^{\infty}(-1)^{n}\frac{u^{2n+1}%
}{n!(2n+1)}%
\end{equation}
which gives the expansion
\begin{equation}
I=\frac{1}{t}-\frac{2g^{4}t}{3}+\frac{8g^{6}t^{2}}{15}-\frac{26g^{8}t^{3}%
}{105}+\frac{16g^{10}t^{4}}{189}+O[t]^{9/2}%
\end{equation}
Thus this gives the following formula for the scalar invariants
\begin{equation}
\label{Laplacian}a_{0}(x,\Delta_{2})=\frac{1}{4\pi}\dim V\,,\ a_{2}%
(x,\Delta_{2})=0\,,\ a_{4}(x,\Delta_{2})=\frac{1}{4\pi}\dim V(-\frac{2g^{4}%
}{3})
\end{equation}
and since $\dim V=2$ and the dimension of spinors is $2$ in dimension $2$ one
gets
\begin{equation}
a_{0}(x,1_{S}\otimes\Delta_{2})=\frac{1}{4\pi}4=\frac{1}{\pi}\,,\ a_{4}%
(x,1_{S}\otimes\Delta_{2})=\frac{1}{4\pi}4(-\frac{2g^{4}}{3})=-\frac{2g^{4}%
}{3\pi}%
\end{equation}
We now need to add the contribution coming from $E$. Note that if one has
\begin{equation}
\mathrm{Trace}(e^{-t(1_{S}\otimes\Delta_{2})})\sim a_{0}t^{-1}+a_{4}t+\ldots
\end{equation}
and if $E$ is a scalar, one gets
\begin{equation}
\mathrm{Trace}(e^{-t(1_{S}\otimes\Delta_{2}-E)})\sim a_{0}t^{-1}+a_{0}%
E+(a_{4}+a_{0}\frac{E^{2}}{2})t+\ldots
\end{equation}
In our case $E$ is not a scalar, one has
\begin{equation}
E=\gamma_{1}\gamma_{2}\otimes(\nabla_{1}\nabla_{2}-\nabla_{2}\nabla_{1}%
)=g^{2}\gamma_{1}\gamma_{2}\otimes(2i\sigma_{3})
\end{equation}
and $E^{2}=4g^{4}\times1_{S\otimes V}$ is a multiple of the identity operator.
The trace of $E$ vanishes and the correction of the $a_{4}$ is the same as if
$E^{2}=4g^{4}$. Thus the relevant combination is
\begin{equation}
a_{4}+a_{0}\frac{E^{2}}{2}=a_{4}+2g^{4}a_{0}=a_{0}\left(  2g^{4}+\frac{a_{4}%
}{a_{0}}\right)
\end{equation}
Now in our case we have $a_{4}=a_{0}(-\frac{2g^{4}}{3})$ and thus
\begin{equation}
2g^{4}+\frac{a_{4}}{a_{0}}=2g^{4}-\frac{2g^{4}}{3}=\frac{4g^{4}}{3}%
\end{equation}
which gives
\begin{equation}
a_{0}(x,D^{2})=\frac{1}{\pi}\,,\ a_{4}(x,D^{2})=\frac{1}{\pi}\frac{4g^{4}}{3}%
\end{equation}
To obtain a $4$-dimensional example we take the product by the flat Dirac in
two dimensions, whose expansion gives
\begin{equation}
\mathrm{Trace}(e^{-tD_{2}^{2}})\sim\frac{2}{4\pi t}%
\end{equation}
where the $2$ comes from the dimension of spinors. Thus for the $4$%
-dimensional Dirac $D_{4}$ with coefficients in the two dimensional trivial
bundle $V$ and connection whose first two components are the $\sigma_{\mu}$
one gets the heat expansion
\begin{equation}
a_{0}(x,D_{4}^{2})=\frac{1}{2\pi^{2}}\,,\ a_{4}(x,D_{4}^{2})=\frac{1}{2\pi
^{2}}\frac{4g^{4}}{3}%
\end{equation}
In fact one obtains the full list of the coefficients $a_{n}$ and the
expansion
\begin{equation}
\frac{1}{2\pi^{2}}\left(  \frac{1}{t^{2}}+\frac{4g^{4}}{3}+\frac{8g^{6}t}%
{15}-\frac{32g^{8}t^{2}}{35}+\frac{1088g^{10}t^{3}}{945}-\frac{9088g^{12}%
t^{4}}{10395}+O[t]^{9/2}\right)
\end{equation}
The above concrete example allows one to check directly that the coefficient
of the $\Omega_{\mu\nu}\Omega_{\mu\nu}$ term in $a_{4}$ is $\frac{1}{12}$
(multiplied by the normalization factor $(4\pi)^{-m/2}$). Indeed, in the
example of the two dimensional Laplacian $\Delta_{2}$, the term in $\frac
{1}{4\pi}\dim V(-\frac{2g^{4}}{3})$ is
\begin{equation}
\label{a4}\frac{2}{4\pi} (-\frac{2g^{4}}{3})= \frac{1}{4\pi} \frac{1}{12}
\mathrm{Trace}(\Omega_{\mu\nu}\Omega_{\mu\nu})
\end{equation}
since there are two $\Omega_{\mu\nu}\Omega_{\mu\nu}$ each equal to $-4g^{2}$,
as $\nabla_{1}\nabla_{2}-\nabla_{2}\nabla_{1}=-2ig^{2}\sigma_{3}$. Thus
$\mathrm{Trace}(\Omega_{\mu\nu}\Omega_{\mu\nu})=-2\times2\times4 g^{4}=-16
g^{4}$. With this one can check directly the coefficients of the $a_{6}$ terms
which we have used. The coefficient of the $E^{3}$ term is $\frac16$
(multiplied by the normalization factor $(4\pi)^{-m/2}$) as is clear by taking
$E$ to be a constant. The coefficient of the term $E\Omega^{2}$ is the same as
the coefficient of the $\Omega^{2}$ term, as is again seen by taking $E$ to be
a scalar and multiplying the two series $(t^{-2}a_{0} + t^{-1}a_{2}+a_{4}+ t
a_{6}+....)(1+tE+t^{2}E^{2}/2+t^{3}E^{3}/6+...)$.

\begin{acknowledgement}
The work of AHC is supported in part by the National Science Foundation grant Phys-0854779.
\end{acknowledgement}

\end{document}